\documentclass[aps,reprint, superscriptaddress, longbibliography]{revtex4-2}

\usepackage{dcolumn}
\usepackage{graphicx}
\usepackage{bm}
\usepackage{amssymb}
\usepackage{amsmath}
\usepackage{hyperref}
\usepackage{comment}
\usepackage{mathtools}
\usepackage{xcolor}
\usepackage{float}
\usepackage{xcolor}
\usepackage[define-L-C-R]{nicematrix}
\usepackage{mathrsfs}
\usepackage{placeins}
\usepackage{multirow}
\hypersetup{colorlinks=true, citecolor=blue,linkcolor=blue,urlcolor=blue}

\newcommand\Eq[1]{Eq.~(\ref{#1})}
\newcommand\Fig[1]{Fig.~\ref{#1}}

\newcommand\white[1]{\color{white}#1}

\begin{document}


\title{Symmetry-protected Topological Phases in Spinful Bosons with a Flat Band
}



\author{Hong Yang}
\author{Hayate Nakano}
\affiliation{
Department of Physics, Graduate School of Science, The University of Tokyo, 7-3-1 Hongo, Bunkyo-ku,  Tokyo 113-0033, Japan
}

\author{Hosho Katsura}
\affiliation{
Department of Physics, Graduate School of Science, The University of Tokyo, 7-3-1 Hongo, Bunkyo-ku, Tokyo 113-0033, Japan
}
\affiliation{
Institute for Physics of Intelligence, The University of Tokyo, 7-3-1 Hongo, Bunkyo-ku, Tokyo 113-0033, Japan
}
\affiliation{
Trans-scale Quantum Science Institute, The University of Tokyo, 7-3-1 Hongo, Bunkyo-ku, Tokyo 113-0033, Japan
}


\date{\today}

\begin{abstract}

We theoretically demonstrate that interacting symmetry-protected topological (SPT) phases can be realized with ultracold spinful bosonic atoms loaded on the lattices which have a flat band at the bottom of the band structure. 
Ground states of such systems 
are not conventional Mott insulators in the sense that
the ground states possess not only spin
fluctuations but also non-negligible charge fluctuations.
The SPT phases in such systems
are determined by both spin and charge fluctuations 
at zero temperature. 
We find that the many-body ground states 
of such systems can be exactly obtained 
in some special cases, and these exact ground states
turn out to serve as 
representative states of the SPT phases.
As a concrete example,
we demonstrate that 
\mbox{spin-1} bosons on a sawtooth chain can be in 
an SPT phase protected by
$\mathbb{Z}_2 \times \mathbb{Z}_2$ spin rotation symmetry
or time-reversal symmetry,
and this SPT phase is a result of spin fluctuations.
We also show that \mbox{spin-3} bosons on a kagome lattice can be in an SPT
phase protected by $D_2$ point group symmetry,
but this SPT phase is, however, a 
result of charge fluctuations.



\end{abstract}

\pacs{}

\maketitle

\section{Introduction}
\mbox{Symmetry-protected topological (SPT)} phases refer to the quantum phases of
those
short-range entangled ground states that can never be smoothly deformed into 
product states
while preserving certain symmetry.
On the other hand, a ground state is classified into a trivial phase if it can be smoothly deformed into 
a product state even when certain symmetry is imposed~\cite{zeng2019quantum}.
A product state stands for a 
tensor product of microscopic states
and possesses no quantum entanglement.
In contrast,
entanglement in the SPT phases 
cannot be smoothly eliminated 
when preserving the symmetry.
The Affleck-Kennedy-Lieb-Tasaki (AKLT) models
provide great insight into the 
SPT phases of interacting bosonic systems.
The AKLT models are a class of 
quantum spin models that can be defined
on arbitrary lattices~\cite{AKLT1987,AKLT1988}.
The models have exact and unique ground states, known as
the valance-bond-solid (VBS) states.
In a simple one dimensional (1D) chain, 
the spin-1 VBS state (i.e., the ground state of the \mbox{spin-1} AKLT model)
represents an SPT phase protected by
any of the following symmetries~\cite{PTBO_2010, PTBO_2012, PhysRevB.80.155131, Tasaki2020}:
(a) $\mathbb{Z}_2 \times \mathbb{Z}_2$ spin rotation symmetry,
(b) time-reversal symmetry,
and (c) inversion symmetry.
This SPT phase is often called the \textit{Haldane phase}.
This Haldane phase is also characterized by a
nonlocal order parameter---the spin string order parameter, 
which quantifies the hidden antiferromagnetic order in the 1D spin-1 VBS state~\cite{Kennedy_Tasaki, Oshikawa_1992, PTBO_2012, Tasaki2020}.
In certain two and higher dimensional lattices,
integer-spin VBS states 
can be in SPT phases if either
translation symmetry or crystalline symmetry
is involved, as we will discuss later.

Ultracold atoms/molecules in optical lattices serve as an ideal platform for realizing topological quantum phases
due to the high tunability of interactions, the viability of building various lattice structures, 
and the feasibility of directly measuring nonlocal order parameters~\cite{Science2017, Science2019}.
Motivated by recent experimental progress,
many theoretical predictions about the existence of the Haldane phase in
lattice systems of 
bosons~\cite{EBHM_PhysRevLett.97.260401, EBMH_PhysRevB.77.245119, EBHM_Rossini_2012, Realizing_spin0, EBHM_PhysRevLett.110.265303, PhysRevLett.113.020401, PhysRevLett.118.120401, PhysRevB.83.155110, PhysRevA.99.012122, SOC_PhysRevLett.115.195302, 
PhysRevB.95.165131, Perturb_BLBQ} 
and fermions~\cite{PhysRevB.81.020408, 
PhysRevB.91.075121, 
PhysRevB.91.155128, 
PhysRevB.92.041120, 
PhysRevB.96.155133, 
PhysRevB.98.045128, 
PhysRevB.99.054414, 
PhysRevLett.122.106402, 
Montorsi2019homogeneous}
have been made.

 Alkali-metal atoms carry integer spins and are thus 
often treated as spinful bosons in experiments~\cite{KAWAGUCHI2012253,* RevModPhys.85.1191}.
Spinful bosons in optical lattices typically have
both spin and charge degrees of freedom (DOF).
Free from the Pauli exclusion principle,
one major difficulty of theoretically studying 
the systems of many-body
spinful bosons lies in their immense Hilbert spaces
(i.e., a huge number of DOF).
Therefore, except for very few rigorous results~\cite{Katsura_Tasaki, PhysRevLett.122.053401},
various approximations or constraints have been employed to simplify the problem 
(i.e., to reduce the Hilbert space dimension
by freezing some DOF).
In particular, to theoretically investigate the 
Haldane phase
of bosonic atoms in 1D,
there have been two main approaches.
One is to study the effective spin Hamiltonians
by focusing on the conventional Mott insulating limit where the charge DOF are frozen~\cite{Perturb_BLBQ}.
For example, the system of Mott insulating spin-1 bosons is effectively described by the
bilinear-biquadratic (BLBQ) model, whose ground state
in 1D has been known to exhibit the Haldane phase 
in a wide parameter region~\cite{Lauchli_BLBQ}.
The other approach is to study models that describe itinerant but spinless bosons.
(A system is said to be itinerant if it has charge DOF.)
In the itinerant case, 
it is generally believed that a sufficiently strong long-range (repulsive) interaction is indispensable
for triggering the SPT phase~\cite{EBHM_PhysRevLett.97.260401, EBMH_PhysRevB.77.245119, EBHM_Rossini_2012, Realizing_spin0, EBHM_PhysRevLett.110.265303, PhysRevLett.113.020401, PhysRevLett.118.120401, PhysRevB.83.155110, PhysRevA.99.012122,PhysRevB.95.165131}.
The mechanism is as follows.
At the filling of one spinless boson per site on average,
if we truncate the particle number on each site to $n = 0, 1, 2$,
one can define pseudo-spin as $\mathcal{S}^z := n-1$,
thus resulting in an effective spin-1 model,
where the long-range repulsion acts as an anisotropic spin exchange interaction~\cite{PTBO_2010, EBMH_PhysRevB.77.245119}. 
However, among bosonic alkali-metal atoms, 
although a relatively strong dipole-dipole interaction plays the role of long-range interaction
in certain situations~\cite{KAWAGUCHI2012253,* RevModPhys.85.1191},
the dipole-dipole interaction is usually
much weaker than 
the short-range $s$-wave collision, and
thus the long-range interaction is typically 
negligible in many experiments~\cite{KAWAGUCHI2012253,* RevModPhys.85.1191}.


In short, despite the fact that itinerant, spinful, and short-range interacting bosonic atoms 
are very common in experiments, 
due to the difficulty of theoretically 
dealing with the huge amount of
DOF,
it remains an open question whether the
SPT phases can be realized in such systems.
Moreover, if the answer is yes, 
what kinds of SPT phases can we get?
We address these issues and argue that, 
when there is a flat band
at the bottom of the band structure (which we dub a \textit{bottom flat band}), 
SPT phases can be realized with short-range interacting
bosons that 
possess both \mbox{unfrozen} spin and charge DOF.
As a result, 
the SPT phases in such systems are characterized by
nontrivial spin or charge entanglement. 

A flat band refers to an energy band 
that is independent of the quasimomentum.
Usually, a flat band in an optical lattice 
is the highest band. However, 
by shaking the optical lattices, 
one can invert the sign of hopping~\cite{PhysRevLett.95.260404, Eckardt_2010,PhysRevB.34.3625, PhysRevLett.67.516, Gro_mann_1992,PhysRevLett.78.2932},
and the flat band thus becomes the lowest band. 
Such lattice shaking techniques have been realized
experimentally~\cite{PhysRevLett.99.220403,PhysRevA.79.013611,PhysRevLett.100.190405,PhysRevLett.102.100403, struck2011quantum}.

Single-body eigenstates of a flat band can usually 
be chosen to be strictly localized on 
a finite number of lattice sites. 
Such eigenstates are termed as \textit{compact localized states} (CLSs)~\cite{PhysRevB.99.045107, PhysRevB.95.115309}.
Different CLSs reside in different patches (regions)
of the lattice.
Short-range interaction ($s$-wave collision)
between two bosons
can happen only
when
their wave functions 
have a finite overlap.
(This is natural, because the short-range interaction does not
occur unless two particles are very close to each other.)
At low temperatures, 
boson wave functions
tend to avoid overlapping each other
in order to lower the system's energy.
Let $X$ be a $d$-dimensional lattice with a bottom flat band and $N$ unit cells.
When $N$ \mbox{spin-$f$} bosons are loaded on $X$,
the wave function overlaps can be minimized
if each of the $N$ CLSs hosts a boson.
In other words,
$N$ bosons are distributed into $N$ different patches.
A boson is free to move around within a patch, 
which gives rise to charge fluctuations in the ground state.
On the other hand,
since all the patches (CLSs) are occupied by bosons
(i.e., the whole lattice is fully ``packed" with bosons),
partial overlaps between neighboring wave functions are
inevitable.
We notice an analogy between the Hamiltonian 
that describes the short-range $s$-wave collision among
spin-$f$ bosons and the spin-$f$ AKLT Hamiltonian.
This analogy implies that the wave function overlaps
will not cost energy, 
if the spins of bosons entangle in a clever way 
similar to a \mbox{spin-$f$} VBS state.
(Intuitively, since the $s$-wave collision
is spin-dependent by its nature, 
when the bosons are in a certain spin state,
the collision between them can be avoided
even if the bosons are very close to each other.)
When certain parameters in the Hamiltonian are fine-tuned,
the above configuration (lattice fully packed with CLSs)
becomes the exact and unique ground state, and
the state turns out to serve as a representative state of 
the symmetry-protected phases of the system.
(In this paper, the term ``symmetry-protected phase" 
refers to either SPT or trivial phase.)
We find that the phases are determined by
the spin or charge fluctuations in the ground state.
In this paper,
we find a large class of models whose 
ground states
can be exactly written down
when certain parameters are properly chosen.
Each model has several on-site and crystalline symmetries.
Depending on the symmetry,
these exact ground states
can be in either SPT 
or trivial phases.
In particular, in terms of crystalline symmetries,
charge fluctuations can play a nontrivial role.



This paper will gradually build up 
a general framework on the SPT phases of spinful bosons
with a flat band,
starting from a simple 1D spin-1 model before progressing towards
general dimensions and general spins.
The remainder of this paper is divided into two parts: Sec.~\ref{Sec_spin-1_BHMSC} and Sec.~\ref{Sec_Generalizations}. 
In Sec.~\ref{Sec_spin-1_BHMSC},
we use spin-1 bosons on the 1D sawtooth chain as a concrete example to demonstrate our argument.
The sawtooth chain has two energy bands,
and the bottom one is flat.
We prove that 
when the interaction between \mbox{spin-1} bosons is fine-tuned,
the ground state is unique and can be 
exactly written down.
The proof is based on the fact that
the ground state can be exactly mapped to
the 1D spin-1 VBS state.
This exact ground state turns out to be in
a Haldane phase.
Beyond the fine-tuned case,
based on perturbation theory and numerical calculations,
we confirmed that the Haldane phase exists in 
a rather broad parameter region.
In Sec.~\ref{Sec_Generalizations},
we discuss the SPT phases with a general setup:
short-range interacting 
spin-$f$ bosons on a bottom-flat-band lattice $X$
in $d$ dimension.
Let $|\text{GS}_{f,X}\rangle$ be the many-body ground state.
Let $|\text{VBS}_{f,X'}\rangle$ be the spin-$f$
VBS state defined on a lattice $X'$ (i.e., the ground state of the spin-$f$ AKLT model on $X'$).
With fine-tuned interactions,
$|\text{GS}_{f,X}\rangle$ can be exactly mapped to
$|\text{VBS}_{f,X'}\rangle$,
provided that
the lattice structures of $X$ and $X'$ satisfy a certain relation.
This proves that $|\text{GS}_{f,X}\rangle$ is the exact and unique ground state of the itinerant spin-$f$ model.
The spin fluctuations of $|\text{GS}_{f,X}\rangle$
are inherited from $|\text{VBS}_{f,X'}\rangle$.
Therefore, with respect to the spin rotation symmetry
or the combination of spin rotation and translation symmetry,
the $d$-dimensional
symmetry-protected phase of $|\text{GS}_{f,X}\rangle$ is identical
to that of $|\text{VBS}_{f,X'}\rangle$.
Spins in $|\text{VBS}_{f,X'}\rangle$
are pinned to the lattice sites and cannot move.
However, $|\text{GS}_{f,X}\rangle$ is not a
conventional Mott-insulating state 
(where a fixed number of bosons stay rigidly on each site), i.e.,
\mbox{spin-$f$} bosons in $|\text{GS}_{f,X}\rangle$
have nonvanishing charge fluctuations.
It turns out that
in terms of crystalline symmetries (i.e., point group or space group symmetries),
both spin and charge fluctuations in $|\text{GS}_{f,X}\rangle$ together
determine its symmetry-protected phase.
Hence, 
the crystalline-symmetry-protected phases of $|\text{GS}_{f,X}\rangle$ and $|\text{VBS}_{f,X'}\rangle$
may not be identical,
because the charge fluctuations may play a nontrivial role
in the former state.
For example, as we will show later, 
interacting spin-3 bosons in the kagome lattice 
can be in an SPT phase
protected by the point group $D_2$ or $D_3$,
and this SPT phase is purely 
a consequence of charge fluctuations at zero temperature.

\smallskip

\section{Spin-1 bosons on a sawtooth chain: an example}
\label{Sec_spin-1_BHMSC}

Let us start from a simple but nontrivial model: 
the spin-1 Bose-Hubbard model on the sawtooth chain (BHMSC).
In Sec.~\ref{Sec_Hamiltonian},
we introduce the spin-1 BHMSC
and the 1D spin-1 bilinear-biquadratic (BLBQ) model.
The ground state of the BLBQ model is exactly solvable at the AKLT point.
In Sec.~\ref{Sec_ExactGroundStates},
we prove that 
in a special case where the interaction between spin-1 bosons is fine-tuned,
the ground state sectors of the spin-1 BHMSC and the AKLT model
can be exactly mapped to each other,
which enables us to write down an exact and unique ground state of the spin-1 BHMSC.
This ground state, 
as we will show in Sec.~\ref{Sec_HaldanePhase},
turns out to serve as a representative state of the Haldane phase.
We find that the Haldane phase in this itinerant spin-1 boson system, 
characterized by both nonvanishing spin and charge string order parameters,
is protected by (a) $\mathbb{Z}_2 \times \mathbb{Z}_2$ symmetry or (b) time-reversal symmetry,
but not (c) inversion symmetry.
In Sec.~\ref{Sec_PerturbationTheory},
perturbation theory builds another bridge between the spin-1 BHMSC and the BLBQ model.
In Sec.~\ref{Sec_NumericalAnalysis}, the
phase diagram of the spin-1 BHMSC is investigated
with numerical calculations based on the variational uniform matrix product state (VUMPS) algorithm~\cite{PhysRevB.97.045145, SciPostPhysLectNotes},
which suggests that the system can be in either a gapped Haldane or gapless critical phase.

\subsection{Hamiltonian} \label{Sec_Hamiltonian}

For spin-1 bosons (such as $^7$Li, $^{23}$Na, $^{41}$K, etc.)
in a lattice system,
let $\hat{a}^{\dagger}_{r,\alpha}$ ($\hat{a}_{r,\alpha}$) be the operator that creates (annihilates) 
a boson at lattice site $r$ with magnetic sublevel $\alpha=-1,0,1$.
The on-site spin operator $\hat{\bm{S}}_{r} = (\hat{S}_{r}^x, \hat{S}_{r}^y, \hat{S}_{r}^z)$
is defined as $\hat{S}_{r}^{z} := \sum_{\alpha,\beta} \hat{a}_{r,\alpha}^\dagger S^{z}_{\alpha,\beta} \hat{a}_{r,\beta}$ 
with $S^{z}_{\alpha,\beta} = \alpha  \delta_{\alpha,\beta}$ being the $z$-component of the spin matrix for spin-1
(and similar definitions for $\hat{S}_{r}^{x}$ and $\hat{S}_{r}^{y}$).
We also define $\hat{n}_{r} := \sum_\alpha \hat{a}_{r,\alpha}^\dagger \hat{a}_{r,\alpha}$ which
counts the particle number on site $r$.
Spin-1 atoms in optical lattices are effectively described by 
the spin-1 Bose-Hubbard model~\cite{Imambekov2003, Tsuchiya2004}
\begin{equation}
\begin{split}
	\hat{H} &= \hat{H}_{\text{hop}} + \hat{H}_{\text{int}},\\
	\hat{H}_{\text{hop}} &= -\sum_{  \langle r,r' \rangle } \sum\limits_{ \alpha=-1 }^1   t_{r,r'} \  \hat{a}^{\dagger}_{r,\alpha}\hat{a}_{r',\alpha}  + \sum_{r} V_r \hat{n}_r,\\
	\hat{H}_{\text{int}} &= \sum_{r}^{} \left( g_{0,r} \hat{P}_{r}^{(0)} + g_{2,r} \hat{P}_{r}^{(2)} \right),
\end{split} \label{H}
\end{equation}
where $\hat{H}_{\text{hop}}$ is the single-body Hamiltonian
which contains both hopping and on-site potential terms,
and $\hat{H}_{\text{int}}$ describes the interactions ($s$-wave collisions)
between spin-1 bosonic atoms~\cite{Ho1998,* ohmi1998bose, KAWAGUCHI2012253,* RevModPhys.85.1191}.
There are two kinds of interactions: 
$\hat{P}_{r}^{(S)}$ stands for the projection operator onto the state with total spin $S=0,2$
for a pair of \mbox{spin-1} bosons at site $r$. 
For example, $\hat{P}_{r}^{(0)} = \hat{b}^\dagger_r \hat{b}_r$,
where 
\begin{equation}
	\hat{b}^\dagger_r :=\frac{1}{\sqrt{6}} ( \hat{a}^\dagger_{r,0}  \hat{a}^\dagger_{r,0} - 2 \hat{a}^\dagger_{r,1}  \hat{a}^\dagger_{r,-1} )\end{equation}
creates a spin singlet.
$S=1$ is forbidden 
because two spin-1 bosons on the same site never form a total spin \mbox{$S=1$} state---such a spin state is antisymmetric. 
The projection operators can be explicitly expressed as 
$\hat{P}_{r}^{(0)} = [ -(\hat{\bm{S}}_{r})^2 + (\hat{n}_{r})^2 + \hat{n}_{r} ] / 6$
and 
$\hat{P}_{r}^{(2)} = [ (\hat{\bm{S}}_{r})^2 + 2(\hat{n}_{r})^2 - 4\hat{n}_{r} ] / 6$
\cite{Ho1998,* ohmi1998bose, KAWAGUCHI2012253,* RevModPhys.85.1191}.
The sum of them yields the ``completeness relation":
\begin{equation}
	\hat{P}_{r}^{(0)} + \hat{P}_{r}^{(2)} = \frac{1}{2} \hat{n}_{r} \left( \hat{n}_{r} -1 \right).
	\label{completeness_relation}
\end{equation}
We assume the interaction strength $g_{S,r} \geqslant 0$
as is the case of long-lived alkali-metal spin-1 condensates~\cite{KAWAGUCHI2012253,* RevModPhys.85.1191}; 
$\hat{H}_{\text{int}}$ is thus positive semidefinite.

\begin{figure}
  \centering
  \includegraphics[width=0.47\textwidth]{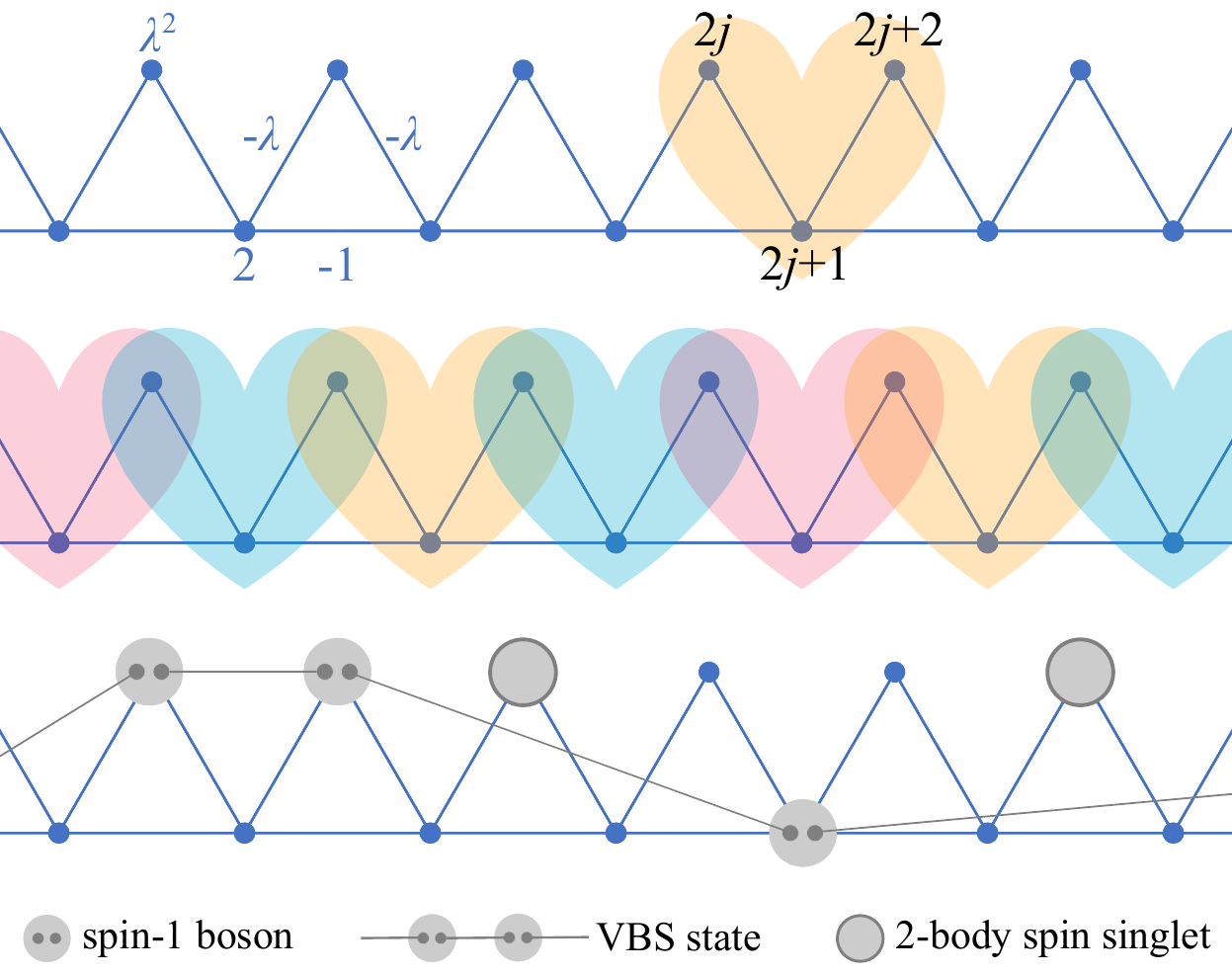}
  \caption{(Upper panel) The sawtooth lattice and one of its zero-energy state.
  The state is localized on three consecutive sites covered by the heart shape.
  The blue characters indicate the values of $t_{r,r'}$ and $V_r$.
  (Middle panel) The state $|\bm{\beta}\rangle$ in \Eq{eq:beta} with a typical choice of $\bm{\beta}$. 
  Three different colors denote three different magnetic sublevels.
  Linearly independent CLSs cover the whole lattice, and two neighboring CLSs overlap on a top site.
  (Lower panel) The ``hidden VBS order" illustrated by a typical component of $|\text{GS}\rangle$ in \Eq{GS}.}
  \label{fig:sawtooth}
\end{figure}

On a sawtooth chain (see \Fig{fig:sawtooth}) with $N$ unit cells ($2N$ sites), 
the single-body Hamiltonian can be written in a compact form as~\cite{PhysRevLett.69.1608, 10.1143/PTP.99.489, PhysRevB.100.214423}
\begin{equation}
\hat{H}_{\text{hop}} = \hat{H}_{\text{saw}} = \sum\limits_{i=1}^N \sum\limits_{\alpha=-1}^1 \hat{A}_{i,\alpha}^\dagger \hat{A}_{i,\alpha}, \label{H_delta_PBC}
\end{equation}
where 
$ \hat{A}_{i,\alpha}^\dagger :=\hat{a}^{\dagger}_{2i-1,\alpha} + \lambda \hat{a}^{\dagger}_{2i,\alpha} + \hat{a}^{\dagger}_{2i+1,\alpha}$
determines the values of $t_{r,r'}$ and $V_r$ in \Eq{H},
and we assume $\lambda\in \mathbb{R}\backslash \{0\}$.
Periodic boundary condition (PBC) has been imposed.
$\hat{H}_{\text{saw}}$ is positive semi-definite,
and it has two energy bands:
a dispersive band with energy $\lambda^2 +2 +2 \cos k >0$ and a flat band with exactly zero energy.
Every eigenstate of the flat band can be chosen to be localized on three sites (see \Fig{fig:sawtooth}):\begin{equation}
	\hat{B}^\dagger_{j, \alpha} := \frac{1}{\sqrt{\lambda^2+2}} (\hat{a}^\dagger_{2j,\alpha} - \lambda \hat{a}^\dagger_{2j+1, \alpha} + \hat{a}^\dagger_{2j+2,\alpha}), \label{B}
\end{equation}
where $\hat{B}^\dagger_{j, \alpha}$ creates a particle in
a zero-energy eigenstate.
In other words, $\hat{B}^\dagger_{j, \alpha}$ is a CLS
creation operator.
An experimental scheme for realizing an 
optical sawtooth chain 
has been proposed~\cite{zhang2015one}.

Note that lattices with a bottom flat band (and CLSs) widely exist; they can actually be constructed systematically, see
Sec.~\ref{Sec_GS_boson_flatband}.

\begin{figure}
	\includegraphics[width=0.35\textwidth]{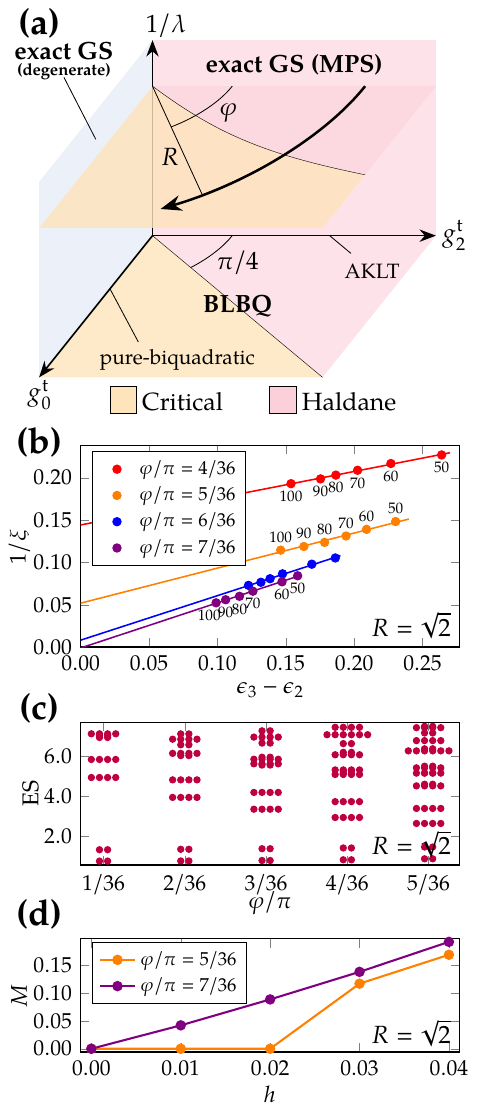}
	\caption{(a) Schematic phase diagram of \mbox{spin-1} bosons on a sawtooth chain 
	 (in the thermodynamic limit \mbox{$N \to \infty$})
	 in the parameter space $(g_0^\text{t}, g_2^\text{t}, 1/\lambda)$. 
	 In the phase diagram we have assumed $g_0^\text{t}, g_2^\text{t}, \lambda \geqslant 0$ and $g_0^\text{b}=g_2^\text{b}= 1/\lambda^2$.
	 In the $g_0^\text{t}=0$ plane, $\hat{H}$ has an exact and unique ground state (GS) given in \Eq{GS}.
	 In the $g_2^\text{t}=0$ plane, the GS is massively degenerate,
	 and the ferromagnetic states in \Eq{ferro_state} are exact ground states.
	 Phase diagram in the $1/\lambda \to 0$ plane, derived by perturbation theory,
	 coincides with the phase diagram of the BLBQ model.
	 Phase diagram in the $1/\lambda =1 $ plane is determined by the numerical calculation based on the VUMPS algorithm.
	 In particular, numerical results along the curved arrow parameterized by $( \sqrt{2} \sin \varphi, \sqrt{2} \cos \varphi, 1)$
	 are shown in (b)--(d).
	 (b) Scaling of the inverse correlation length $1/\xi:= \epsilon_2$ with respect to $\epsilon_3- \epsilon_2$.
	 Numbers near the data points denote the corresponding bond dimensions of each block,
	 see Appendix~\ref{App_MPSansatz}.
	 We can see that a quantum phase transition occurs between $\varphi= 6\pi/36$ and $7\pi/36$.
	 (c) The whole entanglement spectrum (ES) in the Haldane phase region shows the even-fold degeneracy.
	 For clarity, here we present only the lowest part of the ES.
	 (d) Magnetization $M$ with respect to the applied magnetic field $h$ in the $z$-direction.
	 In (c) and (d), the bond dimension of each block is $50$.
	 }
	\label{fig:phase_diagram}
\end{figure}

From now on, the total particle number on the sawtooth chain is assumed to be
the same as the number of unit cells $N$.
For simplicity,
we also assume translation symmetry:
$g_{S,r} \equiv g_S^{\text{t}}$
for top sites ($r=\text{even}$) and $g_{S,r} \equiv g_S^{\text{b}}$ for bottom sites ($r=\text{odd}$).
The phase diagram of the spin-1 BHMSC with respect to $(g_0^\text{t}, g_2^\text{t}, 1/\lambda)$ 
is shown in~\Fig{fig:phase_diagram}(a).

For later purposes, we also introduce the 1D \mbox{spin-1} BLBQ model with PBC, whose Hamiltonian is given by~\cite{Lauchli_BLBQ}
\begin{equation}
	\begin{split}
		\hat{H}_{\text{BLBQ}} &=  \sum_{j=1 }^N \left( \tilde{g}_0 \hat{P}_{j, j+1}^{(0)} + \tilde{g}_1 \hat{P}_{j, j+1}^{(1)} + \tilde{g}_2 \hat{P}_{j, j+1}^{(2)} \right),  \\
	& =J \sum_{ j=1  }^N \bigg[ \cos\theta(\hat{\bm{S}}_j \cdot \hat{\bm{S}}_{j+1} ) 
	+ \sin\theta  (\hat{\bm{S}}_j \cdot \hat{\bm{S}}_{j+1})^2 \bigg] +c, \label{BLBQ}
	\end{split}
\end{equation}
where $\hat{P}_{j, j+1}^{(F)}$ projects the state of two neighboring sites onto the state with total spin $F=0,1,2$. 
Spin operators $\{ \hat{\bm{S}}_{j} \}$ act on the spin-chain Hilbert space spanned by the $S^z$-basis
$\{ |\psi_{\bm{\alpha}} \rangle := | \alpha_1, \alpha_2,..., \alpha_N \rangle \}$.
Parameters $(J\cos\theta,J\sin\theta,c)$ with $J>0$ linearly depend on $(\tilde{g}_0, \tilde{g}_1, \tilde{g}_2)$.
This model is in the Haldane phase when $-\pi/4 < \theta <\pi/4$,
while it is in the critical phase when $\pi/4 \leqslant \theta \leqslant \pi/2$.
At $\theta=\arctan(1/3)$ and $\pi/2$, the model is particularly known as the AKLT model
and the pure-biquadratic model~\cite{Parkinson_1987, Parkinson_1988}, respectively.
For the 1D \mbox{spin-1} AKLT model
\begin{equation}
	\hat{H}_{\text{AKLT}} = \sum_{j=1 }^N  \tilde{g}_2 \hat{P}_{j, j+1}^{(2)}   \quad\quad (\tilde{g}_2>0),
	\label{AKLT_H}
\end{equation}
the ground state $|\text{VBS}\rangle$ under PBC is unique and can be exactly written as a matrix product state (MPS)
\begin{equation}
	|\text{VBS}\rangle = \sum_{\alpha_1,...,\alpha_N} \text{Tr}( M^{\alpha_1} M^{\alpha_2} ... M^{\alpha_N})  |\psi_{\bm{\alpha}} \rangle,
	\label{VBS_AKLT}
\end{equation}
where $M^{\pm 1}:=\mp\sqrt{2}\sigma^{\pm}$, $M^{0}:=\sigma^{z}$, and $\sigma^{\pm, z}$ are Pauli matrices.
$|\text{VBS}\rangle$ in \Eq{VBS_AKLT} is known as the 1D \mbox{spin-1} VBS state, 
which can be graphically represented as in \Fig{fig:VBS_states}(a).
See Sec.~\ref{Sec_VBS} for details.

We notice the analogy between the
$s$-wave collision Hamiltonian $\hat{H}_{\text{int}}$
and the BLBQ Hamiltonian $\hat{H}_{\text{BLBQ}}$.
This will help us the find the exact ground state and
the SPT phase in the spinful, itinerant, and short-range
interacting bosonic systems.


\smallskip

\subsection{Exact ground states} \label{Sec_ExactGroundStates}

Since both $\hat{H}_{\text{saw}}$ and $\hat{H}_{\text{int}}$ are positive semi-definite,
a zero-energy ground state of $\hat{H}$, if exists, must satisfy
(i) $ \hat{H}_{\text{saw}}  | \text{GS} \rangle = 0$ and (ii) $\hat{H}_{\text{int}}  | \text{GS} \rangle =0$.
In accordance with (i), there must be
\begin{equation}
	| \text{GS} \rangle  
	= \sum\limits_{\sum\limits_{j,\mu} n_{j,\mu}=N} x_{\bm{n}} \Bigg( \prod\limits_{j=1}^N \prod\limits_{\mu=-1}^{1} (\hat{B}^\dagger_{j,\mu})^{n_{j,\mu}} \Bigg) |\text{vac}\rangle, \label{gs_general}
\end{equation}
where $x_{\bm{n}} \in \mathbb{C}$, $\bm{n} = (..., n_{i,1} ,n_{i,0}, n_{i,-1}, n_{i+1,1}, ...)$,
and $|\text{vac}\rangle$ is the vacuum state.
Assume $g_0^{\text{b}}, g_2^{\text{b}} >0$,
according to \Eq{completeness_relation} and the positive semidefiniteness of $\hat{P}_{r}^{(S)}$,
one can conclude that
\begin{equation}
\begin{split}
&\hat{H}_{\text{int}} | \text{GS} \rangle = 0 \\
\iff &\left( g_0^{\text{b}} \hat{P}_{2j+1}^{(0)} + g_2^{\text{b}} \hat{P}_{2j+1}^{(2)} \right) | \text{GS} \rangle =0, \ \forall j \\
	\iff & \hat{n}_{2j+1} \left( \hat{n}_{2j+1} -1 \right) | \text{GS} \rangle=0, \ \forall j\\
	\iff &\hat{a}_{2j+1,\alpha} \hat{a}_{2j+1,\beta} | \text{GS} \rangle =0, \ \forall j, \alpha, \beta \\
	\Longrightarrow \ & x_{\bm{n}}=0, \forall \bm{n}  \text{ s.t. } n_{j,+1} + n_{j,0} + n_{j,-1} > 1.
\end{split} \label{uniqueness_of_FPS}
\end{equation}
Equation~(\ref{gs_general}) thus reduces to 
\begin{equation}
	| \text{GS} \rangle = \sum_{\bm{\beta}} y_{\bm{\beta}} |\bm{\beta}\rangle, \label{GS_beta}
\end{equation}
where
$y_{\bm{\beta}} \in \mathbb{C}$, $\bm{\beta} = (\beta_1,...,\beta_N)$, and
\begin{equation}
	|\bm{\beta} \rangle := \Bigg( \prod\limits_{j=1}^N  \hat{B}^\dagger_{j,\beta_j} \Bigg)  |\text{vac}\rangle. \label{eq:beta}
\end{equation}
A typical $|\bm{\beta} \rangle$ is illustrated in \Fig{fig:sawtooth}.
Note that $ |\bm{\beta} \rangle$'s are linearly independent but not orthonormal because 
$K_{jj'} := [\hat{B}_{j, \mu}, \hat{B}^\dagger_{j', \mu}] \neq \delta_{jj'}$.
We define
the ``dual operator" of $\hat{B}_{j, \mu}$ as~\cite{mielke1993}
\begin{equation}
	\hat{C}_{j,\mu} := \sum_{j'} (K^{-1})_{jj'} \hat{B}_{j',\mu} \label{C}
\end{equation}
such that
$[\hat{C}_{j,\mu}, \hat{B}^\dagger_{j', \mu'}] = \delta_{jj'} \delta_{\mu\mu'}$.
Further defining
\begin{equation}
	\langle \widetilde{\bm{\alpha}} | := \langle \text{vac}| \Bigg( \prod_{j=1}^N  \hat{C}_{j,\alpha_j} \Bigg)
\end{equation}
such that $\langle \widetilde{\bm{\alpha}}  | \bm{\beta} \rangle = \delta_{\bm{\alpha} \bm{\beta}}$,
eigenequation $\hat{H} | \text{GS} \rangle = 0$ then implies the matrix equation
$\sum_{\bm{\beta}}  \langle \widetilde{\bm{\alpha}} | \hat{H} | \bm{\beta} \rangle  \   y_{\bm{\beta}} = 0$.
Impressively,
explicit calculation shows that
\begin{equation}
 \langle \widetilde{\bm{\alpha}} | \hat{H} | \bm{\beta} \rangle = \langle \widetilde{\bm{\alpha}} | \hat{H}_{\text{int}} | \bm{\beta} \rangle =  \langle \psi_{\bm{\alpha}}  | \hat{H}_{\text{BLBQ}} | \psi_{\bm{\beta}} \rangle,  \label{Hab=Hab}
 \end{equation} 
provided that we take $\tilde{g}_1 = 0$ and $\tilde{g}_S = 2g^{\text{t}}_S d /(\lambda^2+2)$ in \Eq{BLBQ},
where 
$d>0$ is a coefficient depending on the inverse matrix $K^{-1}$ (matrix $K$ is always invertable). 
Equation~(\ref{Hab=Hab}) indicates that there is a one-to-one correspondence
between the zero-energy states of $\hat{H}$ and $\hat{H}_{\text{BLBQ}}$.
Note that such correspondence does not hold for
eigenstates with nonzero energy,
because $\hat{P}^{(S)}_{r=\text{odd}}  | \bm{\beta} \rangle = 0$
implies that nonzero-energy eigenstates cannot be purely spanned by $\{ | \bm{\beta} \rangle \}$.
It is known that in the following two cases, 
$\hat{H}_{\text{BLBQ}}$ possesses zero-energy ground states:
(1) AKLT point ($\tilde{g}_0 = \tilde{g}_1 = 0$, $\tilde{g}_2>0$)
and (2) pure-biquadratic point ($\tilde{g}_1 = \tilde{g}_2 = 0$, $\tilde{g}_0>0$).

Case (1) maps to $g^{\text{t}}_0=0$ and $ g^{\text{t}}_2>0$ for $\hat{H}$. 
In this case, 
the ground state of $\hat{H}$ is unique, 
which follows from the uniqueness of the ground state of the AKLT model~\cite{AKLT1987,AKLT1988}.
Despite the fact that the $|\text{GS}\rangle$ in \Eq{GS_beta}
is not represented in an orthonormal basis,
the coefficient $y_{\bm{\beta}}$ is identical to that of the 1D VBS state in \Eq{VBS_AKLT}: 
\begin{equation}
	y_{\bm{\beta}} = \text{Tr}(M^{\beta_1} M^{\beta_2}...M^{\beta_N}). \label{y_beta}
\end{equation}
Further expanding $\hat{B}^\dagger_{j,\beta_j}$ in terms of $\hat{a}^{\dagger}$'s,
we can see that $|\text{GS}\rangle$ is a
superposition of states of the form
\begin{equation}
	(-\lambda)^{b} \sum_{\bm{\beta}} \text{Tr}(M^{\beta_1}  M^{\beta_2}...M^{\beta_N}) 
	\hat{a}^\dagger_{r_1,\beta_1} 
	\hat{a}^\dagger_{r_2,\beta_2}...
	\hat{a}^\dagger_{r_N,\beta_N} \label{hidden_VBS_sawtooth_chain},
\end{equation}
where the integer $b$ depends on $\{r_1,...,r_N\}$.
In \Eq{hidden_VBS_sawtooth_chain},
we note that as long as two particles occupy the same top site $\ell$,
there is the identity 
\begin{equation}
	\sum_{\beta_j,\beta_{j+1}} M^{\beta_j} M^{\beta_{j+1}}
    \hat{a}^\dagger_{\ell, \beta_{j}} \hat{a}^\dagger_{\ell, \beta_{j+1}} = \sqrt{6} \ \hat{b}^\dagger_{\ell}  I_{2}, \label{singlet_identity}
\end{equation}
where $\ell=2j$ or $2j+2$ 
and $I_{2}$ is a 2-by-2 identity matrix.
Equation~(\ref{singlet_identity}) implies that \Eq{hidden_VBS_sawtooth_chain}
has ``hidden VBS order", i.e., if we ignore all the vacant sites
and sites occupied by a spin singlet, the remaining bosons form a perfect VBS state, see~\Fig{fig:sawtooth}.
This enables us to express $|\text{GS}\rangle$ in an orthonormal Fock basis as
\begin{equation}
	|\text{GS}\rangle 	
	= \sum_{\tau_1,...,\tau_{2N}=-1}^3 \text{Tr} \left( 
	 \prod\limits_{j=1}^{N} ( F^{\tau_{2j-1}} E^{\tau_{2j}}  )
	\right) 
	\left( \prod\limits_{r=1}^{2N} \hat{d}_{r,\tau_r}^\dagger \right)
	| \text{vac}  \rangle
	,  \label{GS}
\end{equation}
where $\hat{d}_{r,\tau}^\dagger := \hat{a}_{r, \tau}^\dagger$ for $\tau=-1,0,1$,
while $\hat{d}_{r,2}^\dagger := \hat{b}_{r}^\dagger$ and $\hat{d}_{r,3}^\dagger := 1$, and
\begin{equation}
\begin{split}
	\sum_{\tau=-1}^3 F^{\tau} \hat{d}_{r,\tau}^\dagger &= \frac{1}{\sqrt{\lambda^2+2}}
	\left(
	\begin{matrix} 
	I_{2} & -\lambda\sum_{\alpha} M^\alpha \ \hat{a}^\dagger_{r,\alpha}\\ 
	0 & I_{2}  
	\end{matrix}
	\right) , \\
	\sum_{\tau=-1}^3 E^{\tau} \hat{d}_{r,\tau}^\dagger &= 
	\left(
	\begin{matrix} 
	\sum_\alpha M^\alpha \ \hat{a}^\dagger_{r,\alpha} & \sqrt{6} \ \hat{b}^\dagger_{r} I_2 \\
	I_{2} & \sum_{\alpha} M^\alpha \ \hat{a}^\dagger_{r,\alpha}
	\end{matrix}
	\right) .
	\end{split} \label{matrices}
\end{equation}
Matrices $F^{\tau}$ and $E^{\tau}$ are determined from \Eq{matrices};
the matrix product state (MPS) in \Eq{GS} is injective~\footnote{ 
One can verify that the largest absolute eigenvalue of the
transfer matrix $\sum_{\tau,\tau'} F^{\tau} E^{\tau'} \otimes F^{\tau} E^{\tau'}$ is non-degenerate.
This is equivalent to the statement that the MPS in \Eq{GS} is injective. 
}.
Using \Eq{matrices}, one can easily see that 
the ground state in \Eq{GS} is indeed a
superposition of 1D VBS states 
decorated with two-body singlets and/or vacant sites.

Case (2) maps to $g^{\text{t}}_2=0$ and $ g^{\text{t}}_0>0$ for $\hat{H}$. 
It is obvious that the ferromagnetic states
\begin{equation}
	\Big(\sum_{r=1}^{2N} \hat{S}_r^{-} \Big)^{k} \ \Big(\prod_{j=1}^N \hat{B}_{j,1}^\dagger \Big) |\text{vac}\rangle,  \quad  k=0,1,...,2N
	\label{ferro_state}
\end{equation}
with total spin $S_{\text{tot}}=N$ are exact ground states of $\hat{H}$.
The spin-1 pure-biquadratic chain 
$\hat{H}_{\text{PB}} = \sum_{j=1 }^N  \tilde{g}_0 \hat{P}_{j, j+1}^{(0)}    (\text{with } \tilde{g}_0>0)$
is integrable, 
and 
there are numerous ground states with $S_{\text{tot}}$ ranging from $0$ to $N-1$ that are degenerate with
$(\sum_j \hat{S}_j^-)^k |\psi_{(1,1,...,1)}\rangle$~\cite{dimer-trimer, Eggert, Parkinson_1987,Parkinson_1988}.
The absence of a ferromagnetic phase in \Fig{fig:phase_diagram}(a) can thus be understood from such degeneracy:
after adding interaction $ \sum_{r=\text{even}} g_2^\text{t} \hat{P}_{r}^{(2)}$ (with  
$g_2^\text{t}>0$) that disfavors the ferromagnetic states,
states with smaller $S_{\text{tot}}$ are picked up as the ground states.


\subsection{The Haldane phase}  \label{Sec_HaldanePhase}

In this section we investigate the properties of the MPS $|\text{GS}\rangle$.
Let $G$ be the symmetry group of $\hat{H}$
and $\hat{U}(q)$ be a symmetry operation (on the Hilbert space) corresponding to the group element $q\in G$,
i.e., $[\hat{H}, \hat{U}(q)]=0$.
Subjected to $q$, the unique ground state transforms as $|\text{GS}\rangle \to \hat{U}(q) |\text{GS}\rangle$, 
while the matrices in \Eq{GS} transform as~\footnote{\label{footnote2}Equation~(\ref{uniqueness_of_MPS}) is usually proved in the canonical form; see Ref.~\cite{PhysRevLett.100.167202} or Theorem~7 in Ref.~\cite{PerezGarcia2007}. However, the equation holds regardless of the form of an injective MPS; see Ref.~\cite{Tasaki2020} or Sec.~7.3 in Ref.~\cite{Fannes1992}.}
\begin{equation}
	F^{\tau_{2j-1}} E^{\tau_{2j}}  \to \ \mathrm{e}^{\mathrm{i}\phi_q} \ u_q^\dagger \ F^{\tau_{2j-1}} E^{\tau_{2j}} \ u_q,  \label{uniqueness_of_MPS}
\end{equation}
where $\{u_q\}_{q \in G}$ are unitary matrices which are used to classify the 1D SPT phases~\cite{PTBO_2010, Tasaki2020, PhysRevB.83.035107}.

The group $\mathbb{Z}_2 \times \mathbb{Z}_2 =\{1, \hat{U}(x), \hat{U}(y), \hat{U}(z) \}$
is a symmetry group of $\hat{H}$, where $\hat{U}(\delta):=\exp(-\mathrm{i}\pi \sum_r \hat{S}_{r}^{\delta})$
is the spin rotation about the $\delta=x,y,z$-axis.
The Hamiltonian is also invariant under
time-reversal $\hat{U}(\text{TR}) := \hat{U}(y)\hat{K}$ (where $\hat{K}$ is a complex conjugation operator),
space inversion $\hat{U}(\mathcal{I})$,
spin rotation together with inversion $\hat{U}(z\mathcal{I}):=\hat{U}(z)\hat{U}(\mathcal{I})$,
and pseudo-spin rotation together inversion 
$\hat{U}(n\mathcal{I}):=\exp[-\mathrm{i}\pi \sum_r (\hat{n}_r-1)] \hat{U}(\mathcal{I})$.

For $\hat{U}(\delta)$ and $\hat{U}(\text{TR})$,
we can define their respective topological indices 
using the corresponding unitary matrices in \Eq{uniqueness_of_MPS}
as $\mathcal{Q}_{\mathbb{Z}_2 \times \mathbb{Z}_2} :=\text{Tr}(u_x u_z u_x^\dagger u_z^\dagger)/\chi$
and $\mathcal{Q}_{\text{TR}} := \text{Tr}(u_{\text{TR}}   u_{\text{TR}}^*) /\chi$, 
where $\chi$ is the bond dimension of the MPS~\cite{TopoIndex_PhysRevB.86.125441}.
It is known that $\mathcal{Q}_{\mathbb{Z}_2 \times \mathbb{Z}_2}$ equals $-1$ for the Haldane phase protected by
$\mathbb{Z}_2 \times \mathbb{Z}_2$ symmetry while $1$ for the trivial phase,
similarly for $\mathcal{Q}_{\text{TR}}$~\cite{PTBO_2010}.
When $0<|\lambda|<\infty$, the system has inversion symmetry with respect to every lattice site.
However, the site-centered inversion symmetry cannot protect SPT phases.
The groups 
$\{ 1, \hat{U}(\mathcal{I})\}$, $\{ 1, \hat{U}(z\mathcal{I})\}$, and $\{ 1, \hat{U}(n\mathcal{I})\}$
can protect SPT phases only when $\hat{U}(\mathcal{I})$ is a bond-centered inversion;
see also Ref.~\cite{PhysRevLett.114.177204} and 
Appendix A of Ref.~\cite{PhysRevX.7.011020}.
When the bond-centered inversion symmetry is present, we can similarly define 
$\mathcal{Q}_{\mathcal{I}} := \text{Tr}(u_{\mathcal{I}} u^*_{\mathcal{I}}) / \chi$,
$\mathcal{Q}_{z\mathcal{I}} := \text{Tr}(u_{z\mathcal{I}} u^*_{z\mathcal{I}}) / \chi$, 
and $\mathcal{Q}_{n\mathcal{I}} := \text{Tr}(u_{n\mathcal{I}} u^*_{n\mathcal{I}}) / \chi$,
which are quantized to $+1$ and $-1$ for trivial and Haldane phases, 
respectively~\cite{PTBO_2010, PTBO_2012, TopoIndex_PhysRevB.86.125441}.
The state $|\text{GS}\rangle$ at $\lambda=0$ and $|\lambda| \to \infty$ has bond-centered inversion symmetry.
At $|\lambda| \to \infty$, $| \text{GS} \rangle$ reduces to \Eq{VBS_AKLT}.
At $\lambda=0$, although $|\text{GS}_{\lambda=0}\rangle $ is not the unique ground state of $\hat{H}|_{\lambda=0}$,
one can (in principle) always find a parent Hamiltonian that has $|\text{GS}_{\lambda=0}\rangle$ as the unique ground state~\cite{perez2007peps},
and hence
the state itself is still worth studying.
Actually, it turns out that $|\text{GS}_{\lambda=0}\rangle$
can be viewed as a spinful generalization 
of the Haldane insulator state in spinless bosons;
see Appendix~\ref{App_HI}.
Table~\ref{table:u} summarizes the unitary matrices $\{u_q\}$
with respect to different symmetry operations on $|\text{GS}\rangle$. 
It is then clear that the Haldane phase of $|\text{GS}_{0<|\lambda|<\infty}\rangle$ is protected by 
$\mathbb{Z}_2 \times \mathbb{Z}_2$ symmetry or time-reversal symmetry.
Interestingly,
when the inversion symmetry is involved,
$|\text{GS}_{\lambda=0}\rangle$ and 
$|\text{GS}_{|\lambda| \to \infty}\rangle$
can be in different phases.
This difference originates from the charge fluctuations in $|\text{GS}_{\lambda=0}\rangle$.
We claim that in general, charge fluctuations
can play a nontrivial role in the SPT orders protected 
by crystalline symmetries,
see Sec.~\ref{Sec_GS_SPT} for details.

\begin{table*}[t]
\caption{Unitary matrices in \Eq{uniqueness_of_MPS} with respect to various symmetry operations on $|\text{GS}\rangle$.
In accordance with the values of $\mathcal{Q}_{\mathbb{Z}_2 \times \mathbb{Z}_2}$, $\mathcal{Q}_{\text{TR}}$,
$\mathcal{Q}_{\mathcal{I}}$, $\mathcal{Q}_{z\mathcal{I}}$, and $\mathcal{Q}_{n\mathcal{I}}$,
the two shadowed matrices denote trivial phases, while the other matrices denote the SPT phase.
N/A means the symmetry group cannot give an SPT/trivial classification.}
\begin{ruledtabular}
\begin{tabular}{cccccc}
 &
$ \hat{U}(\delta) \ \ (\delta=x,y,z)$&
$ \hat{U}(\text{TR}) $&
$ \hat{U}(\mathcal{I}) $&
$ \hat{U}(z\mathcal{I}) $&
$ \hat{U}(n \mathcal{I}) $\\
\colrule
$\lambda=0$&
$
u_\delta= \left( 
\begin{array}{cc}
 \sigma^{\delta}  & 0 \\
 0 & \sigma^{\delta}  \\
\end{array}
\right)
$&
$
u_{\text{TR}}= \left( 
\begin{array}{cc}
 \sigma^{y}  & 0 \\
 0 & \sigma^{y}  \\
\end{array}
\right)
$&
\colorbox{blue!25}{$
u_{\mathcal{I}}= \left( 
\begin{array}{cc}
 0  & -\sigma^y \\
\sigma^y & 0  \\
\end{array}
\right)
$}&
$u_{z\mathcal{I}}=\left( 
\begin{array}{cc}
 0  & -\sigma^{x} \\
 \sigma^{x} & 0  \\
\end{array}
\right)$&
$
u_{n \mathcal{I}}= \left( 
\begin{array}{cc}
 0  & \sigma^{y} \\
 \sigma^{y} & 0  \\
\end{array}
\right)
$\\
$0<|\lambda|<\infty$&
$
u_\delta= \left( 
\begin{array}{cc}
 \sigma^{\delta}  & 0 \\
 0 & \sigma^{\delta}  \\
\end{array}
\right)
$&
$
u_{\text{TR}}= \left( 
\begin{array}{cc}
 \sigma^{y}  & 0 \\
 0 & \sigma^{y}  \\
\end{array}
\right)
$&
N/A&
N/A&
N/A\\
$|\lambda|\to \infty$&
$u_\delta= \sigma^{\delta}$&
$u_{\text{TR}}= \sigma^{y}$&
$u_{\mathcal{I}}=\sigma^y$&
\colorbox{blue!25}{$ u_{z\mathcal{I}} = \sigma^x $}&
$u_{n \mathcal{I}}=\sigma^y$
\end{tabular}  
\end{ruledtabular} \label{table:u}
\end{table*}

Using the exact MPS in \Eq{GS} and \Eq{matrices}, 
various quantities that characterize the Haldane phase can be calculated analytically.
For example,
the spin string order parameter
$\mathcal{O}^{\delta}:= - \lim_{L \to \infty} \lim_{N \to \infty} \frac{1}{\langle \text{GS} | \text{GS} \rangle}
   \langle \text{GS} | (\hat{S}_{r}^{\delta} + \hat{S}_{r+1}^{\delta} ) 
   \exp[\mathrm{i}\pi \sum_{k=r+2}^{r+2L-1} \hat{S}_{k}^{\delta}] (\hat{S}_{r+2L}^{\delta} + \hat{S}_{r+2L+1}^{\delta})| \text{GS} \rangle $
and the charge string order parameter
$\mathcal{C}:= - \lim_{L \to \infty} \lim_{N \to \infty}
\frac{1}{\langle \text{GS} | \text{GS} \rangle}
   \langle \text{GS} | \ (\hat{n}_{r} + \hat{n}_{r+1} -1 ) \ 
   \exp\{\mathrm{i}\pi [  \sum_{k=r+2}^{r+2L-1} \hat{n}_{k}  - (L-1)  ] \} \ (\hat{n}_{r+2L} + \hat{n}_{r+2L+1} -1 )\ | \text{GS} \rangle $
are found to be
\begin{equation}
\begin{split}
	\mathcal{O}^{\delta} &=
    \frac{16 \big[ 9 \lambda ^6+\left(5 Q+48\right) \lambda ^2+3 \left(Q+11\right) \lambda ^4+24 \big]^2}{ Q^2 \left(3 \lambda ^2+Q+6\right)^2 (Q+3\lambda^2)^2 },\\
    \mathcal{C} &= \frac{24 \left(3 \lambda ^2+2\right)^2}{Q^2 \big[ 3 \lambda ^4+\left(Q+12\right) \lambda ^2+2 \left(Q+5\right) \big] },
\end{split}
\end{equation}
where $Q:=\sqrt{9 \lambda ^4+36 \lambda ^2+24}$.
It can be shown that both string order parameters are nonzero:
$ 4/(\sqrt{6}+3)^2 < \mathcal{O}^{\delta} < 4/9$ and
$0< \mathcal{C} < 0.207$.
For the open boundary condition (OBC),
we can show the existence of both spin and charge edge states;
see Appendix.~\ref{App_edge}.
It is known that the (seemingly unrelated) spin string order and edge state 
are unified in the context of hidden $\mathbb{Z}_2 \times \mathbb{Z}_2$ symmetry breaking.
In spin chains, this can be seen 
with the Kennedy-Tasaki transformation~\cite{Kennedy_Tasaki,Oshikawa_1992,PTBO_2012,Tasaki2020}.
In spin-$f$ itinerant systems ($f=$ integer), the Kennedy-Tasaki transformation is also applicable;
see Appendix.~\ref{App_KT}.

The ``hidden VBS order" is a unique feature for the Haldane phase in systems with both spin and charge fluctuations.
It is closely related to both string order parameters $\mathcal{O}^{\delta}$ and $\mathcal{C}$.
Since vacant and doubly occupied sites have zero spin,
the ``hidden VBS order" immediately implies the hidden antiferromagnetic order measured by $\mathcal{O}^{\delta}$.
On the other hand, $\mathcal{C}$ measures to what extent
the VBS states are diluted in the background of vacant and doubly occupied sites.

In the presence of both translation symmetry and $\mathbb{Z}_2 \times \mathbb{Z}_2$ symmetry,
four distinct SPT phases can exist~\cite{PhysRevB.84.075135},
and one of them is represented by $|\text{GS}\rangle$.
The other three can be realized by unitary transformations of $\hat{H}_{\text{hop}}$,
see Appendix~\ref{App_translation} for details.

\subsection{Perturbation theory} \label{Sec_PerturbationTheory}

Beyond the cases where the ground state is exactly solvable,
the phase of $\hat{H}$ can still be determined analytically 
when $|\lambda|$ is large enough. See \Fig{fig:phase_diagram}(a).
In the limit $|\lambda| \to \infty$,
if we assume $g_{0}^\text{b}$ and $g_{2}^\text{b}$ are around the magnitude of $\lambda^2$,
the unperturbed ground state will have each bottom site occupied by exactly one particle.
In this case, perturbation theory tells us that
the low-energy effective Hamiltonian of $\hat{H}$
is given by $\hat{H}_{\text{BLBQ}}$ in \Eq{BLBQ}
with $J \propto \lambda^{-2}$ and
\begin{equation}
	\theta = \arctan \left( \frac{1}{3} \cdot \frac{3 g_0^{\text{t}} +4 g_0^{\text{t}} \lambda ^2/g_2^{\text{t}}+2 \lambda ^2}{g_0^{\text{t}}+2 \lambda ^2} \right) \geqslant \arctan \frac{1}{3},
	\label{theta}
\end{equation}
where $J$ and $\theta$ are independent of $g_S^{\text{b}}$.
From \Eq{theta}, we know that
the effective model is in the Haldane phase
when $0 \leqslant g_0^{\text{t}} < g_2^{\text{t}}$
while in the critical phase 
when $0 < g_2^{\text{t}} \leqslant g_0^{\text{t}}$.
In particular, 
$g_0^{\text{t}}=0$ and $g_2^{\text{t}}=0$
corresponds to the AKLT and pure-quadratic point,
respectively.

\subsection{Numerical analysis}  \label{Sec_NumericalAnalysis}

Beyond the three special planes in \Fig{fig:phase_diagram}(a) where 
either exact ground states can be found or perturbation theory works,
the phase diagram of the spin-1 BHMSC can in general be determined by numerical calculations.
In the thermodynamic limit $N \to \infty$,
we find the phase diagram in the $\lambda=1$ plane 
in~\Fig{fig:phase_diagram}(a) with the VUMPS algorithm~\cite{PhysRevB.97.045145, SciPostPhysLectNotes}. 
Due to the fact that the total number of particles 
and unit cells are the same,
matrices in the MPS ansatz used in the algorithm are assumed to be block-banded~\cite{PhysRevB.97.235155}.
Also, the maximum particle number on each site is truncated to three.
See Appendix~\ref{App_MPSansatz} for details of the MPS ansatz.
Let $\epsilon_i := -\ln |\lambda_i|$, where $\lambda_i$ is the $i$th largest absolute eigenvalue of the transfer matrix,
and $|\lambda_1|$ is normalized to~1.
When the bond dimension $\chi$ is extrapolated to infinity,
the correlation length $\xi:= 1/\epsilon_2 $ 
diverges for gapless phases
while it converges to a finite value for gapped phases.
This fact is known to be well reflected in
the scaling relation of
$1/\xi(\chi)$ with respect to
$ \epsilon_3(\chi) -\epsilon_2(\chi)$~\cite{PhysRevX.8.041033, PhysRevB.78.024410},
as shown in~\Fig{fig:phase_diagram}(b).
In the region of the gapped phase in~\Fig{fig:phase_diagram}(a), 
we find that
$\mathcal{Q}_{\mathbb{Z}_2 \times \mathbb{Z}_2} = \mathcal{Q}_{\text{TR}} =-1$, 
which suggests that the gapped phase is the Haldane phase.
The Haldane phase is characterized by
an even-fold degenerate entanglement spectrum~\cite{PTBO_2010}, see~\Fig{fig:phase_diagram}(c).
The ground state magnetization $M := \lim_{N \to \infty} \langle \sum_{r=1}^{2N} \hat{S}_r^z \rangle / N$
is calculated after adding $-h \sum_{r} \hat{S}_r^z$ to $\hat{H}$,
where $h$ is the magnetic field; see~\Fig{fig:phase_diagram}(d).
In the gapless region, $M$ grows almost linearly with $h$,
which suggests that the gapless phase is the critical phase~\cite{PhysRevB.83.184433}.
In the Haldane phase region, however, $M$ is expected to exhibit a zero plateau for small $h$~\cite{PhysRevB.83.184433},
which is indeed the case as in~\Fig{fig:phase_diagram}(d). 
Note that the phase boundary in the $\lambda=1$ plane
is curved; see Appendix~\ref{App_MPSansatz} 
for numerical evidence.

\subsection{Short summary for the spin-1 bosons on a sawtooth chain}

To demonstrate how the Haldane phase emerges in 
short-range interacting spinful bosons loaded on
lattices with a bottom flat band,
we have used the spin-1 BHMSC as an example.
We show that this system has some deep connections with the BLBQ model.
In particular, in a special case, 
by an exact mapping to the ground state of the AKLT model,
we obtain the exact and unique ground state of the spin-1 BHMSC.
This exact ground state turns out to serve as a representative state of the Haldane phase. 
The phase diagram of this model is obtained by perturbation theory and numerical calculations based on the VUMPS algorithm,
and we find that the Haldane phase exists in a rather wide parameter region.

We expect that, even if the bottom band is not perfectly flat,
the nature of the many-body ground states
should remain unchanged as long as the interaction strength
is sufficiently strong.
Such robustness of the ground states
has been rigorously proved in some classes of 
Hubbard models 
with a nearly flat band~\cite{Tasaki2020,PhysRevB.100.214423}.

\section{General theory} \label{Sec_Generalizations}

The sawtooth chain is not special in the sense that
there are many other lattices possessing a bottom flat band,
it is thus natural to expect that the SPT phases can be
realized with spinful bosons loaded on these lattices.
Our approach in the previous section can be generalized.
In this section, we present a general theory for
the SPT phases of spin-$f$ bosons with a bottom flat band. 
We first show in Sec.~\ref{Sec_VBS} that the AKLT model 
and VBS state can be generalized to higher spins 
and higher dimensional lattices.
Let $|\text{VBS}_{f,X'}\rangle$ be the exact and unique ground state of 
the spin-$f$ AKLT model defined on a lattice $X'$.
On the other hand,
bottom-flat-band lattices can be constructed systematically.
Let $|\text{GS}_{f,X}\rangle$ be the ground state of
$N$ spin-$f$ bosons on a bottom-flat-band lattice $X$
with $N$ unit cells.
In Sec.~\ref{Sec_GS_boson_flatband}, 
we show that with fine-tuned parameters,
$|\text{GS}_{f,X}\rangle$ can be exactly mapped to 
$|\text{VBS}_{f,X'}\rangle$, provided that $f$, $X$, and $X'$
satisfy a certain relation.
This means that $|\text{GS}_{f,X}\rangle$ is the exact and unique ground state of the itinerant spin-$f$ model.
In Sec.~\ref{Sec_GS_SPT}, 
with various $f$ and $X$,
we classify the quantum phases of $|\text{GS}_{f,X}\rangle$'s
from the viewpoint of SPT orders.
In particular,
we find that in terms of crystalline symmetries,
not only spin fluctuations but also 
charge fluctuations
in $|\text{GS}_{f,X}\rangle$
determine its symmetry-protected phase.

\subsection{Generalized AKLT models and VBS states} \label{Sec_VBS}

It is known that VBS states can be constructed on any lattice in any dimensions~\cite{AKLT1988, kirillov2009valence, Katsura_2010}.
In this article, we consider only bosonic spin-$f$ VBS states ($f=$ integer).
Let $X'=(\Lambda_{X'}, \mathscr{B}_{X'})$ be a lattice (graph) where 
$\Lambda_{X'}$ is the set of sites (vertices) and $\mathscr{B}_{X'}$ is the set of bonds (edges).
A bond is defined by two sites $\{ \bm{j} ,\bm{j}' \}$ with $\bm{j} ,\bm{j}' \in \Lambda_{X'}$.
We assume that every site in $X'$ is directly connected to $2f$ other sites, i.e.,
$\big|\big\{ \bm{j'} \in \Lambda_{X'} | \{ \bm{j} ,\bm{j}' \} \in \mathscr{B}_{X'} \big\}\big| = 2f, \forall \bm{j}$.
(In other words, $X'$ is a regular graph of degree $2f$.)
When there is a spin-$f$ degree of freedom (DOF) residing in every site of $X'$,
an AKLT-type quantum spin model can be defined on $X'$ as
\begin{equation}
	\hat{H}_{\text{AKLT}}^{f,X'} := \tilde{g}_{2f} \sum_{ \{ \bm{j}, \bm{j}' \}\in \mathscr{B}_{X'} } \hat{P}_{\bm{j}, \bm{j}'}^{(2f)}
	\quad \quad (\tilde{g}_{2f}>0),  \label{generlized_AKLT_model}
\end{equation}
where the operator $\hat{P}_{\bm{j}, \bm{j}'}^{(2f)}$ projects 
the state of two \mbox{spin-$f$'s} on two sites $\bm{j}, \bm{j}'$ 
onto the state with total spin $2f$.

\begin{figure}
  \centering
  \includegraphics[width=0.47\textwidth]{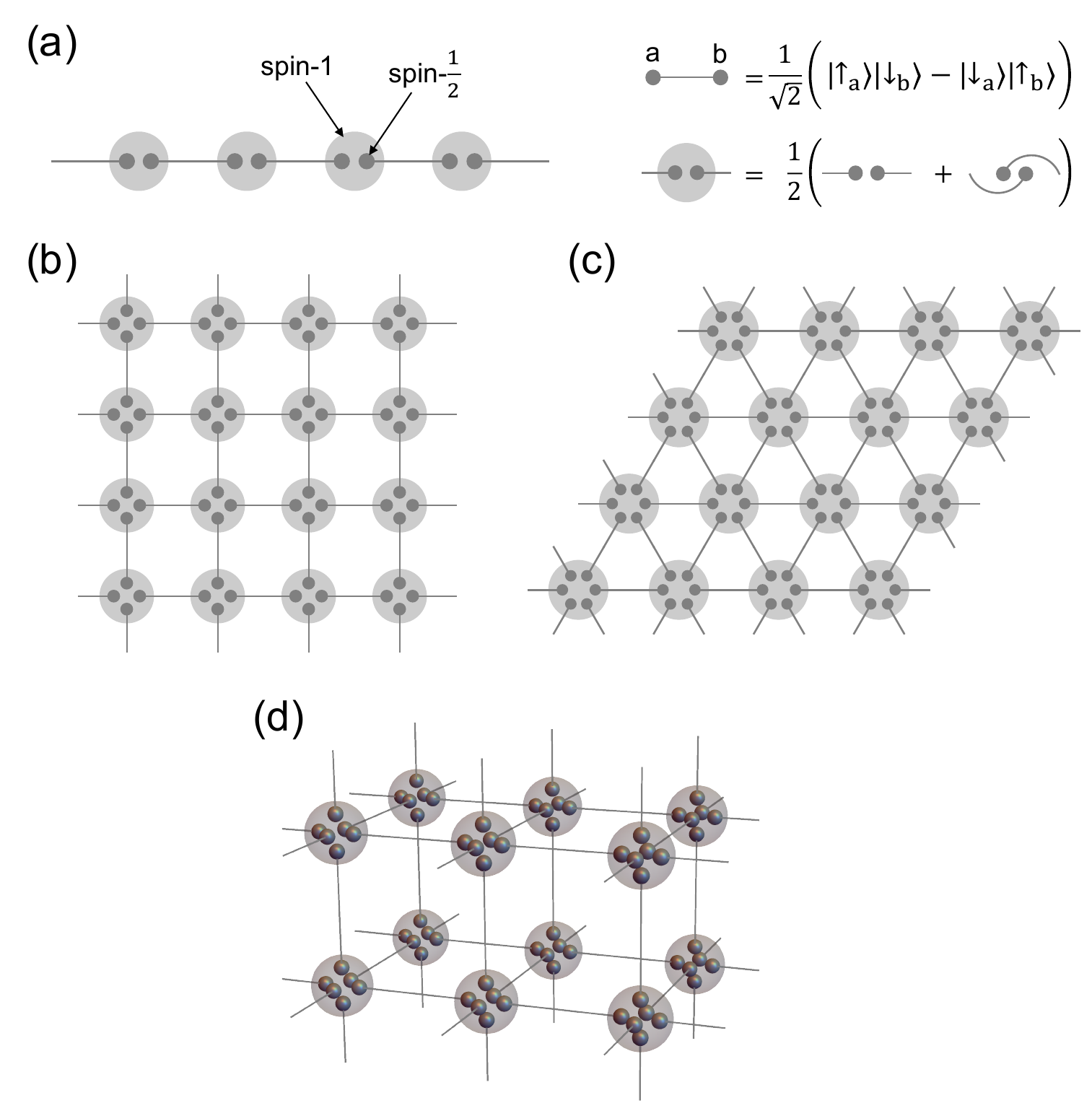}
  \caption{Examples of bosonic VBS states. Each spin-$f$ is regarded as a composite state of $2f$ spin-$1/2$'s,
  and a spin singlet is formed between two spin-$1/2$'s on neighboring sites.
  (a) 1D spin-1 VBS state, whose expression is given in \Eq{VBS_AKLT}. (b) 2D spin-2 VBS state on a square lattice, denoted as $|\text{VBS}_{2,\Box}\rangle$. (c) 2D spin-3 VBS state on a triangular lattice, denoted as $|\text{VBS}_{3,\triangle}\rangle$. (d) 3D spin-3 VBS state on a cubic lattice, denoted as $|\text{VBS}_{3,\includegraphics[width=0.012\textwidth]{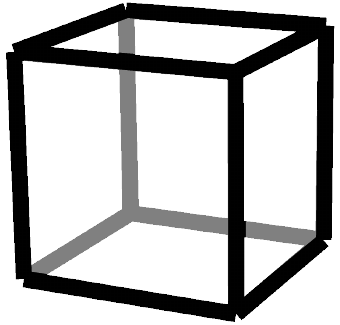}}\rangle$.}
  \label{fig:VBS_states}
\end{figure}

It has been proved that when $| \Lambda_{X'} | = N < \infty$,
$\hat{H}_{\text{AKLT}}^{f,X'}$ has an exact and unique ground state~\cite{Tasaki2020,kennedy1988two},
known as a VBS state:
\begin{equation}
	|\text{VBS}_{f,X'}\rangle = \sum_{\alpha_1,...,\alpha_N=-f}^f S_{\alpha_1,...,\alpha_N} \ |\psi_{\alpha_1,...,\alpha_N}\rangle,  \label{VBS_fX}
\end{equation}
where $\{ | \psi_{\alpha_1, \alpha_2,..., \alpha_N} \rangle \}$ is the spin $S^z$-basis and
the coefficient $S_{\alpha_1,...,\alpha_N}$ encodes short-range entanglement between the spins.
When $X'$ is the simple 1D linear chain
with $f=1$, 
$\hat{H}_{\text{AKLT}}^{f,X'}$ reduces to \Eq{AKLT_H},
and its ground state is the 1D spin-1 VBS state in \Eq{VBS_AKLT},
and $S_{\alpha_1,...,\alpha_N} = y_{\bm{\alpha}}$ in \Eq{y_beta}. 
The structure of this VBS state can be understood as follows:
as shown in \Fig{fig:VBS_states}(a), each spin-$1$ is viewed as a composite state of two spin-$1/2$'s,
and a pair of spin-$1/2$'s on two neighboring sites forms a spin singlet.
For $\hat{H}_{\text{AKLT}}^{f,X'}$ on a general $X'$,
the ground state $|\text{VBS}_{f,X'}\rangle$ can be constructed in the same manner:
each \mbox{spin-$f$} is regarded as a composite state of $2f$ \mbox{spin-$1/2$'s},
and a singlet is formed between two spin-$1/2$'s 
in every bond $\{ \bm{j}, \bm{j}' \}\in \mathscr{B}_{X'}$~\cite{AKLT1988, kirillov2009valence}.
Some other graph representations of VBS states 
in 2D and 3D
are given in
Figs.~\ref{fig:VBS_states}(b)-\ref{fig:VBS_states}(d).

\subsection{Ground states of spin-$f$ bosons with a bottom flat band} \label{Sec_GS_boson_flatband}

Many kinds of atoms carry integer spins,
among which alkali-metal atoms are often used in experiments~\cite{KAWAGUCHI2012253,* RevModPhys.85.1191}.
Alkali-metal atoms have two hyperfine levels, and each level carries integer spin $f$,
see Table~\ref{tab:atom_table}.
Due to the hyperfine interaction, the level with smaller $f$ has lower energy.
Therefore, alkali-metal atoms stay in lower hyperfine level 
when they are optically trapped without external pumping.
For example, as shown in Table~\ref{tab:atom_table}, $^{87}$Rb atoms are often regarded as spin-1 bosons,
while they can indeed be spin-2 bosons if one 
pumps them into the $f=2$ hyperfine level~\cite{KAWAGUCHI2012253,* RevModPhys.85.1191}.

\begin{table}[H]
\begin{center}
\begin{tabular}{ccc}
\hline
\hline
atom                                               & & $f$      \\
\hline
$^{^{\white{1}}7}$Li, $^{23}$Na, $^{41}$K, $^{87}$Rb  & & $1$, $2$ \\
$^{^{\white{1}}25}$Na, $^{79,83}$Rb, $^{131}$Cs               & & $2$, $3$ \\
$^{^{\white{1}}135, 137, 139, 141}$Cs                                        & & $3$, $4$ \\
$^{^{\white{1}}119}$Cs, $^{207,209,211,213}$Fr                                        & & $4$, $5$ \\
\hline
\hline
\end{tabular} 
\end{center} 
\caption{Some alkali-metal isotopes and their hyperfine spins $f$. These isotopes are stable or long-lived (compared to typical experimental cycle time of around 10 seconds) at least in lower hyperfine levels~\cite{KAWAGUCHI2012253,* RevModPhys.85.1191}.}
\label{tab:atom_table}
\end{table}

\begin{figure}
  \centering
  \includegraphics[width=0.3\textwidth]{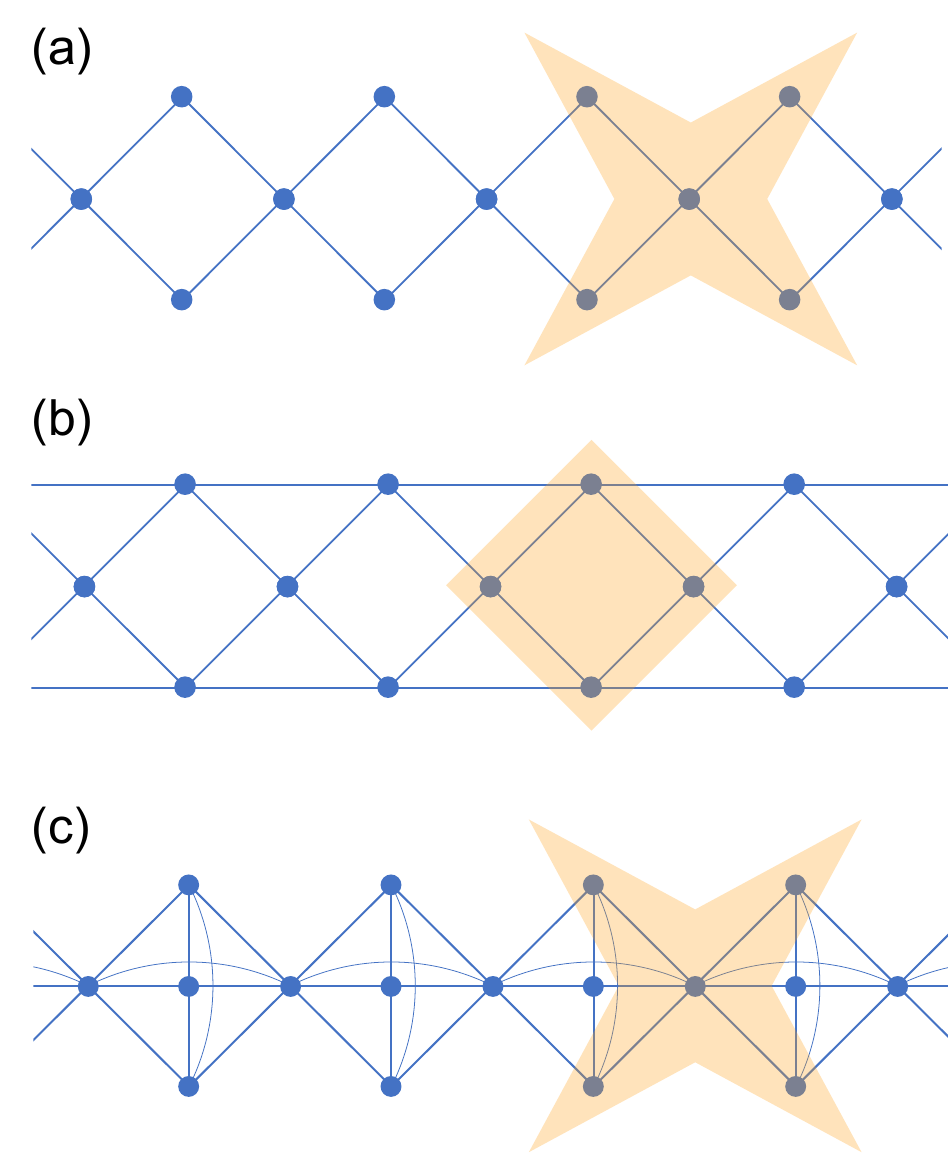}
  \caption{Other examples of 1D lattices that can be used to construct nontrivial ground states with spin-1 bosons. 
  Allowed hopping process between two sites is illustrated by a bond.
  (a) Diamond chain. A $\pi$ flux threads each plaquette~\cite{PhysRevLett.99.026404}.
  A CLS covers five sites, as denoted by the four-pointed star. (b) Kagome ladder, an example of the line graph construction~\cite{Mielke_1992, Katsura_2010}. A CLS is denoted by the square.
  (c) Pyramid chain, an example of the cell construction~\cite{Tasaki2020}. A CLS is denoted by the four-pointed star. }
  \label{fig:1D_FB_lattices}
\end{figure}
\begin{figure}
  \centering
  \includegraphics[width=0.48\textwidth]{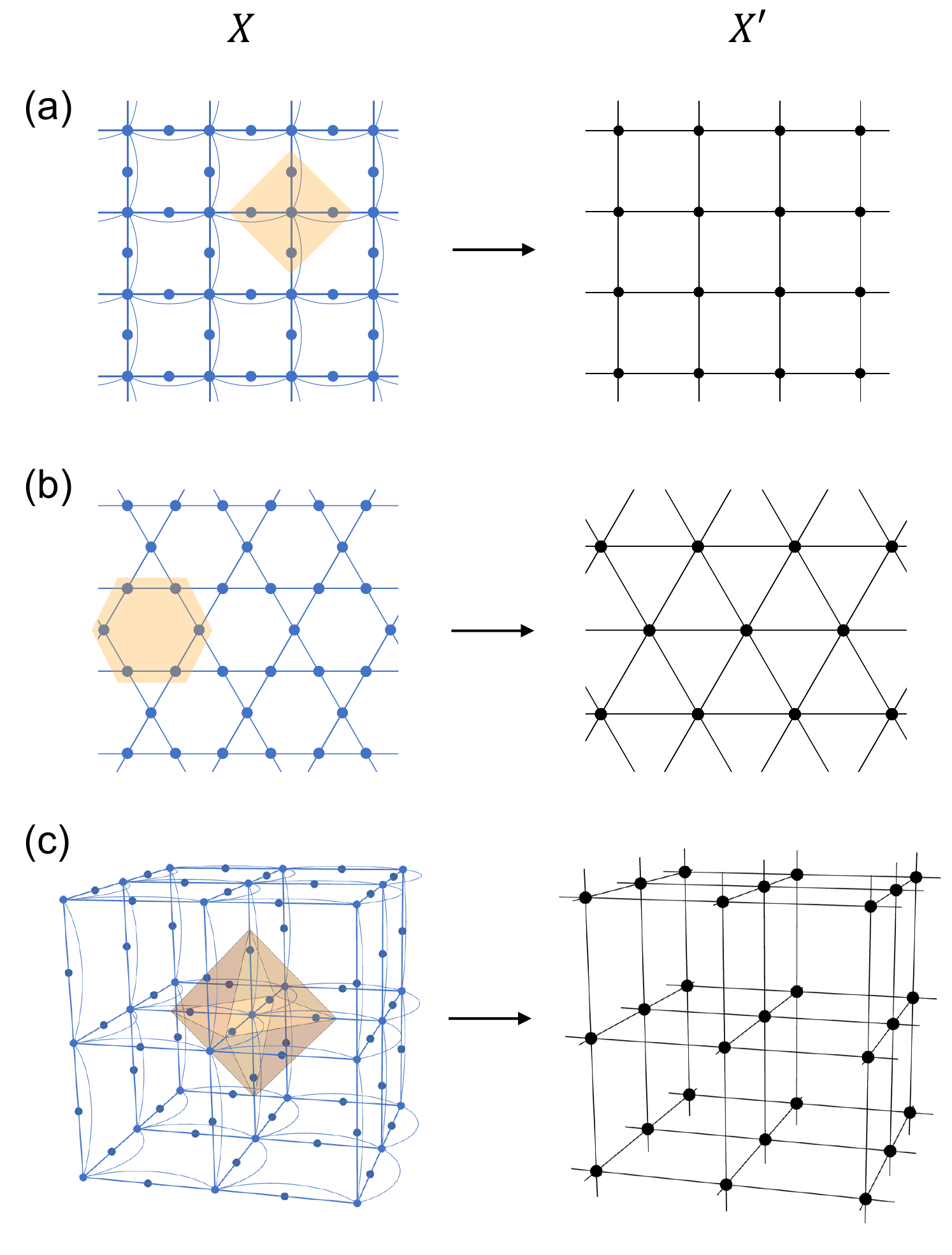}
  \caption{Examples of higher dimensional lattices $X$ with a bottom flat band (left column) and 
  their corresponding lattices $X'$ where the AKLT models are defined (right column). 
  In $X$, allowed hopping processes are illustrated by bonds.
  In $X'$, a bond represents interaction between two spins.
  (a)~2D Tasaki lattice (left), an example of the cell construction~\cite{PhysRevLett.69.1608, 10.1143/PTP.99.489}. A CLS is localized on five sites, as pictured by the square. 
  All the CLSs are related to each other by lattice translation vectors.
  Every CLS overlaps with four other CLSs, thus $f$ should be $4/2=2$.
  The corresponding AKLT model lives on a square lattice (right). 
  We symbolize the 2D Tasaki lattice as \includegraphics[width=0.014\textwidth]{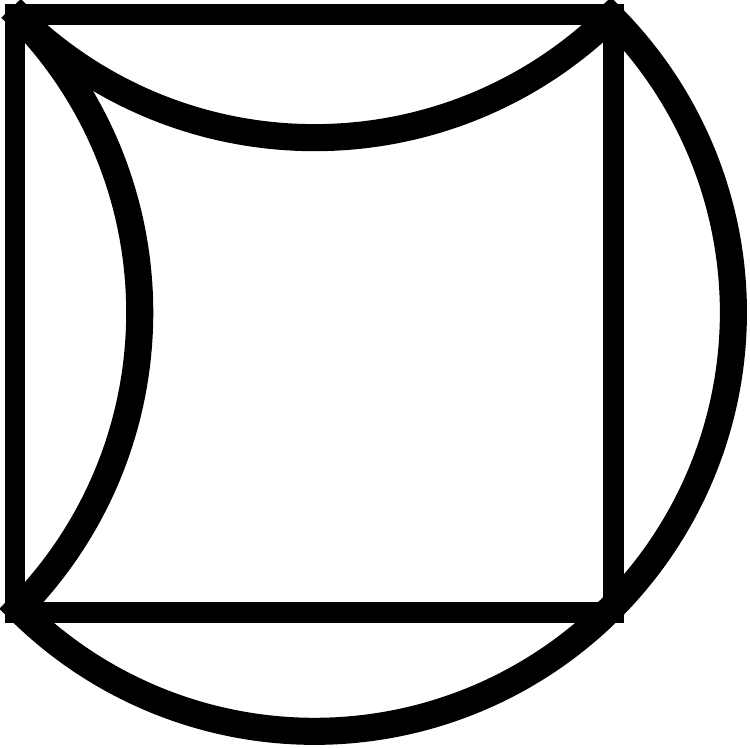} and the square lattice
  as $\square$.
  (b)~Kagome lattice (left), an example of the line graph construction~\cite{Mielke_1992}. 
  A CLS is localized on a hexagon. Every CLS overlaps with $2f=6$ other CLSs, 
  and the corresponding AKLT model lives on a triangular lattice (right).
  We symbolize the kagome lattice as \includegraphics[width=0.015\textwidth]{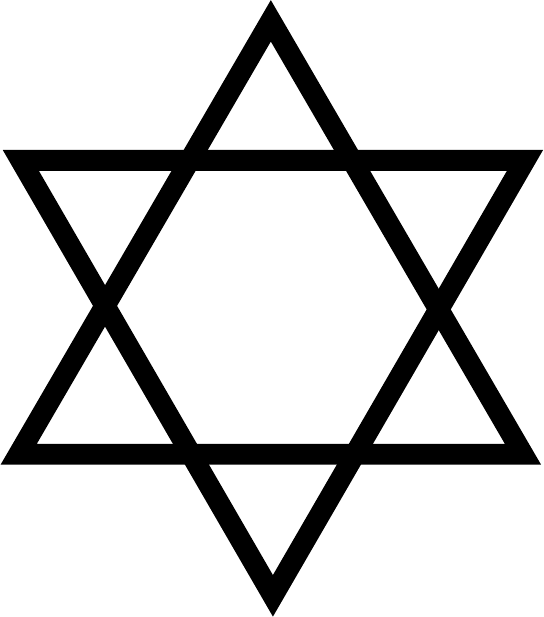} and the triangular lattice
  as $\triangle$.
  An optical kagome lattice has been realized experimentally~\cite{PhysRevLett.108.045305}.
  (c)~3D Tasaki lattice (left), another example of the cell construction~\cite{PhysRevLett.69.1608, 10.1143/PTP.99.489}. 
  A CLS is localized on seven sites, as covered by the octahedron. 
  Every CLS overlaps with $2f=6$ other CLSs, 
  the corresponding AKLT model thus lives on a cubic lattice (right).
  We symbolize the cubic lattice as \includegraphics[width=0.015\textwidth]{cube_symbol.pdf}.}
  \label{fig:2D_FB_lattices}
\end{figure}

We have seen that the single-particle CLSs play a crucial role in constrcting
the many-body ground state.
The CLSs exist not only in the sawtooth chain
but also in all the finite-range hopping lattices possessing a flat band~\cite{PhysRevB.99.045107, PhysRevB.95.115309}.
In fact, there are various systematic approaches to construct flat-band lattices~\cite{PhysRevLett.69.1608, 10.1143/PTP.99.489, MIELKE1993443, Mielke_1992, Mielke_1991, PhysRevB.99.045107, PhysRevA.94.043831, PhysRevB.95.115135, dias2015origami, tanaka2020extension, Tasaki2020, LiuZheng:77308},
among which Tasaki's 
cell construction~\cite{PhysRevLett.69.1608, 10.1143/PTP.99.489} and 
Mielke's line graph construction~\cite{MIELKE1993443, Mielke_1992, Mielke_1991}
always yield a bottom flat band~\footnote{In fact, the sawtooth chain can be produced by
either the cell construction or the line graph construction.}.
These systematic constructions generate infinitely many kinds of lattices in $d \geqslant 1$ dimensions,
such as those shown in \Fig{fig:1D_FB_lattices} for $d=1$ and~\Fig{fig:2D_FB_lattices} for $d=2,3$.

Let $X=(\Lambda_{X}, \mathscr{B}_{X})$
be a bottom-flat-band lattice where 
$\Lambda_{X}$ is the set of sites and $\mathscr{B}_{X}$ is the set of bonds.
A bond is defined by two sites $\{ \bm{r} ,\bm{r}' \}$ with $\bm{r} ,\bm{r}' \in \Lambda_{X}$.
Let $\hat{H}_{\text{hop}}^{f,X}$ be a single-body Hamiltonian for spin-$f$ bosons on $X$:
\begin{equation}
	\hat{H}_{\text{hop}}^{f,X} =  -\sum_{  \{ \bm{r} ,\bm{r}' \}\in \mathscr{B}_{X} } \sum\limits_{ \alpha=-f }^f   t_{\bm{r} ,\bm{r}'} \  \hat{a}^{\dagger}_{\bm{r},\alpha}\hat{a}_{\bm{r}',\alpha}  
	+ \sum_{ \bm{r}\in \Lambda_{X} } V_{\bm{r}} \hat{n}_{\bm{r}},
\end{equation}
where $\hat{a}^{\dagger}_{\bm{r},\alpha}$ creates a boson with magnetic sublevel $\alpha$ at site $\bm{r}$.
Let $N$ be the total number of unit cells in $X$.
The assumption that $X$ has a bottom flat band means
that the single-particle ground state degeneracy of $\hat{H}_{\text{hop}}^{f,X}$ is \mbox{$N(2f+1)$}.
The corresponding CLSs are localized on $N$ different positions
and are related to each other by lattice translation vectors~\footnote{Ignore 
spin for the moment, i.e., take $f=0$.
In $d=1$, $N$ different CLSs can always be chosen to be linearly independent.
However, in $d>1$ with PBC, these $N$ CLSs can be linearly dependent in some cases, such as in kagome lattice~\cite{PhysRevB.99.045107}.
Nevertheless, they can still be linearly independent in $d>1$ with OBC.}.
The shapes of some CLSs are shown in \Fig{fig:1D_FB_lattices} and \ref{fig:2D_FB_lattices}.
Let $( \hat{B}^{f,X}_{j, \alpha } )^\dagger $ 
be the creation operator of a CLS, where $j=1,2,...,N$ labels different positions.
A fully packed state (FPS) on $X$ is defined as a product of
$N$ CLSs:
$( \hat{B}^{f,X}_{1, \alpha_1 } )^\dagger ( \hat{B}^{f,X}_{2, \alpha_2 } )^\dagger ... ( \hat{B}^{f,X}_{N, \alpha_N } )^\dagger |\text{vac} \rangle$.
In an FPS, the lattice is ``fully packed" by $N$ particles.
For example, $| \bm{\beta} \rangle$ in \Eq{eq:beta} is an FPS in the sawtooth chain.

We now consider another lattice $X'=(\Lambda_{X'}, \mathscr{B}_{X'})$ with $|\Lambda_{X'}| = N$, and each site $\bm{j}\in \Lambda_{X'}$ represents a CLS in the FPS of $X$.
Two sites in $X'$ are directly connected iff the two corresponding CLSs in the FPS (partially) overlap.
For example, as shown in \Fig{fig:2D_FB_lattices}, 
if $X$ is the 2D (3D) Tasaki lattice, $X'$ will be the square (cubic) lattice,
while if $X$ is the kagome lattice, $X'$ will then be the triangular lattice.
In the following, we require that
$f$ and $X$ are chosen such that $X'$ satisfies the condition
$\big|\big\{ \bm{j'} \in \Lambda_{X'} | \{ \bm{j} ,\bm{j}' \} \in \mathscr{B}_{X'} \big\}\big| = 2f, \forall \bm{j}\in \Lambda_{X'}$.
Define $ \Lambda_X^{[k]} \subset \Lambda_X$ as a set of sites where $k$ CLSs in the FPS overlap.
For example, in the sawtooth chain $\Lambda_X = \Lambda_X^{[1]} \cup \Lambda_X^{[2]}$, 
where $\Lambda_X^{[1]}$ is the set of all the bottom sites and $\Lambda_X^{[2]}$ is all the top sites.
We further require that 
every site $\bm{r} \in \Lambda_{X}$ is shared by no more than two CLSs in the FPS,
i.e., $\Lambda_X^{[k]} = \emptyset$ for $k>2$.
We then define the spin model $\hat{H}_{\text{AKLT}}^{f,X'}$ on $X'$, as introduced in Sec.~\ref{Sec_VBS}.


The $s$-wave interaction between two spin-$f$ bosons at position $\bm{r}$ is given by
$\sum_{S=0,2,...,2f} g_{S, \bm{r}} \ \hat{P}_{\bm{r}}^{(S)}$,
where the SO(3)-invariant operator $\hat{P}_{\bm{r}}^{(S)}$ 
projects the state onto total spin $S = \text{even}$
and satisfies the ``completeness relation''
$\sum_{S} \hat{P}_{\bm{r}}^{(S)} = \hat{n}_{\bm{r}} (\hat{n}_{\bm{r}}-1)/2$~\cite{KAWAGUCHI2012253,* RevModPhys.85.1191}.
(For interaction between alkali-metal atoms, it is sufficient to consider the short-range $s$-wave scattering~\cite{KAWAGUCHI2012253,* RevModPhys.85.1191}.)
\mbox{Spin-$f$} bosons in optical lattices are described by the \mbox{spin-$f$} Bose-Hubbard model.
On the lattice $X$, the model is given by
\begin{equation}
	\begin{split}
		\hat{H}^{f,X} &:= \hat{H}_{\text{hop}}^{f,X} + \hat{H}_{\text{int}}^{f,X},\\
		\hat{H}_{\text{int}}^{f,X} &:= \sum_{\bm{r} \in \Lambda_X} \sum_{S=0,2,...,2f} g_{S, \bm{r}} \ \hat{P}_{\bm{r}}^{(S)}.
	\end{split}  \label{H^X}
\end{equation}
The $s$-wave scattering Hamiltonian is reminiscent of
the AKLT Hamiltonians.
If $N$ spin-$f$ bosons are loaded on $X$ and $g_{2f,\bm{r} }>0$ and $g_{S<2f,\bm{r} }=0$ for all $\bm{r} \in \Lambda_X^{[2]}$,
following Sec.~\ref{Sec_ExactGroundStates},
the zero-energy ground states of $\hat{H}^{f,X}$ in \Eq{H^X} and $\hat{H}_{\text{AKLT}}^{f, X'}$ in \Eq{generlized_AKLT_model}
can thus be exactly mapped to each other,
just as \Eq{Hab=Hab}.
See Appendix~\ref{App_Uniqueness} for discussions of
the uniqueness of the ground state of $\hat{H}^{f,X}$.

Let us see some concrete examples.
In $d=1$ dimension, besides the sawtooth chain,
spin-1 bosons can be loaded
on the lattices in \Fig{fig:1D_FB_lattices} as well.
In $d=2$, the 2D Tasaki lattice matches \mbox{spin-2} bosons, 
and the corresponding spin-2 AKLT model lives on a square lattice with the VBS ground state in
\Fig{fig:VBS_states}(b).
The kagome lattice is suitable for \mbox{spin-3} bosons,
while the corresponding AKLT model has the spin-3 VBS ground state on a triangular lattice as shown
\Fig{fig:VBS_states}(c).
On the other hand, \mbox{spin-3} bosons are also compatible with the 3D Tasaki lattice,
which corresponds to a 3D spin-3 VBS state in \Fig{fig:VBS_states}(d).
In fact, the Tasaki lattice can be constructed in any dimension, and the sawtooth chain can actually be regarded as the 1D Tasaki lattice~\cite{PhysRevLett.69.1608, 10.1143/PTP.99.489}.
In general, the $d$-dimensional Tasaki lattice matches spin-$d$ bosons, 
and the corresponding AKLT model lives on a $d$-dimensional hypercubic lattice.

\begin{figure}
  \centering
  \includegraphics[width=0.45\textwidth]{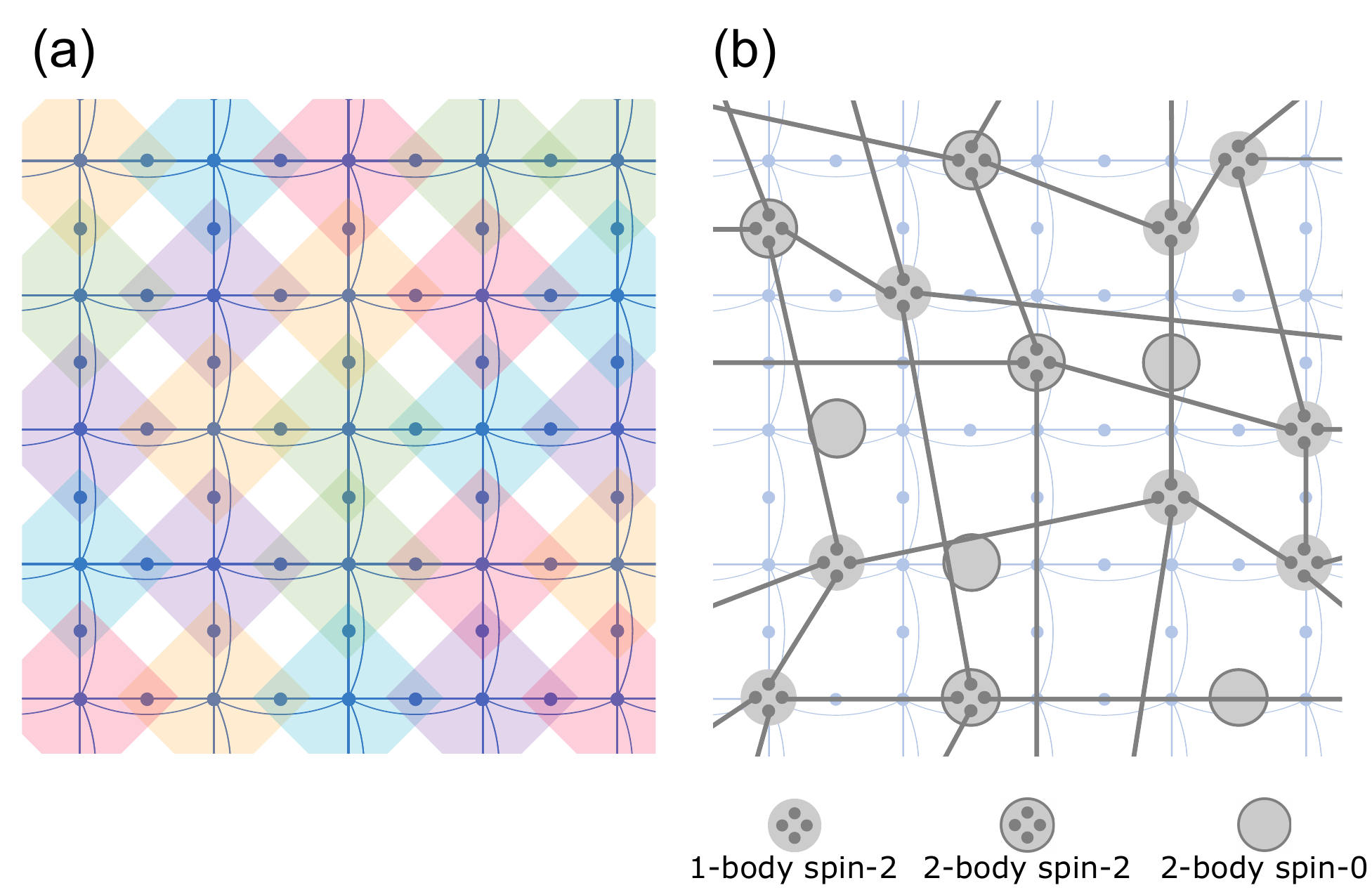}
  \caption{Components of 
  $|\text{GS}_{2,\includegraphics[width=0.011\textwidth]{2DTasaki_symbol.pdf}}\rangle$,
  the exact ground state of spin-2 bosons on the 2D Tasaki lattice.
  (a) Visualization of \Eq{FPS_30} with a typical choice of $\alpha_1,...,\alpha_N$. The five possible values of $\alpha_j$ are represented by five different colors.
  Equation~(\ref{FPS_30}) is a fully packed state (FPS).
  (b) Visualization of \Eq{eq:hidden_VBS_2D_Tasaki} with a typical choice of $\bm{r}_1,...,\bm{r}_N$.
  The state exhibits ``hidden VBS order": if we ignore all two-body spin singlets, the remaining particles form a VBS state.
  The ground state can be viewed as a linear combination of such states with different charge configurations.}
  \label{fig:hidden_VBS_2D}
\end{figure}

Let $|\text{GS}_{f,X}\rangle$ be the exact and unique ground state of $\hat{H}^{f,X}$.
In terms of the Fock basis, $|\text{GS}_{f,X}\rangle$ reads
\begin{equation}
\begin{split}
	|\text{GS}_{f,X}\rangle = &\sum_{\alpha_1,...,\alpha_N=-f}^f S_{\alpha_1,...,\alpha_N} \sum_{\bm{r}_1,...,\bm{r}_N} C_{\bm{r}_1,...,\bm{r}_N}  \\
	&\times \hat{a}^\dagger_{\bm{r}_1,\alpha_1}...\hat{a}^\dagger_{\bm{r}_N,\alpha_N}  |\text{vac}\rangle,
\end{split} \label{GS_X}
\end{equation}
where $S_{\alpha_1,...,\alpha_N}$ and $C_{\bm{r}_1,...,\bm{r}_N}$ are coefficients that correspond
to different spin and charge configurations, respectively.
Summing over the charge DOF first gives an FPS:
\begin{equation}
\begin{split}
	&\sum_{\bm{r}_1,...,\bm{r}_N} C_{\bm{r}_1,...,\bm{r}_N} \ \hat{a}^\dagger_{\bm{r}_1,\alpha_1}...\hat{a}^\dagger_{\bm{r}_N,\alpha_N}  |\text{vac}\rangle \\
	=&	( \hat{B}^{f,X}_{1, \alpha_1 } )^\dagger ( \hat{B}^{f,X}_{2, \alpha_2 } )^\dagger ... ( \hat{B}^{f,X}_{N, \alpha_N } )^\dagger |\text{vac} \rangle.
\end{split}
\label{FPS_30}
\end{equation}
$|\text{GS}_{f,X}\rangle$ is a linear combination of FPSs with different spin configurations, 
and the coefficients $S_{\alpha_1,...,\alpha_N}$ in \Eq{GS_X} and \Eq{VBS_fX} are identical.
However, if we sum over the spin DOF first:
\begin{equation}
	\sum_{\alpha_1,...,\alpha_N=-f}^f S_{\alpha_1,...,\alpha_N} \ \hat{a}^\dagger_{\bm{r}_1,\alpha_1}...\hat{a}^\dagger_{\bm{r}_N,\alpha_N}  |\text{vac}\rangle,  \label{eq:hidden_VBS_2D_Tasaki}
\end{equation}
we will get a state with ``hidden VBS order".
$|\text{GS}_{f,X}\rangle$ can then be alternatively viewed as a linear combination of such states with different charge configurations.
The above analysis suggests that the FPS and ``hidden VBS order" are 
two different but equivalent pictures of understanding the structure of the ground states,
and they together reflect spin and charge fluctuations at zero temperature.
Figure \ref{fig:hidden_VBS_2D} gives examples of
an FPS and a state with ``hidden VBS order"
of spin-2 bosons on the 2D Tasaki lattice.


Let us note that
although an exact ground state $|\text{GS}_{f,X}\rangle$ is a result of fine-tuned interactions,
one can readily believe that the quantum phase 
represented by $|\text{GS}_{f,X}\rangle$ (to be discussed in Sec.~\ref{Sec_GS_SPT})
exists in rather broad parameter regions,
as supported by the evidences shown 
in Sec.~\ref{Sec_PerturbationTheory} and \ref{Sec_NumericalAnalysis} for the sawtooth chain.


\subsection{Classifying the ground states from the \mbox{viewpoint of} SPT phases} \label{Sec_GS_SPT}

As the unique ground state, $|\text{GS}_{f,X}\rangle$ preserves all the symmetries of the system.
One can always think of $|\text{GS}_{f,X}\rangle$ as a representative state of a
certain disordered, gapped, short-range entangled, and symmetry-protected quantum phase.
In order to classify the phases represented by $|\text{GS}_{f,X}\rangle$ with various $f$ and $X$,
there are two main questions that we need to answer.
First, what is the phase of
the corresponding VBS state
$|\text{VBS}_{f,X'}\rangle$?
Second, are the two states $|\text{GS}_{f,X}\rangle$ and $|\text{VBS}_{f,X'}\rangle$ in exactly the same phase?

Recall that for $f=1$ and $X$ being the sawtooth chain, the answers to the two questions
have been completely listed in Table~\ref{table:u}.
The two states \Eq{GS} and \Eq{VBS_AKLT} are in the same phase except when
the inversion symmetry is involved.
In $d>1$ dimensions, however, 
regarding the first question, given an arbitrary $f$ and $X'$,
there is so far no complete answer about the phase of
$|\text{VBS}_{f,X'}\rangle$
in terms of all of its symmetry groups.
Nevertheless, it has been known that with on-site symmetry~\footnote{
``On-site symmetry'' is also called ``internal symmetry''.
It refers to a global symmetry that can be factorized site-by-site, 
and the symmetry operation on each site is
an endomorphism of the on-site Hilbert space.
The $\mathbb{Z}_2 \times \mathbb{Z}_2$ spin rotation is an on-site symmetry because, for example,
$\hat{U}(z)= \prod_{r} \exp(-\mathrm{i}\pi \hat{S}_r^{z}) $, and 
$\hat{S}_r^{z}$ acts only on the local Hilbert space.
} alone,
$|\text{VBS}_{f,X'}\rangle$ always represents a trivial phase in $d>1$ dimensions,
while the combination of 
certain on-site and spatial symmetry
can give an SPT/trivial classification,
as will be discussed in Sec.~\ref{subsub_LSM}.
In addition,
we will show in Sec.~\ref{subsub_crystalline} that 
crystalline symmetries alone can also give an SPT/trivial classification
for $|\text{VBS}_{f,X'}\rangle$.
Regarding the second question,
we claim that $|\text{GS}_{f,X}\rangle$ and $|\text{VBS}_{f,X'}\rangle$ are always in the same phase
protected by on-site symmetry alone or the
 combination of on-site and translation symmetry,
 see Sec.~\ref{subsub_smooth}.
However, their phases 
should be investigated on a case-by-case basis
when crystalline symmetries come into play,
see Sec.~\ref{subsub_crystalline}\&\ref{subsub_LSM}.
In particular, we find that the
charge fluctuations in $|\text{GS}_{f,X}\rangle$
can play a nontrivial role in the SPT orders protected 
by crystalline symmetries.
In the following, for simplicity, we focus only on 
several concrete examples.
The analysis, however, applies to general cases.

\subsubsection{Smooth path argument} \label{subsub_smooth}

\begin{figure}[h]
  \centering
  \includegraphics[width=0.13\textwidth]{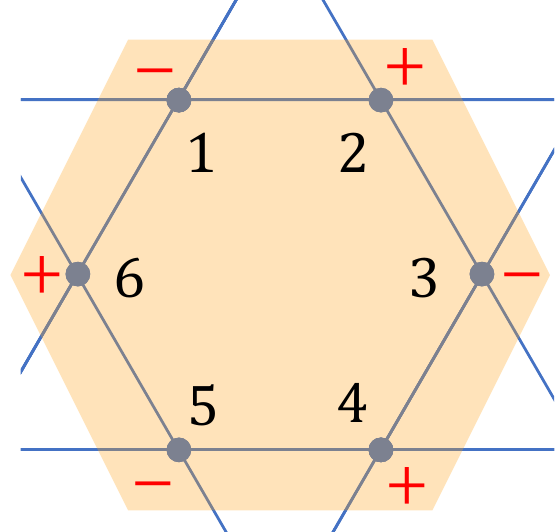}
  \caption{A CLS in the kagome lattice. 
  Sign of the amplitude alternates from site 1 to 6.}
  \label{fig:kagome_CLS}
\end{figure}

In terms of the combination of SO(3) spin rotation 
and translation symmetry
[denote the symmetry group as SO(3)$\times$trn],
the \mbox{spin-2} VBS state on a square lattice 
$|\text{VBS}_{2,\Box}\rangle$
is in an SPT phase~\footnote{
See Sec~\ref{subsub_LSM} or Refs.~\cite{PhysRevB.87.155114, PhysRevX.6.041068, CZX, zeng2019quantum, PhysRevB.94.235159}.
When the translation symmetry is indispensable to protect an SPT phase, such a phase is often called a weak SPT phase.},
while the \mbox{spin-3} VBS state on a triangular lattice 
$|\text{VBS}_{3,\triangle}\rangle$
represents a trivial phase~\footnote{Consider
$|\text{VBS}_{3,\triangle}\rangle$ defined on 
a half-infinite plane. On its 1D boundary, 
every site hosts two ``dangling'' \mbox{spin-$1/2$'s},
as shown in the figure:
\includegraphics[width=0.05\textwidth]{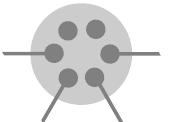}.
Since the six \mbox{spin-$1/2$'s} on the same site form a totally symmetric \mbox{spin-3} degree of freedom,
the two ``dangling'' \mbox{spin-$1/2$'s} have to form a symmetric \mbox{spin-1} degree of freedom. 
One can add perturbations at the boundary that couple
these \mbox{spin-1's} through, for example, the translation invariant \mbox{spin-1} AKLT Hamiltonian. 
This perturbation thus results in a gapped edge state without breaking the combination of SO(3) and translation symmetry, and hence the state $|\text{VBS}_{3,\triangle}\rangle$ is in a trivial phase protected by such symmetry.
In this sense, we say that the edge state of 
$|\text{VBS}_{3,\triangle}\rangle$
can be trivially gapped out.
As for $|\text{VBS}_{2,\Box}\rangle$,
its edge state cannot be gapped out while preserving
the $\text{SO(3)}\times\text{trn}$ symmetry,
see Appendix~\ref{App_LSM} or
Refs.~\cite{CZX, zeng2019quantum, PhysRevB.94.235159}.}.
A key observation is that 
$|\text{GS}_{f,X}\rangle$ can always be
smoothly deformed to $|\text{VBS}_{f,X'}\rangle$
without breaking the SO(3)$\times$trn symmetry.
Therefore, the two states are in the same phase.
For example, 
let 
$\hat{B}^{3,\includegraphics[width=0.011\textwidth]{kagome_symbol2.pdf}}_{1, \alpha }$
be one of the CLS operators on the kagome lattice,
whose exact form is given by
\begin{equation}
	 \hat{B}^{3,\includegraphics[width=0.012\textwidth]{kagome_symbol2.pdf}}_{1, \alpha } 
	 =  \frac{1}{\sqrt{6}} \sum_{x=1}^6 (-1)^x \hat{a}_{x,\alpha}, \label{eq:CLS_kgm}
\end{equation}
where the six sites labeled by $x$ form vertices of a hexagon, as shown in \Fig{fig:kagome_CLS}.
We then define a $\lambda$-deformed CLS operator as
\begin{equation}
	 \hat{B}^{3,\includegraphics[width=0.012\textwidth]{kagome_symbol2.pdf}}_{1, \alpha } (\lambda)
	 = \frac{1}{\sqrt{\lambda^2+5}}\left(  \sum_{x=1}^5 (-1)^x \hat{a}_{x,\alpha} + \lambda \hat{a}_{6,\alpha} \right),
\end{equation}
which satisfies 
$\hat{B}^{3,\includegraphics[width=0.011\textwidth]{kagome_symbol2.pdf}}_{1, \alpha } (1)= \hat{B}^{3,\includegraphics[width=0.011\textwidth]{kagome_symbol2.pdf}}_{1, \alpha }$
and 
$\lim_{\lambda \to \infty} \hat{B}^{3,\includegraphics[width=0.011\textwidth]{kagome_symbol2.pdf}}_{1, \alpha } (\lambda)= \hat{a}_{6,\alpha}$.
By applying lattice translation vectors, we can get all the other
$\hat{B}^{3,\includegraphics[width=0.011\textwidth]{kagome_symbol2.pdf}}_{j, \alpha }(\lambda) $
with $j=2,...,N$.
Now consider the state defined on the kagome lattice
\begin{equation}
	|\text{GS}_{3,\includegraphics[width=0.012\textwidth]{kagome_symbol2.pdf}}(\lambda)\rangle
	= \sum_{\alpha_1,...,\alpha_N=-3}^3 S_{\alpha_1,...,\alpha_N} 
	\prod_{j=1}^N
	\left(\hat{B}^{3,\includegraphics[width=0.012\textwidth]{kagome_symbol2.pdf}}_{j, \alpha_j }(\lambda)\right)^\dagger
	|\text{vac}\rangle, \label{GS_3kgm_smooth}
\end{equation}
where $\{S_{\alpha_1,...,\alpha_N}\}$ are chosen such that
$|\text{GS}_{3,\includegraphics[width=0.011\textwidth]{kagome_symbol2.pdf}}(1)\rangle=
|\text{GS}_{3,\includegraphics[width=0.011\textwidth]{kagome_symbol2.pdf}}\rangle$
is the original ground state of 
$\hat{H}^{3,\includegraphics[width=0.011\textwidth]{kagome_symbol2.pdf}}$.
One can then easily see that 
$\lim_{\lambda \to \infty}|\text{GS}_{3,\includegraphics[width=0.011\textwidth]{kagome_symbol2.pdf}}(\infty)\rangle
=|\text{VBS}_{3,\triangle}\rangle$ and
the state 
$|\text{GS}_{3,\includegraphics[width=0.011\textwidth]{kagome_symbol2.pdf}}(\lambda)\rangle$
remains SO(3)$\times$trn symmetric and short-range entangled for $1<\lambda<\infty$.
Therefore, $|\text{GS}_{3,\includegraphics[width=0.011\textwidth]{kagome_symbol2.pdf}}\rangle$
and $|\text{VBS}_{3,\triangle}\rangle$ are smoothly connected and are in the same trivial phase protected by
SO(3)$\times$trn.
For an arbitrary $f$ and $X$,
a smooth path between
$|\text{GS}_{f,X}\rangle$ and $|\text{VBS}_{f,X'}\rangle$
can always be explicitly constructed by
smoothly deforming every CLS in $X$ to one single site
while preserving the SO(3) or SO(3)$\times$trn symmetry,
and thus the two states always represent the same phase 
protected by the symmetry.
Let 
$|\text{GS}_{2,\includegraphics[width=0.012\textwidth]{2DTasaki_symbol.pdf}}\rangle$
be the exact ground state of spin-2 bosons in the 2D Tasaki lattice;
for the above reason,
$|\text{GS}_{2,\includegraphics[width=0.012\textwidth]{2DTasaki_symbol.pdf}}\rangle$
and
$|\text{VBS}_{2,\Box}\rangle$
are in the same SPT phase protected by SO(3)$\times$trn.

Table~\ref{tab:smooth} summarizes current results.
The purpose of this section (Sec.~\ref{subsub_smooth})
is to demonstrate the smooth path argument.
Details behind the on-site$\times$trn symmetry will be
discussed later (Sec.~\ref{subsub_LSM} \& Appendix~\ref{App_LSM}).

\begin{table}[]
\begin{tabular}{l|cc}
\hline \hline
                                                                                     & \ SO(3)  \                  & \ $\text{SO(3)}\times \text{trn}$              \\ \hline
$|\text{VBS}_{2,\Box}\rangle$                                                        & \multirow{2}{*}{\ trivial\ } & \multirow{2}{*}{\ \textbf{SPT}} \\
$|\text{GS}_{2,\includegraphics[width=0.011\textwidth]{2DTasaki_symbol.pdf}}\rangle$ &                          &                               \\ \hline
$|\text{VBS}_{3,\triangle}\rangle$                                                   & \multirow{2}{*}{\ trivial\ } & \multirow{2}{*}{\ trivial}      \\
$|\text{GS}_{3,\includegraphics[width=0.011\textwidth]{kagome_symbol2.pdf}}\rangle$  &                          &                               
\\ \hline \hline
\end{tabular}
\caption{In terms of on-site symmetry or on-site$\times$trn symmetry,
$|\text{GS}_{f,X}\rangle$ and $|\text{VBS}_{f,X'}\rangle$
are always in the same phase.} \label{tab:smooth}
\end{table}

\FloatBarrier

\subsubsection{SPT/trivial phases protected by crystalline symmetries alone} \label{subsub_crystalline}

For general crystalline symmetries, however,
the smooth path argument does not always apply.
For simplicity, we consider only point group symmetries in $d=2$ dimension
in this article.
Let $G$ be a point group of a Hamiltonian 
with a unique gapped ground state $|\Psi\rangle$.
Let $\hat{U}(q)$ be the symmetry operation (on the Hilbert space) corresponding to the group element $q\in G$.
Subjected to $q$, the unique ground state transforms as,
\begin{equation}
	|\Psi\rangle \to \hat{U}(q) |\Psi\rangle
	= \mathrm{e}^{\mathrm{i}\theta_{q}} |\Psi\rangle,
	\label{eq:1st_rep}
\end{equation} 
where the phase factors $\{\mathrm{e}^{\mathrm{i}\theta_{q}}\}_{q\in G}$ form a 1D representation of $G$,
and different 1D representations
label different phases protected by the point group $G$~\footnote{
All 1D representations of $G$ form an Abelian group,
which is the first cohomology group $H^1[G,U(1)]$.
In \mbox{$d=1,2$} dimensions, $H^1[G,U(1)]$ is believed to give a complete classification of phases protected by the point group $G$.
In $d=3$ dimension, extra indices are needed for 
a complete classification. 
See Refs.~\cite{PhysRevX.7.011020, PhysRevB.96.205106,PhysRevX.8.011040} for general classification theories.}.
When $\{\mathrm{e}^{\mathrm{i}\theta_{q}}\}_{q\in G}$ is a
trivial representation, that is, $\mathrm{e}^{\mathrm{i}\theta_{q}}=1$ for all $q\in G$,
$|\Psi\rangle$ is in a trivial phase.
On the other hand, $|\Psi\rangle$ is in an SPT phase if 
$\{\mathrm{e}^{\mathrm{i}\theta_{q}}\}_{q\in G}$ is a
nontrivial representation~\cite{PhysRevX.7.011020,PhysRevB.96.205106,  PhysRevX.8.011040}.
It is important to be aware that
for point group symmetries alone in $d<3$ dimensions,
the SPT/trivial classifications
become meaningless
when there are microscopic DOF lying
precisely at symmetry centers.
See Sec.~IB of Ref.~\cite{PhysRevB.96.205106}
or Appendix~A of Ref.~\cite{PhysRevX.7.011020}
for details.
In other words, it is only legal to
put the symmetry centers in vacuum.

\begin{figure}[h]
  \centering
  \includegraphics[width=0.49\textwidth]{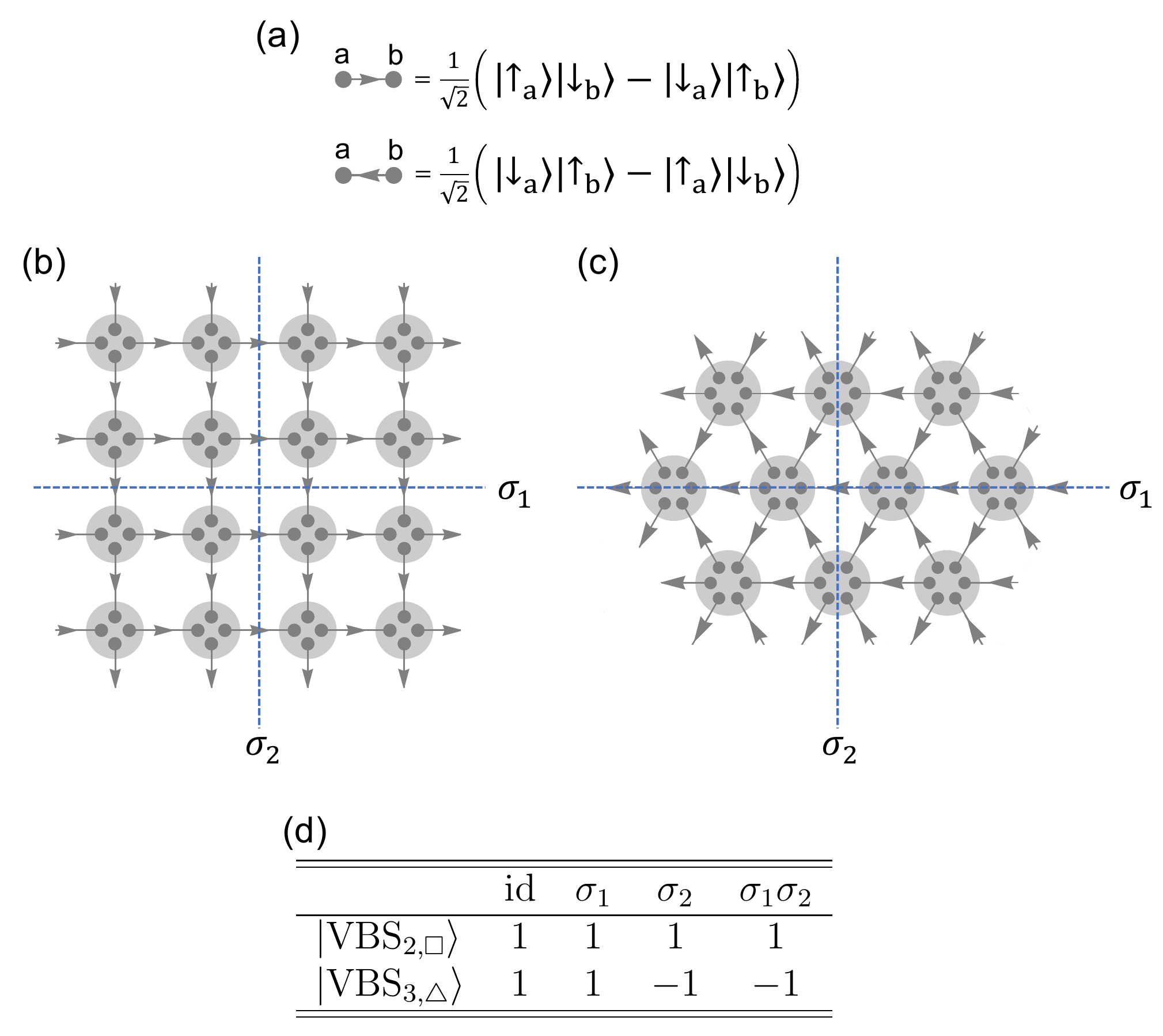}
  \caption{(a) A direction can be assigned to each singlet bond. Two opposite directions differ by a minus sign.
  (b) $|\text{VBS}_{2,\Box}\rangle$ is invariant under reflections about mirror planes $\sigma_1$ and $\sigma_2$. 
  We assume the symmetry center lies in a plaquette.
  (c) $|\text{VBS}_{3,\triangle}\rangle$ is invariant under reflections about mirror planes $\sigma_1$ and $\sigma_2$.
  (d) $|\text{VBS}_{2,\Box}\rangle$ results in a trivial representation of $D_2=\{\mathrm{id},\sigma_1, \sigma_2, \sigma_1 \sigma_2\}$ (plaquette-centered), while $|\text{VBS}_{3,\triangle}\rangle$ results in a nontrivial representation of $D_2$ (bond-centered).}
  \label{fig:D2_VBS}
\end{figure}

In the graph representation of VBS states,
we can assign an arbitrary direction to each singlet bond,
because a singlet state is antisymmetric.
Reversing the direction of a singlet bond 
is equivalent to adding a minus sign;
see \Fig{fig:D2_VBS}(a).
Let us consider the point group $D_2$ as a simple example.
Elements of $D_2$ are generated by two perpendicular 
mirror planes $\sigma_1$ and $\sigma_2$, as shown in \Fig{fig:D2_VBS}(b)-(c).
Consider the states
$|\text{VBS}_{2,\Box}\rangle$
and $|\text{VBS}_{3,\triangle}\rangle$
in the thermodynamic limit. 
$|\text{VBS}_{2,\Box}\rangle$ is $D_2$-symmetric around the center of a plaquette, the center of a bond, or a site.
$|\text{VBS}_{3,\triangle}\rangle$ is 
$D_2$-symmetric around the center of a bond or a site.
As emphasized above, to classify their phases, 
it is illegal to 
put the symmetry center on a site. 
For $|\text{VBS}_{2,\Box}\rangle$ 
with a plaquette-centered $D_2$ symmetry,
there are an
even number~\footnote{
In the
thermodynamic limit, 
it may be subtle to ask if the number is even or odd.
Nevertheless, it is always possible to 
identify the phase of a finite-size system.
The fact is that, the symmetry-protected phase of a finite-size system 
should be identical to that of an infinite system.
The reason is as follows.
All the symmetry-protected phases
are about local properties of the system, 
since there is only short-range entanglement 
in the bulk~\cite{zeng2019quantum}.
In fact,
according to Refs.~\cite{PhysRevX.7.011020, PhysRevB.96.205106},
the point-group-symmetry-protected phases of a 1D or 2D system are determined by 
the properties of a local region around the symmetry center, and the
size of the local region  
roughly agrees with the correlation length.
Local properties are obviously not affected by the those degrees of freedom that are infinitely far away.}
of singlet bonds
being reversed by a mirror reflection [see \Fig{fig:D2_VBS}(b)],
thus $\hat{U}(q) |\text{VBS}_{2,\Box}\rangle
	= |\text{VBS}_{2,\Box}\rangle$, $\forall q \in D_2$.
However, for a bond-centered $D_2$ symmetry,
with respect to the mirror plane perpendicular to 
the central bond, there are an
odd number of singlet bonds being reversed,
which results in a nontrivial representation of $D_2$.
We see that in this example, 
the phase depends on 
the position of the symmetry center~\footnote{How can the same point group 
in the same system results in
two distinct phases by only choosing 
a different symmetry center? 
The reason is that the plaquette-centered symmetry and the bond-centered symmetry are inequivalent 
in the sense that one symmetry alone does not imply the other.
In the presence of translation symmetry, point groups centered in inequivalent positions
are included in a larger space group (or wallpaper group). 
It is reasonable to say that the state $|\text{VBS}_{2,\Box}\rangle$ 
is in an SPT phase protected by the wallpaper group $p4m$,
since bond-centered $D_2$ is a subgroup of $p4m$. 
See Ref~\cite{PhysRevB.96.205106} for the theory of
wallpaper-group-protected phases.}.
For some reason that will be clear later, 
we consider only the plaquette-centered symmetry for $|\text{VBS}_{2,\Box}\rangle$
in the remnant of this article.
For $|\text{VBS}_{3,\triangle}\rangle$, 
the bond-centered symmetry is the only legal choice.
As shown in \Fig{fig:D2_VBS}(c), 
the mirror reflection $\sigma_2$ reverses 
an odd number of bonds, we thus have
$\hat{U}(\sigma_2) |\text{VBS}_{3,\triangle}\rangle
	= -|\text{VBS}_{3,\triangle}\rangle$.
As listed in \Fig{fig:D2_VBS}(d),
$|\text{VBS}_{2,\Box}\rangle$ results in a trivial representation of plaquette-centered $D_2$ and is therefore in a trivial phase,
while $|\text{VBS}_{3,\triangle}\rangle$ is in an SPT phase protected by $D_2$.
Note that from the above discussion, 
one might naively think that
one single mirror plane alone (point group $D_1=\mathbb{Z}_2^P$)
is sufficient to distinguish the SPT from the trivial phase, which is indeed true in $d=1,3$ dimensions~\cite{PTBO_2012, PhysRevX.7.011020}.
However, in the $d=2$ dimension,  $D_1$ symmetry alone
can give only a trivial phase; see Ref.~\cite{PhysRevB.96.205106}.

\begin{figure}[h]
  \centering
  \includegraphics[width=0.5\textwidth]{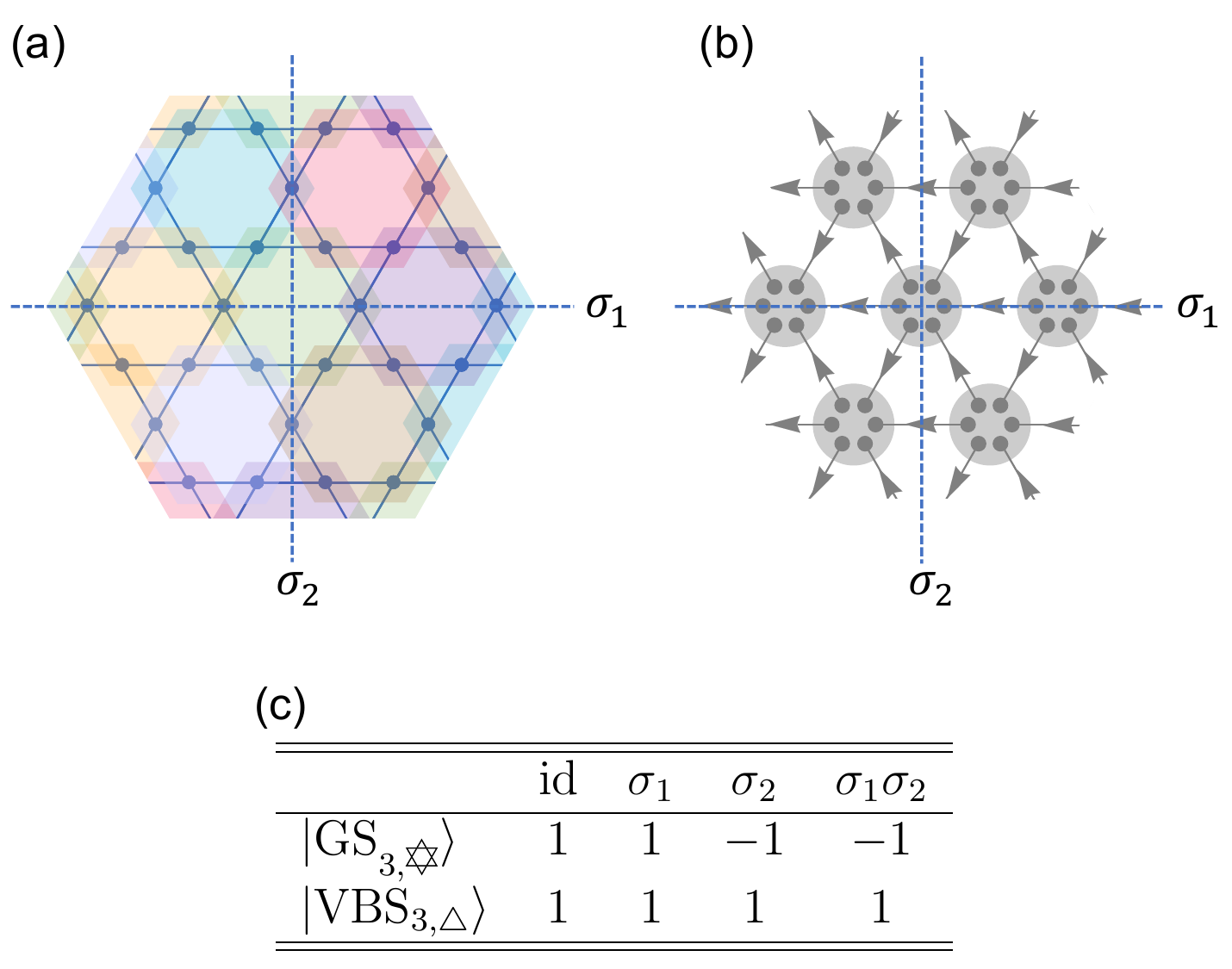}
  \caption{(a) An FPS on the kagome lattice. 
  ($|\text{GS}_{3,\includegraphics[width=0.011\textwidth]{kagome_symbol2.pdf}}\rangle$ is $D_2$ symmetric, 
  and it
  is a superposition of FPSs with different spin configurations.)
  We require that the symmetry center of $D_2$ lies
  at the geometric center of a hexagon.
  (b) 
  For $|\text{VBS}_{3,\triangle}\rangle$,
  the symmetry center
  should lie at a site (spin)
  in order to be compatible with (a).
  (c) 1D representations of $D_2$ 
  associated with
  $|\text{GS}_{3,\includegraphics[width=0.011\textwidth]{kagome_symbol2.pdf}}\rangle$
  and
  $|\text{VBS}_{3,\triangle}\rangle$.
  Note that the latter state transforms trivially,
  but this does not mean that it is in a trivial phase.}
  \label{fig:D2_GS_VBS}
\end{figure}

In general,
a smooth path between
$|\text{GS}_{f,X}\rangle$ and $|\text{VBS}_{f,X'}\rangle$
that preserve on-site$\times$trn symmetry
may or may not break crystalline symmetries.
For example,
by smoothly deforming every CLS into the single site 
at its center,
$|\text{GS}_{2,\includegraphics[width=0.012\textwidth]{2DTasaki_symbol.pdf}}\rangle$
reduces to
$|\text{VBS}_{2,\Box}\rangle$
while preserving the plaquette-centered $D_2$ symmetry,
see \Fig{fig:hidden_VBS_2D}(a).
(Though $|\text{GS}_{2,\includegraphics[width=0.012\textwidth]{2DTasaki_symbol.pdf}}\rangle$
also has site-centered $D_2$ symmetry, such symmetry does not give a phase classification.)
For $|\text{GS}_{3,\includegraphics[width=0.011\textwidth]{kagome_symbol2.pdf}}\rangle$,
however, the smooth path described by \Eq{GS_3kgm_smooth}
breaks the $D_2$ symmetry.
In general, when we are not able to
find a path that is both
crystalline-symmetry-preserving
and smooth,
such a path either is too complicated to be explicitly found
or simply does not exist. 
Nevertheless, it is always possible to
investigate the 
crystalline-symmetry-protected
phase of 
$|\text{GS}_{f,X}\rangle$
case-by-case.
We again use 
$|\text{GS}_{3,\includegraphics[width=0.011\textwidth]{kagome_symbol2.pdf}}\rangle$
as an example.
As shown in \Fig{fig:D2_GS_VBS}(a),
we put the symmetry center at
the geometric center of a hexagonal plaquette.
A CLS in the kagome lattice can actually be regarded as a
zero-dimensional SPT phase protected by $D_2$,
because, for example,
according to \Eq{eq:CLS_kgm} and \Fig{fig:kagome_CLS},
$\hat{U}(\sigma_2) \hat{B}^{3,\includegraphics[width=0.011\textwidth]{kagome_symbol2.pdf}}_{1, \alpha } \hat{U}(\sigma_2) 
=- \hat{B}^{3,\includegraphics[width=0.011\textwidth]{kagome_symbol2.pdf}}_{1, \alpha }$.
The many-body ground state 
$|\text{GS}_{3,\includegraphics[width=0.011\textwidth]{kagome_symbol2.pdf}}\rangle$
is a fully packing of CLSs
with entangled spin DOF.
The spin configurations of
$|\text{GS}_{3,\includegraphics[width=0.011\textwidth]{kagome_symbol2.pdf}}\rangle$,
which is inherited from $|\text{VBS}_{3,\triangle}\rangle$,
transforms trivially,
as shown in Figs.~\ref{fig:D2_GS_VBS}(b) and \ref{fig:D2_GS_VBS}(c).
[Note that \Fig{fig:D2_GS_VBS}(b) does not imply that 
$|\text{VBS}_{3,\triangle}\rangle$
is in a trivial phase protected by $D_2$, as the symmetry is site-centered.]
Nevertheless, $|\text{GS}_{3,\includegraphics[width=0.011\textwidth]{kagome_symbol2.pdf}}\rangle$
yields a nontrivial representation of $D_2$
thanks to how CLSs transform.
We thus see that 
$|\text{GS}_{3,\includegraphics[width=0.011\textwidth]{kagome_symbol2.pdf}}\rangle$
is in an SPT phase protected by $D_2$;
the SPT phase is purely a result of 
charge fluctuations at zero temperature,
as the spin DOF contribute trivially.
Similarly, 
with the symmetry center in \Fig{fig:D2_GS_VBS}(a),
one can also show that 
$|\text{GS}_{3,\includegraphics[width=0.011\textwidth]{kagome_symbol2.pdf}}\rangle$
represents an SPT phase protected by point group $D_3$
or $D_6$,
while $D_3$ and $D_6$ are not a proper symmetries 
for the phase of
$|\text{VBS}_{3,\triangle}\rangle$, since the VBS state is $D_3$ or $D_6$ invariant
only about a site.
Once again, the SPT phase
of $|\text{GS}_{3,\includegraphics[width=0.011\textwidth]{kagome_symbol2.pdf}}\rangle$
protected by $D_3$ or $D_6$
originates from the charge fluctuations of each CLS.
Results of the current section are summarized in
Table~\ref{tab:pgSPT}.


\begin{table}[]
\begin{tabular}{l|cc|cc}
\hline \hline
      & \multicolumn{1}{l}{$|\text{VBS}_{2,\Box}\rangle$} & \multicolumn{1}{l|}{$|\text{GS}_{2,\includegraphics[width=0.012\textwidth]{2DTasaki_symbol.pdf}}\rangle$} & \multicolumn{1}{l}{$|\text{VBS}_{3,\triangle}\rangle$} & \multicolumn{1}{l}{$|\text{GS}_{3,\includegraphics[width=0.011\textwidth]{kagome_symbol2.pdf}}\rangle$} \\ \hline
$D_1$ & trivial                                           & trivial                                                                                                   & trivial                                                & trivial                                                                                                 \\
$D_2$ & trivial                                           & trivial                                                                                                   & \textbf{SPT}                                                    & \textbf{SPT}                                                                                                     \\
$D_3$ & N/A                                           & N/A                                                                                                   & N/A                                           & \textbf{SPT}                                                                                            \\
$D_4$ & trivial                                              & trivial                                                                                                       & N/A                                                    & N/A                                                                                            \\ \hline \hline
\end{tabular}
\caption{Classifying the phases of 
2D bosonic states
in terms of point group symmetries.
Symmetry centers of the points groups 
are placed in vacuum.
In addition, $D_2$ and $D_4$ are assumed to be 
plaquette-centered symmetries
for $|\text{VBS}_{2,\Box}\rangle$.
N/A means either that the state does not have such symmetry
or that the symmetry can only be site-centered.} 
\label{tab:pgSPT}
\end{table}

For a specific symmetry, 
to identify the phase
of $| \text{GS}_{f,X} \rangle$,
we can first try to find a both symmetry-preserving and 
smooth path that
connects $| \text{GS}_{f,X} \rangle$ to
$| \text{VBS}_{f,X'} \rangle$,
provided that the phase of $| \text{VBS}_{f,X'} \rangle$
is already known.
(For point group symmetries, 
we require that 
there are no microscopic DOF
lying at the symmetry center
all along the path~\footnote{The state in \Eq{GS} with $ 0 < |\lambda| < \infty $
is inversion symmetric only about a site.
In Table~\ref{table:u}, we can see that this path smoothly connects two
distinct phases
even when the (site-centered) inversion symmetry is preserved.}.)
When such a path cannot be explicitly found,
it is either too complicated to be found or  
simply absent.
Nevertheless, for point group symmetries,
based on \Eq{eq:1st_rep},
one can always classify the phase of 
$| \text{GS}_{f,X} \rangle$
without the help of the smooth path argument.

We would like to mention some related research.
The ``fragile Mott insulator" studied in
Ref~\cite{PhysRevLett.105.166402}
can be understood as SPT phases 
protected by point group symmetries.
In contrast,
the ``featureless Mott insulator" 
of spinless bosons studied in
Refs.~\cite{PhysRevLett.110.125301, kimchi2013featureless}
should be classified into
trivial phases protected by point group symmetries.

\FloatBarrier

\subsubsection{SPT phases protected by $\text{on-site} \times \text{crystalline}$ symmetries:
a result of the Lieb-Schultz-Mattis (LSM) theorems}
\label{subsub_LSM}

\begin{table}[h]
\begin{tabular}{l|cc|cc}
\hline \hline
                                              & \multicolumn{1}{l}{$|\text{VBS}_{2,\Box}\rangle$} & \multicolumn{1}{l|}{$|\text{GS}_{2,\includegraphics[width=0.011\textwidth]{2DTasaki_symbol.pdf}}\rangle$} & \multicolumn{1}{l}{$|\text{VBS}_{3,\includegraphics[width=0.013\textwidth]{cube_symbol.pdf}}\rangle$} & \multicolumn{1}{l}{$|\text{GS}_{3,\text{3DTas}}\rangle$} \\ \hline
$\mathbb{Z}_2 \times \mathbb{Z}_2\times\text{trn}$  & \textbf{SPT}                                      & \textbf{SPT}                                                                                              & \textbf{SPT}                                                                                          & \textbf{SPT}                                             \\
$\text{TR}\times\text{trn}$                                & \textbf{SPT}                                      & \textbf{SPT}                                                                                              & \textbf{SPT}                                                                                          & \textbf{SPT}                                             \\
$\mathbb{Z}_2 \times \mathbb{Z}_2\times D_1 $ & \textbf{SPT}                                      & \textbf{SPT}                                                                                              & trivial                                                                                               & trivial                                                  \\
$\text{TR}\times D_1$                                & \textbf{SPT}                                      & \textbf{SPT}                                                                                              & trivial                                                                                               & trivial                                                  \\ \hline \hline
\end{tabular}
\caption{Classifying the phases of 2D and 3D bosonic states
in terms of some $\text{on-site}\times\text{crystalline}$ symmetries.
$D_1=\mathbb{Z}_2^P$ in this table
refers to mirror reflection symmetry 
\textit{along an array of sites}.
All the SPT phases on the table are a result of LSM theorems.
The spin-3 VBS state on a cubic lattice
$|\text{VBS}_{3,\includegraphics[width=0.013\textwidth]{cube_symbol.pdf}}\rangle$
represents a trivial phase in terms of $\mathbb{Z}_2 \times \mathbb{Z}_2\times D_1 $
or $\text{TR}\times D_1$, because its surface state can be
trivially gapped out without breaking these two symmetries.
$|\text{GS}_{3,\text{3DTas}}\rangle$ refers to the exact ground state of spin-3 bosons on the 3D Tasaki lattice,
and it is smoothly connected to $|\text{VBS}_{3,\includegraphics[width=0.013\textwidth]{cube_symbol.pdf}}\rangle$
while preserving the four symmetries on the table.
}
\label{tab:LSM}
\end{table}

There can also be nontrivial interplay between 
on-site and crystalline symmetries.
For instance, the $\text{SO(3)}\times\text{trn}$ symmetry
introduced in Sec.~\ref{subsub_smooth} is one such example.
In fact, 
$\text{SO(3)}\times\text{trn}$ 
is sufficient but not necessary
to protect the SPT phase of $|\text{VBS}_{2,\Box}\rangle$
: the subgroup
\mbox{$\mathbb{Z}_2 \times \mathbb{Z}_2 \times \text{trn}$}
is enough.
Besides,
$|\text{VBS}_{2,\Box}\rangle$
is in an SPT phase protected by 
the combination of 
time-reversal (TR)
and \textit{site-centered} mirror reflection (= point group $D_1=\mathbb{Z}_2^P$) symmetry.
Other such symmetries include $\text{TR}\times\text{trn}$,
$\mathbb{Z}_2 \times \mathbb{Z}_2\times D_1 $, and so on, 
see Table~\ref{tab:LSM}.
In fact, the reason why certain 
$\text{on-site}\times\text{spatial}$ symmetries
can give SPT/trivial classifications in $d>1$ dimensions 
is closely related to the LSM theorems,
see Appendix~\ref{App_LSM} for details. 

As explained in Sec.~\ref{subsub_smooth},
$|\text{GS}_{f,X}\rangle$ is always smoothly 
connected to $|\text{VBS}_{f,X'}\rangle$
while preserving \mbox{$\text{on-site}\times\text{trn}$} symmetries.
However, for \mbox{$\text{on-site}\times\text{point group}$}
or \mbox{$\text{on-site}\times\text{space group}$} symmetries,
a symmetry-preserving smooth path may not exist.
(Luckily, for $\mathbb{Z}_2 \times \mathbb{Z}_2\times D_1 $
and $\text{TR}\times D_1$, 
$|\text{GS}_{2,\includegraphics[width=0.011\textwidth]{2DTasaki_symbol.pdf}}\rangle$
and $|\text{GS}_{3,\text{3DTas}}\rangle$
are smoothly connected to their 
corresponding VBS states.)
In the case where a symmetry-preserving smooth path
cannot be found, the phase of $|\text{GS}_{f,X}\rangle$ 
can always be judged
by examining that 
if one can trivially gap out the edge state without out
breaking the symmetry:
$|\text{GS}_{f,X}\rangle$ is in an SPT phase if its
edge state cannot be trivially gapped out.

\section{Discussion}

We show that the SPT phases can be realized with 
short-range interacting
spinful bosons that are loaded on the lattices with a bottom flat band.
Such systems are described by the 
spinful Bose-Hubbard models.
The ground states of such systems have both spin and charge fluctuations.
The single-body eigenstates of a flat band can usually be
chosen to be strictly localized on finite number of sites,
known as compact localized states (CLSs).
When $N$ \mbox{spin-$f$} bosons are loaded on a bottom-flat-band lattice $X$ with $N$ unit cells,
at low temperatures, the particles' wave functions 
tend to avoid overlapping each other in order to minimize the system's energy.
In particular, when the interaction 
strength between spin-$f$ bosons
is fine-tuned, 
in the ground state $|\text{GS}_{f,X}\rangle$,
$N$ bosons exactly occupy $N$ CLSs on different patches.
We make use of the analogy between
the Hamiltonian that describes the $s$-wave collision among
spin-$f$ bosons and the spin-$f$ AKLT Hamiltonian.
This analogy enables us to exactly map 
$|\text{GS}_{f,X}\rangle$
onto $|\text{VBS}_{f,X'}\rangle$, where the latter state
is the spin-$f$ VBS state on the lattice $X'$.
This implies that $|\text{GS}_{f,X}\rangle$ is the exact and unique many-body ground state of the \mbox{spin-$f$}
Bose-Hubbard model.
The choice of $f$ and $X'$ is determined 
by the geometry of $X$.
Note that bottom-flat-band lattices $X$ can be 
constructed systematically,
see Sec.~\ref{Sec_GS_boson_flatband}.

Over the years, 
exact results have proved to be highly valuable in 
quantum and statistical physics.
Our work features the exact many-body ground states
$|\text{GS}_{f,X}\rangle$
of spinful itinerant systems.
The spin fluctuations of $|\text{GS}_{f,X}\rangle$
is inherited from $|\text{VBS}_{f,X'}\rangle$.
Therefore, with respect to the spin rotation symmetry
or the \mbox{$\text{spin rotation} \times \text{translation}$} symmetry,
the symmetry-protected phase of $|\text{GS}_{f,X}\rangle$ is identical
to that of $|\text{VBS}_{f,X'}\rangle$.
However, unlike $|\text{VBS}_{f,X'}\rangle$, the state
$|\text{GS}_{f,X}\rangle$ also possesses nonvanishing 
charge fluctuations,
and in terms of crystalline symmetries,
both spin and charge fluctuations in $|\text{GS}_{f,X}\rangle$ together
determine its symmetry-protected phase.
Hence, as explained in Sec.~\ref{subsub_crystalline},
one cannot simply conclude that
the crystalline-symmetry-protected phase of $|\text{GS}_{f,X}\rangle$ is also inherited from
 $|\text{VBS}_{f,X'}\rangle$,
because charge fluctuations may play a nontrivial role
in the former state,
which is indeed the case for spin-3 bosons in the kagome lattice.
Although our analysis in \mbox{$d>1$} dimensions is based on the exact ground states $|\text{GS}_{f,X}\rangle$ (as a consequence of fine-tuned parameters),
we expect that 
just like what has been shown in the \mbox{spin-1} BHMSC,
the SPT phases survive in wider parameter regions, 
and $|\text{GS}_{f,X}\rangle$ just serves as a representative state of the phases.


We having been ignoring the long-range
dipole-dipole interaction (DDI),
and this can be justified in many 
alkali-metal atom experiments.
In fact, several kinds of
transition-metal atoms can also be regarded as
spinful bosons,
such as $^{52}$Cr (spin-3), $^{164}$Dy (spin-8), 
and $^{168}$Er (spin-6)~\cite{KAWAGUCHI2012253,* RevModPhys.85.1191}.
Interestingly, these transition-metal atoms 
have very strong magnetic DDI~\cite{PhysRevLett_Cr,PhysRevLett_Dy,PhysRevLett_Er}.
It is also known that 
even for bosonic alkali-metal atoms,
the magnetic DDI
can have a significant effect 
in certain cases~\cite{KAWAGUCHI2012253,* RevModPhys.85.1191}.
When taking the DDI into account (in addition to the
short-range $s$-wave collision),
it is probably impossible to exactly write down
the many-body ground states.
Nevertheless, we expect that the DDI induces new phases, 
such as charge density wave and supersolid,
due to its long-range nature.
Hunting new phases, including the SPT phases, 
in itinerant 
spinful bosonic systems with DDI
will be an interesting future direction.
Note that systems with magnetic DDI no longer have
spin rotation symmetry.
Instead, the magnetic DDI is 
invariant under simultaneous rotation 
in both spin and real spaces
$\hat{V}^{\delta}(\theta):=\exp[-\mathrm{i}\theta \sum_{\bm{r}} (\hat{S}_{\bm{r}}^{\delta} + \hat{L}_{\bm{r}}^{\delta} )]$,
where $\hat{L}_{\bm{r}}^{\delta}$ is the orbital angular momentum operator in
the $\delta(=x,y,z)$-direction
for particles at position $\bm{r}$~\cite{KAWAGUCHI2012253,* RevModPhys.85.1191, PhysRevLett.96.080405}.
In other words, such systems 
conserve total angular momentum in free space.
When constrained on a lattice, the systems
can still preserve some 
discrete rotation symmetries,
though such rotation symmetries are not on-site.
In the future studies,
it is worth investigating how such symmetries classify 
the SPT phases.

Finally, we would like to make some remarks 
about the flat band.
The flat band has been gaining much attention these years because 
it was found to give rise to various collective phenomena,
such as ferromagnetism~\cite{10.1143/PTP.99.489, mielke1993} and superconductivity~\cite{LiuZheng:77308, cao2018unconventional}, 
in quantum many-body systems. 
In this paper, we
discover that the flat band can also be an origin 
of interacting SPT phases. 
We believe this work stimulates future research
on the relation between flat bands and topological
quantum physics.

Interestingly, there exists many kinds of lattices where
the flat band appears in the middle or top
of the band structure~\cite{PhysRevB.99.045107, PhysRevA.94.043831, PhysRevB.95.115135, dias2015origami}.
One can certainly follow the scheme in this paper
to construct the many-body eigenstates in these lattices
with the help of the CLSs of the flat bands.
The resulting many-body eigenstates, 
due to their short-range entangled nature,
are actually \textit{quantum many-body scars}~\cite{turner2018weak, PhysRevLett.122.040603, 
PhysRevB.102.241115, PhysRevResearch.2.043267, PhysRevLett.124.180604},
which lead to weak ergodicity breaking of the systems.
Exploring 
quantum many-body scars in spinful atoms with a flat band
will be another intriguing future direction.









\begin{acknowledgements}
We acknowledge stimulating discussions with Hal Tasaki, Synge Todo, and Linhao Li.
H.~Y. was supported by Grant-in-Aid for JSPS Research Fellowship for Young Scientists (DC1) No.~20J20715.
H.~N. was supported by the Advanced Leading Graduate Course for Photon Science (ALPS) at the University of Tokyo.
H.~K. was supported in part by JSPS Grant-in-Aid for Scientific Research on Innovative Areas No.~JP20H04630, JSPS KAKENHI Grant No.~JP18K03445, and the Inamori Foundation. 
\end{acknowledgements}

\appendix

\section{The Haldane insulator phase} \label{App_HI}

\begin{figure}[h]
  \centering
  \includegraphics[width=0.43\textwidth]{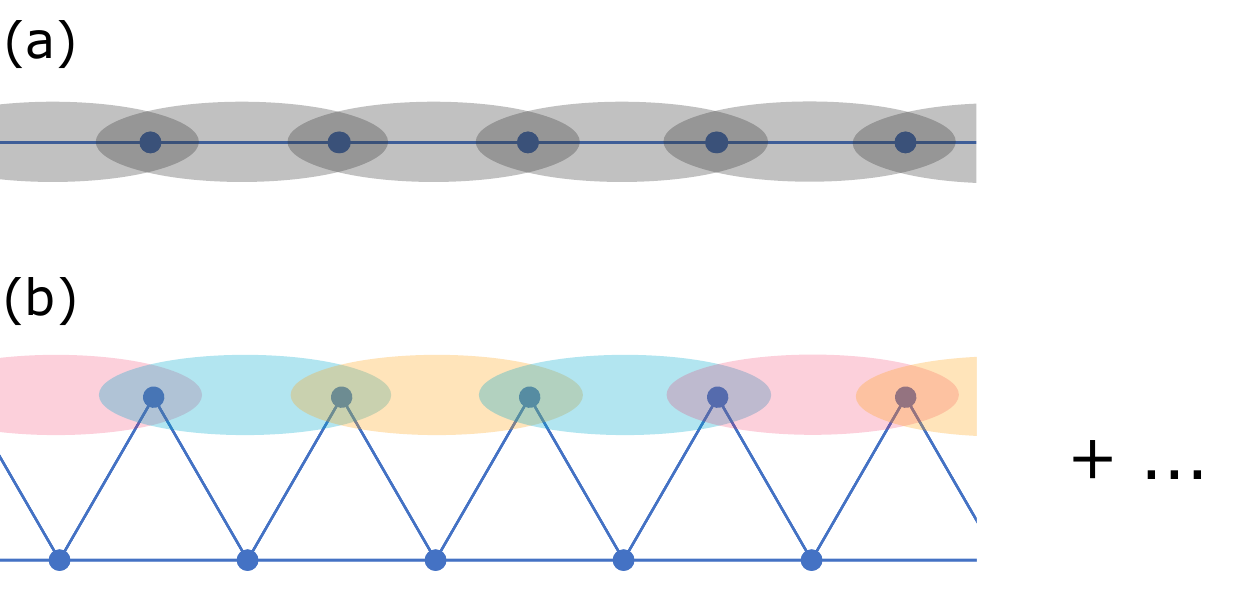}
  \caption{(a) The Haldane insulator phase in the 1D system of spinless bosons. Every ellipse represents a single particle.
  (b) The state $|\text{GS}\rangle$ at $\lambda=0$. Three different colors denote three different spin states,
  and all allowed spin configurations are summed up (with coefficients).}
  \label{fig:lambda=0}
\end{figure}

Spinless bosons in optical lattices with dipole-dipole interaction are described by the extended Bose-Hubbard model.
The nature of the Haldane insulator phase in the 1D extended Bose-Hubbard model can be captured by the following state~\cite{EBMH_PhysRevB.77.245119}:
\begin{equation}
	| \Psi_{\text{HI}} \rangle = \prod_{j} (\hat{a}_j^\dagger + \hat{a}_{j+1}^\dagger) | \text{vac} \rangle,
\end{equation}
where $\hat{a}_j^\dagger$ creates a spinless boson at site $j$;
see \Fig{fig:lambda=0}(a).
The state $| \Psi_{\text{HI}} \rangle$ represents a trivial phase in the sense of inversion symmetry,
but it represents a Haldane phase
protected by the combination of 
pseudo-spin rotation and the inversion symmetry, i.e., the group $\{ 1, \hat{U}(n\mathcal{I})\}$
with $\hat{U}(n\mathcal{I}):=\exp[-\mathrm{i}\pi \sum_r (\hat{n}_r-1)] \hat{U}(\mathcal{I})$.
This is the same for $|\text{GS}_{\lambda=0}\rangle$, see Table~\ref{table:u}.
In addition, the Haldane phase of $|\text{GS}_{\lambda=0}\rangle$ is protected by
other symmetries related to the spin DOF. In this sense,
we can say the state $|\text{GS}_{\lambda=0}\rangle$ represents a
spinful Haldane insulator phase.
Note that both $| \Psi_{\text{HI}} \rangle$ and $|\text{GS}_{\lambda=0}\rangle$
exhibit perfect hidden charge order, i.e., 
vacant sites and doubly occupied sites appear alternatively if we ignore all the
singly occupied sites.

\section{Spin and charge edge states} \label{App_edge}
\begin{figure}
  \centering
  \includegraphics[width=0.4\textwidth]{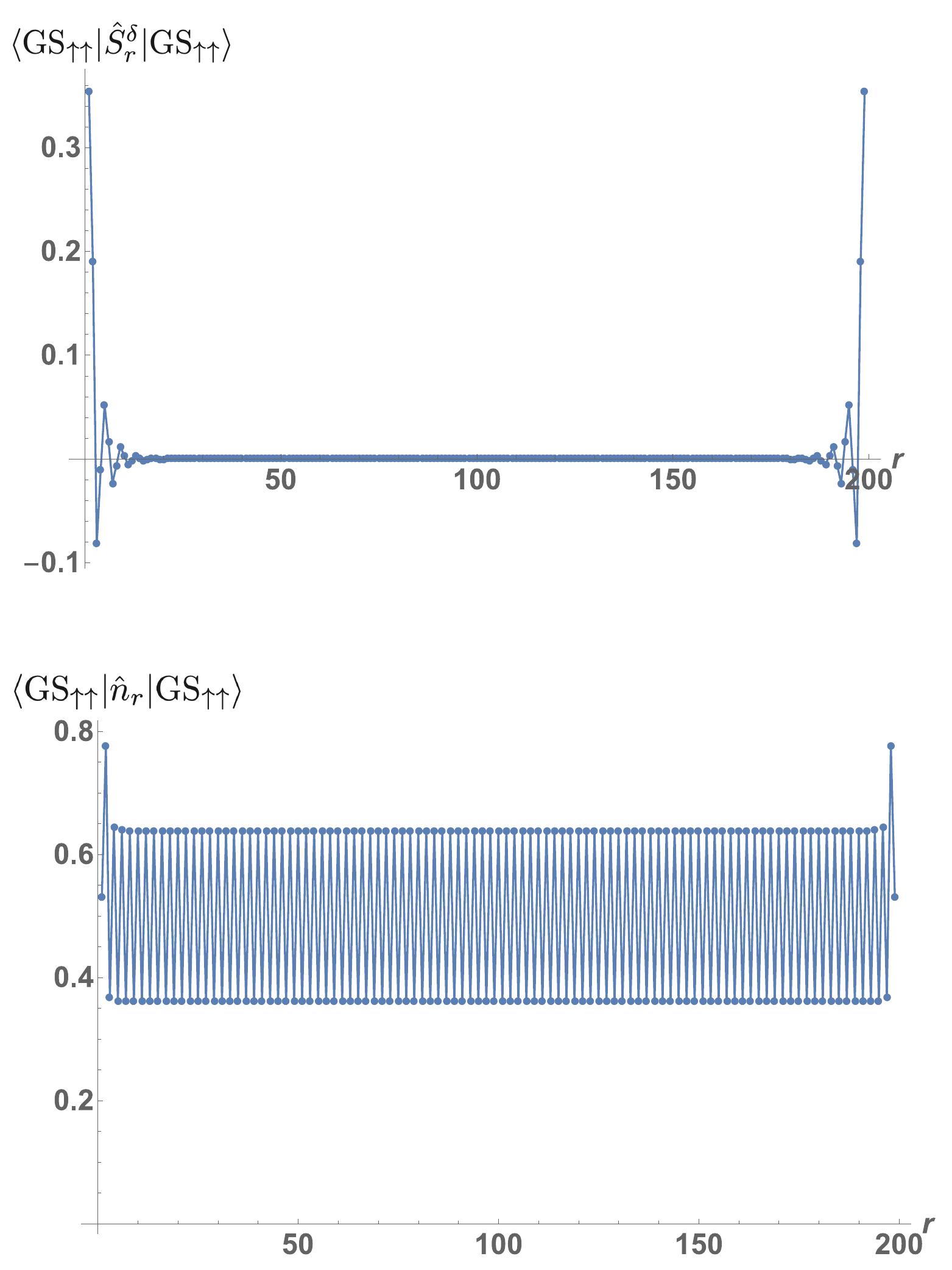}
  \caption{Spin and charge edge states. 
  We have assumed that the sawtooth chain ends with bottom sites at both ends 
  and taken $N=100$ and $\lambda=1$.}
  \label{fig:edge_states}
\end{figure}

For spin-1 BHMSC with OBC, $g^{\text{t}}_0=0$ and $ g^{\text{t}}_2>0$,
there are four degenerate ground states $ | \text{GS}_{\uparrow \uparrow} \rangle$,
$ | \text{GS}_{\uparrow \downarrow} \rangle$, $ | \text{GS}_{\downarrow \uparrow} \rangle$,
and $ | \text{GS}_{\downarrow \downarrow} \rangle$,
which correspond to four independent edge spin-$1/2$ states.
The analytical forms of 
$\langle \text{GS}_{\uparrow \uparrow} | \hat{S}_{r}^{\delta}  | \text{GS}_{\uparrow \uparrow} \rangle$,
$\langle \text{GS}_{\uparrow \uparrow} | \hat{n}_{r} | \text{GS}_{\uparrow \uparrow} \rangle$, etc.
are rather complicated and will not be presented.
Instead, their plots are shown in~\Fig{fig:edge_states}.
The spin edge state decays more slowly than the charge one,
because the spin correlation length is longer than the charge correlation length:
\begin{equation}
	\begin{split}
		\xi_{\text{spin}} &= \left( \ln \frac{3 \lambda ^2+\sqrt{9 \lambda ^4+36 \lambda ^2+24}+6}{\lambda ^2+\sqrt{\lambda ^4+4 \lambda ^2+24}+2} \right)^{-1},\\
		\xi_{\text{charge}} &=\left( \ln \frac{3 \lambda ^2+\sqrt{9 \lambda ^4+36 \lambda ^2+24}+6}{3 \lambda ^2-\sqrt{9 \lambda ^4+36 \lambda ^2+24}+6}  \right)^{-1}.
	\end{split}
\end{equation}

\begin{figure}
  \centering
  \includegraphics[width=0.3\textwidth]{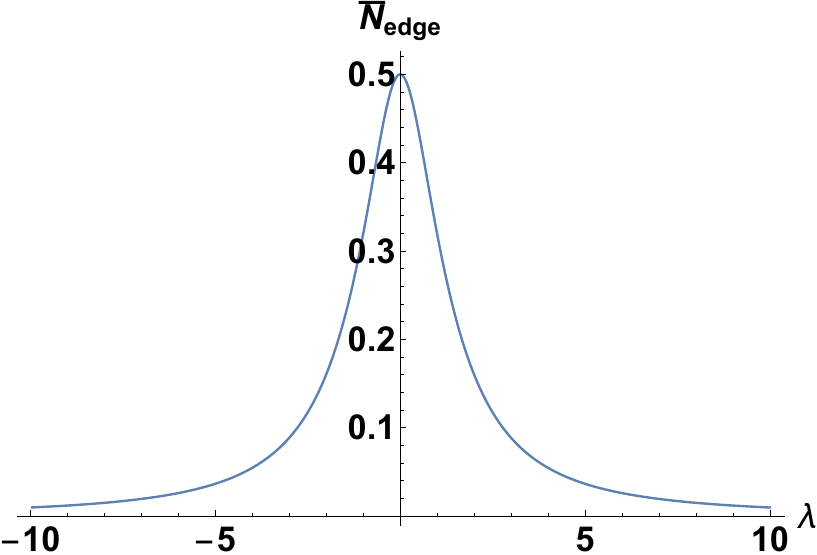}
  \caption{The reduced edge particle number $\bar{N}_{\text{edge}}$ as a function of $\lambda$.}
  \label{fig:N_edge}
\end{figure}

There is always a spin-$1/2$ localized at each edge.
We define the reduced edge particle number as
$\bar{N}_{\text{edge}} :=  \sum_{r=1}^\infty  (  \lim_{N \to \infty} 
	\langle \text{GS}_{\uparrow \uparrow} | \hat{n}_{r} | \text{GS}_{\uparrow \uparrow} \rangle
	- 1/2)$.
$\bar{N}_{\text{edge}} $ is a function of $\lambda$,
and $\bar{N}_{\text{edge}}=1/2$ at $\lambda=0$.
See~\Fig{fig:N_edge}.
Note that $\bar{N}_{\text{edge}}$ is independent of the four degenerate ground states.

Let us note that for crystalline-symmetry-protected topological phases in $d<3$ dimensions, there are no anomalous edge states~\cite{PTBO_2010,PhysRevB.96.205106}. For example, both
$|\text{GS}_{3,\includegraphics[width=0.011\textwidth]{kagome_symbol2.pdf}}\rangle$ and $|\text{VBS}_{3,\triangle}\rangle$ are protected by $D_2$ symmetry (see Sec.~\ref{Sec_Generalizations}),
 and they do not have anomalous edge states.

\section{Kennedy-Tasaki transformation for integer-spin itinerant systems}  \label{App_KT}

The Kennedy-Tasaki transformation is a nonlocal unitary transformation defined 
on an open chain of length $L$
as~\cite{Kennedy_Tasaki,Oshikawa_1992,PTBO_2012,Tasaki2020}
\begin{equation}
	\hat{U}_{\mathrm{KT}}:=\prod_{j =1}^L \exp \left[i \pi\left(\sum_{i=1}^{j-1} \hat{S}_{i}^{z}\right) \hat{S}_{j}^{x}\right],
\end{equation}
and it is hermitian:
$\hat{U}_{\mathrm{KT}}^\dagger = \hat{U}_{\mathrm{KT}}$. 
$\hat{U}_{\mathrm{KT}}$ is also invariant under $\mathbb{Z}_2 \times \mathbb{Z}_2$ spin rotation. 
For an arbitrary integer-spin chain,
let $\hat{h}_j$ be a local Hamiltonian, and
the sufficient and necessary condition for $ \hat{U}_{\mathrm{KT}} \hat{h}_j \hat{U}_{\mathrm{KT}} $
to be also local is that $\hat{h}_j$ is $\mathbb{Z}_2 \times \mathbb{Z}_2$ invariant~\cite{Oshikawa_1992,PTBO_2010}.
This statement can be extended to integer-spin itinerant systems.
For example, the spin-1 Bose-Hubbard model has SO(3) spin rotation symmetry,
and the on-site interaction $\hat{P}^{(S)}_r$ is invariant under $\hat{U}_{\mathrm{KT}}$,
while the SO(3)-invariant hopping transforms as
\begin{equation}
	\begin{split}
		&\hat{U}_{\mathrm{KT}}\  \left(\sum_{\alpha=0,\pm1} \hat{a}^\dagger_{i,\alpha} \hat{a}_{j,\alpha} + \text{h.c.}  \right) \ \hat{U}_{\mathrm{KT}} \\
	=&  \mathrm{e}^{\mathrm{i} \pi \sum\limits_{v=i}^{j-1} \hat{S}_{v}^{z}}   a^\dagger_{i,0} \hat{a}_{j,0}  + \text{h.c.} \\
	&+\frac{1}{2} \mathrm{e}^{ \mathrm{i} \pi \sum\limits_{v=i+1}^{j} \hat{S}_{v}^{x} }
	 \bigg[
	\Big(  \mathrm{e}^{\mathrm{i} \pi \sum\limits_{u=i}^{j-1} \hat{S}^z_u}  +1 \Big) \left(a^\dagger_{i,+} \hat{a}_{j,+} + a^\dagger_{i,-} \hat{a}_{j,-} \right) \\
	&+ \Big(  \mathrm{e}^{\mathrm{i} \pi \sum\limits_{u=i}^{j-1} \hat{S}^z_u}  -1 \Big) \left(\hat{a}^\dagger_{i,+} a_{j,-} + \hat{a}^\dagger_{i,-} \hat{a}_{j,+} \right)
	\bigg] + \text{h.c.}.
	\end{split}
\end{equation}
We can see that the transformed hopping has $\mathbb{Z}_2 \times \mathbb{Z}_2$ symmetry
and is still local if the original hopping is local.

\section{Translation symmetry and the Haldane phases} \label{App_translation}

\begin{figure}[H]
  \centering
  \includegraphics[width=0.4\textwidth]{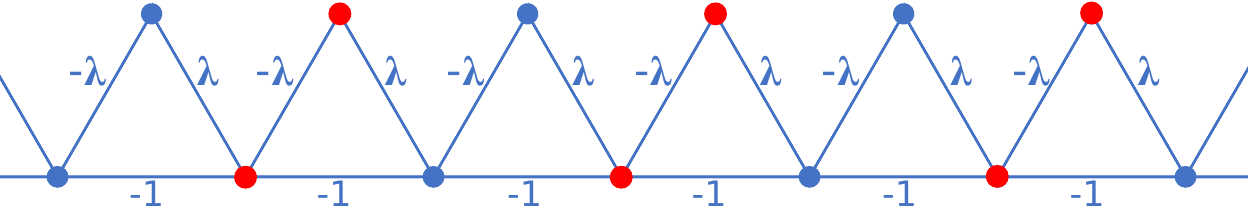}
  \caption{Hopping constant $t_{r,r'}$ for magnetic sublevel $\pm 1$ in $\hat{W}^\gamma \hat{H}_{\text{hop}} \hat{W}^\gamma $ when $\gamma=z$. 
  Operator $\hat{W}^\gamma$ acts on the red sites.
  One can see that the transformation $ \hat{W}^\gamma $ preserves the translation symmetry.}
  \label{fig:translation_inv}
\end{figure}

In the presence of both translation symmetry and $\mathbb{Z}_2 \times \mathbb{Z}_2$ symmetry,
$\{ \phi_q \}_{q \in \mathbb{Z}_2 \times \mathbb{Z}_2}$ in~\Eq{uniqueness_of_MPS}
forms a 1D representation of the group $\mathbb{Z}_2 \times \mathbb{Z}_2$.
In this case,
all the phases of gapped states that do not break the two
symmetries are classified by a pair of indices $(\omega, \gamma)$ where $\omega \in \mathcal{Q}_{\mathbb{Z}_2 \times \mathbb{Z}_2}$
and $\gamma$ labels different 1D representations of $\mathbb{Z}_2 \times \mathbb{Z}_2$~\cite{PhysRevB.84.075135}; 
see Table~\ref{1DZ2Z2}.
Let us again use the sawtooth chain as an example.
It is easy to see that
the state $| \text{GS} \rangle$ in~\Eq{GS} corresponds to the row $\gamma=1$ in Table~\ref{1DZ2Z2}.
We now show that other three SPT phases labeled by $(\omega=-1, \gamma=x,y,z)$ can be obtained by slightly modifying
$\hat{H}_{\text{hop}}$. 
Define
\begin{equation}
	\hat{W}^\gamma := \prod_{k=1}^{N/4} \mathrm{e}^{-\mathrm{i}\pi (\hat{S}_{4k-1}^{\gamma} + \hat{S}_{4k}^{\gamma})},
	\quad \gamma=x,y,z.
\end{equation}
The operator $\hat{W}^\gamma$ acts on the red sites pictured in~\Fig{fig:translation_inv}. 
Interaction $\hat{H}_{\text{int}}$ is invariant under $\hat{W}^\gamma$, thus
$\hat{H}^\gamma :=\hat{W}^\gamma \hat{H} \hat{W}^\gamma = \hat{W}^\gamma \hat{H}_{\text{hop}} \hat{W}^\gamma + \hat{H}_{\text{int}}$.
The transformation does not break the translation symmetry.
The unique ground state of $\hat{H}^\gamma$ is given by
\begin{equation}
\begin{split}
	| \text{GS}^\gamma \rangle &= \hat{W}^\gamma | \text{GS} \rangle \\
	&= \sum_{\tau_1,...,\tau_{2N}=-1}^3 \text{Tr} \left( 
	  F^{\tau_{1}} E^{\tau_{2}} \Sigma^\gamma F^{\tau_{3}} E^{\tau_{4}} \Sigma^\gamma ...	\right) \times \\
	&\quad \left( \prod\limits_{r=1}^{2N} \hat{d}_{r,\tau_r}^\dagger \right)
	| \text{vac}  \rangle,
\end{split}
\end{equation}
where $\Sigma^\gamma := \text{diag}(\sigma^\gamma, \sigma^\gamma)$.
Under the $\mathbb{Z}_2 \times \mathbb{Z}_2$ spin rotation, matrices in $| \text{GS}^\gamma \rangle$ transforms as
\begin{equation}
	F^{\tau_{2j-1}} E^{\tau_{2j}} \Sigma^\gamma \to \ \mathrm{e}^{\mathrm{i}\phi_q} \ u_q^\dagger \ F^{\tau_{2j-1}} E^{\tau_{2j}} \Sigma^\gamma \ u_q.
\end{equation}
Unitary matrices $\{u_q\}$ above are also given in Table~\ref{table:u}.
Explicit calculation yields the other three different 1D representations of $\mathbb{Z}_2 \times \mathbb{Z}_2$ in Table~\ref{1DZ2Z2}.
We thus have the desired SPT phases.

\begin{table}[H]
\begin{center}
\begin{tabular}{ l| c c c c }
\hline \hline
          & $\phi_1$ & $\phi_x$ & $\phi_y$ & $\phi_z$ \\ \hline 
$\gamma=1$ & $+1$         & +1         & +1         & +1         \\ 
$\gamma=x$ & +1         & +1         & $-1$        & $-1$        \\ 
$\gamma=y$ & +1         & $-1$        & +1         & $-1$        \\ 
$\gamma=z$ & +1         & $-1$        & $-1$        & +1         \\ 
\hline \hline
\end{tabular}
\end{center}
\caption{Four different 1D representations of $\mathbb{Z}_2 \times \mathbb{Z}_2$.}
\label{1DZ2Z2}
\end{table}

\section{MPS ansatz for numerical calculations} \label{App_MPSansatz}

The MPS ansatz used in our numerical calculation
can be regarded as a generalization of \Eq{GS}. 
For the matrices in \Eq{GS}, we can assign a pair of quantum numbers to each block as
\begin{equation}
	\begin{split}
		\mathcal{F} &:= \sqrt{\lambda^2+2} \sum_{\tau=-1}^3 F^{\tau} \hat{d}_{r,\tau}^\dagger =
\begin{pNiceArray}{C|C}[first-row,first-col]
& 0 & 1\\
1/2 & I_2 & -\lambda \sum_\alpha M^\alpha \hat{a}^\dagger_{r,\alpha}\\
\hline
3/2 & 0 & I_2\\
\end{pNiceArray},\\ 
 \\
\mathcal{E} &:= \sum_{\tau=-1}^3 E^{\tau} \hat{d}_{r,\tau}^\dagger=
\begin{pNiceArray}{C|C}[first-row,first-col]
& 1/2 & 3/2\\
0 & \sum_\alpha M^\alpha \hat{a}^\dagger_{r,\alpha} & \sqrt{6}\hat{b}^\dagger_r I_2 \\
\hline
1 & I_2 & \sum_\alpha M^\alpha \hat{a}^\dagger_{r,\alpha}\\
\end{pNiceArray}. \label{eq:app_FE}
	\end{split}
\end{equation}
The quantum numbers are assigned by the following rule:
\begin{align}
\begin{NiceArray}{CC}
& n+m - 1/2\\
& \uparrow\\
n \rightarrow 
&\text{a nonzero block that creates } m\text{ particles} 
\end{NiceArray}
\end{align}
For example, the upper right block of $\mathcal{E}$ creates two particles,
thus the block is labeled by $(0,3/2)$.
For the product of $2N$ matrices, we replace $1/2$ in the above rule with $2N/2=N$, such that
\begin{align}
    \underbrace{\mathcal{EFEF}\ldots \mathcal{EF}}_{2N} = \begin{pNiceArray}{C|C}[first-row,first-col]
& 0 & 1\\
0 & \mathcal{X}_{11} & \mathcal{X}_{12}\\
\hline
1 & \mathcal{X}_{21} & \mathcal{X}_{22}\\
\end{pNiceArray},\label{eq:2n}
\end{align}
where $\mathcal{X}_{11}$ and $\mathcal{X}_{22}$ are $2 \times 2$ blocks which create $N$ particles,
while $\mathcal{X}_{12}$ and $\mathcal{X}_{21}$ create $N+1$ and $N-1$ particles, respectively.
In the thermodynamic limit $N \to \infty$, \Eq{eq:2n} gives an MPS 
where the particle number equals the number of unit cells.

The exact ground state given by \Eq{eq:2n} has bond dimension $\chi=4$.
In general, however, we need to use an MPS with larger bond dimension to better approximate
the true ground state. We can thus generalize \Eq{eq:app_FE} to
a block-banded form:
\begin{widetext}
	\begin{equation}
		  \tilde{\mathcal{F}} = 
    \begin{pNiceMatrix}[first-row,first-col]
    & 0 & \cdots & m & m+1 & \cdots & n-1\\
    1/2 & \mathcal{X}^{(1)}_{[0]} & \cdots & \mathcal{X}^{(1)}_{[m]} & 0 & \cdots & 0\\
    3/2 & 0 & \ddots &  & \ddots & \ddots & \vdots\\
    \vdots &\vdots & \ddots & \ddots & & \ddots & 0\\
    \vdots &\vdots & & \ddots & \ddots & & \mathcal{X}_{[m]}^{(n - m)}\\
    \vdots &\vdots & & &\ddots & \ddots & \vdots \\
    n-\dfrac{1}{2}& 0 & \cdots &\cdots &\cdots & 0 &\mathcal{X}_{[0]}^{(n)} 
    \end{pNiceMatrix}, \quad
  \tilde{\mathcal{E}} = 
  \begin{pNiceMatrix}[first-row,first-col]
    &1/2 & \cdots & m-\dfrac{1}{2} & m+\dfrac{1}{2} & \cdots & n-\dfrac{1}{2}\\
    0& \mathcal{Y}^{(1)}_{[1]} & \cdots   & \mathcal{Y}^{(1)}_{[m]} & 0 & \cdots & 0\\
    1& \mathcal{Y}^{(1)}_{[0]} & \ddots  & & \ddots & \ddots & \vdots\\
    2 &0 & \ddots & \ddots & &\ddots & 0\\
    \vdots &\vdots & \ddots & \ddots & \ddots  & & \mathcal{Y}_{[m]}^{(n - m +1)}\\
    \vdots &\vdots & & \ddots  & \ddots & \ddots  & \vdots \\
    n-1&0 & \cdots & \cdots  & 0 &\mathcal{Y}_{[0]}^{(n-1)} &\mathcal{Y}_{[1]}^{(n)}
    \end{pNiceMatrix},\label{eq:suMPS}
\end{equation}
\end{widetext}
where $\mathcal{X}_{[k]}$ and $\mathcal{Y}_{[k]}$ denote $d\times d$ blocks that create $k$ particles,
and the maximum particle number on each site is truncated to $m$.
The bond dimension is $\chi=dn$.
The MPS $|\Psi( \tilde{\mathcal{E}},  \tilde{\mathcal{F}} ) \rangle := \text{Tr}(...\tilde{\mathcal{E}} \tilde{\mathcal{F}} \tilde{\mathcal{E}} \tilde{\mathcal{F}}...) |\text{vac}\rangle$ is called a
symmetric uniform MPS (suMPS)~\cite{PhysRevB.97.235155}.
In our numerical calculations, we fix $m=3$, $n=4$ and vary $d$.
We optimize the suMPS by minimizing its energy expectation value,
and the optimization is done by the VUMPS algorithm.

Now we justify the particle number truncation $m=3$.
The energy variance of the MPS $|\Psi( \tilde{\mathcal{E}},  \tilde{\mathcal{F}} ) \rangle$ is measured by
\begin{equation}
	\begin{split}
		 \sigma^2 &:= \frac{1}{N}\left(  \langle \Psi( \tilde{\mathcal{E}}, \tilde{\mathcal{F}} ) |  \hat{H}^2 |\Psi( \tilde{\mathcal{E}},  \tilde{\mathcal{F}} ) \rangle -  \langle \Psi( \tilde{\mathcal{E}},  \tilde{\mathcal{F}} ) |  \hat{H} |\Psi( \tilde{\mathcal{E}},  \tilde{\mathcal{F}} ) \rangle^2 \right)\\
    &= \frac{1}{N} \langle \Psi( \tilde{\mathcal{E}},  \tilde{\mathcal{F}} ) | \hat{H}\left(1 - \hat{P}_{\tilde{\mathcal{E}},  \tilde{\mathcal{F}}}\right)\hat{H} |\Psi( \tilde{\mathcal{E}},  \tilde{\mathcal{F}} ) \rangle,
	\end{split} \label{eq:sigma^2}
\end{equation}
where $\hat{P}_{\tilde{\mathcal{E}},  \tilde{\mathcal{F}}}:= |\Psi( \tilde{\mathcal{E}},  \tilde{\mathcal{F}} ) \rangle \langle \Psi( \tilde{\mathcal{E}}, \tilde{\mathcal{F}} ) |$.
Let
\begin{align}
    \mathbb{H}_r^{\leqslant m} = \mathrm{span}\left( \bigg\{ \prod\limits_{\alpha=-1}^1 (\hat{a}^\dagger_{r,\alpha})^{n_{\alpha}}|{\mathrm{vac}} \rangle_r \bigg|  \sum_{\alpha} n_{\alpha} \leqslant m \bigg\} \right)
\end{align}
be the local truncated Hilbert space and $\hat{P}_{\forall \leqslant m}$ be the projection operator onto 
the total truncated Hilbert space
$\mathbb{H}^{\forall\leqslant m} \coloneqq \bigotimes_{i=r}^{2N} \mathbb{H}_r^{\leqslant m}$.
The Hamiltonian in $\mathbb{H}^{\forall\leqslant m}$ reads 
$\hat{H}_{\forall\leqslant m} := \hat{P}_{\forall\leqslant m} \hat{H} \hat{P}_{\forall\leqslant m}$.
Equation~(\ref{eq:sigma^2}) can thus be rewritten as
\begin{equation}
	\begin{split}
		\sigma^2 &= \frac{1}{N}   \langle \Psi( \tilde{\mathcal{E}}, \tilde{\mathcal{F}} ) |  {\hat{H}_{\forall \leqslant m} \left(1 - \hat{P}_{\tilde{\mathcal{E}},  \tilde{\mathcal{F}}} \right)   \hat{H}_{\forall \leqslant m}}   |\Psi( \tilde{\mathcal{E}},  \tilde{\mathcal{F}} ) \rangle \\
		& \quad + \frac{1}{N}  \langle \Psi( \tilde{\mathcal{E}}, \tilde{\mathcal{F}} ) | {\hat{H} \left( 1- \hat{P}_{\forall \leqslant m} \right) \hat{H}} |\Psi( \tilde{\mathcal{E}},  \tilde{\mathcal{F}} ) \rangle.  \label{eq:sigma^2_2terms}
	\end{split}
\end{equation}
The first term above can be viewed as the variance in $\mathbb{H}^{\forall\leqslant m}$, 
and it quantifies the effect of finite bond dimension.
Similar to spin or fermion systems, the first term can be calculated efficiently~\cite{PhysRevB.97.045145, PhysRevB.97.045125}.
On the other hand, the second term quantifies the effect of truncation.
Note that although $\hat{H} ( 1- \hat{P}_{\forall \leqslant m} ) \hat{H}$ contains $\mathcal{O}(N^2)$ nonlocal terms, 
only $\mathcal{O}(N)$ local terms return nonzero values when sandwiched by $|\Psi( \tilde{\mathcal{E}},  \tilde{\mathcal{F}} ) \rangle$.
Table~\ref{tab:variance} shows that for $m=3$, at least near the Haldane-critical phase transition point,
the effect of particle number truncation is about $100$ times smaller than the effect of finite bond dimension.

\begin{table}[h]
\begin{center}
\begin{tabular}{c|cc}
\hline \hline
$\varphi/\pi $ & first & second \\ \hline
 1/36 & $1.144\times 10^{-4}$ & $8.082\times 10^{-5}$ \\
 2/36 & $3.463\times 10^{-4}$ & $1.190\times 10^{-4}$ \\
 3/36 & $9.122\times 10^{-4}$ & $1.267\times 10^{-4}$ \\
 4/36 & $1.957\times 10^{-3}$ & $1.142\times 10^{-4}$ \\
 5/36 & $3.543\times 10^{-3}$ & $9.620\times 10^{-5}$ \\
 6/36 & $5.702\times 10^{-3}$ & $7.819\times 10^{-5}$ \\
 7/36 & $7.348\times 10^{-3}$ & $6.284\times 10^{-5}$ \\
 8/36 & $8.614\times 10^{-3}$ & $5.209\times 10^{-5}$ \\
 9/36 & $8.956\times 10^{-3}$ & $4.874\times 10^{-5}$ \\
\hline \hline
\end{tabular}    
\end{center}
\caption{Examples of the actual values of the first and second terms in \Eq{eq:sigma^2_2terms} with $m=3$, $d=100$, $\lambda=1$, and $R=1$.
The Haldane-critical phase transition happens between $\varphi=6/36$ and $7/36$; see \Fig{fig:phase_diagram}(b).}
\label{tab:variance}
\end{table}

\begin{figure}[h]
  \centering
  \includegraphics[width=0.4\textwidth]{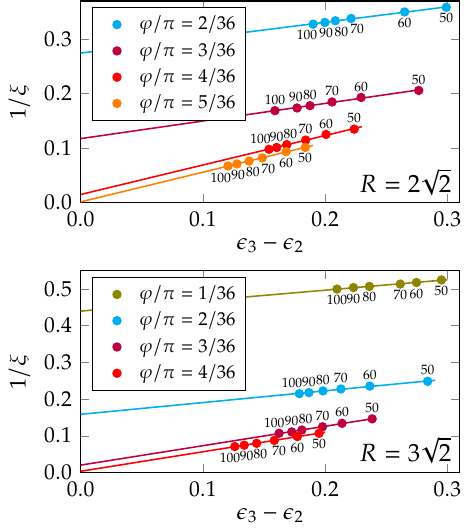}
  \caption{Scaling of the inverse correlation length $1/\xi:= \epsilon_2$ with respect to $\epsilon_3- \epsilon_2$.
	 Numbers near the data points denote the corresponding bond dimensions $d$ of each block.	 
	 Along the path with $R=2\sqrt{2}$,
	 we see that a quantum phase transition occurs between $\varphi= 4\pi/36$ and $5\pi/36$.
	 On the other hand, along the path with $R=3\sqrt{2}$,
	 a phase transition occurs between $\varphi= 3\pi/36$ and $4\pi/36$.}
  \label{fig:scaling_appendix}
\end{figure}

Finally, we provide further
numerical evidence that determines the shape of the 
phase boundary in the $\lambda=1$ plane in~\Fig{fig:phase_diagram}(a).
The transfer matrix is defined as the sum of
Kronecker products
$\sum_{\tau,\tau'} \tilde{E}^\tau \tilde{F}^{\tau'} \otimes \tilde{E}^{\tau}  \tilde{F}^{\tau'}$,
where
$\tilde{E}^\tau:= \langle \tau | \tilde{\mathcal{E}} |\text{vac}\rangle $ and
$\tilde{F}^\tau:= \langle \tau | \tilde{\mathcal{F}} |\text{vac}\rangle $
with $\{ | \tau \rangle \}$ being a basis in the Fock space.
Let $\epsilon_i = -\ln |\lambda_i|$, where $\lambda_i$ is the $i$th largest absolute eigenvalue of the transfer matrix,
and $|\lambda_1|$ is normalized to~1.
As we change the bond dimension $\chi$,
we calculate the scaling of the inverse correlation length
$1/\xi:= \epsilon_2 $
with respect to
$ \epsilon_3 -\epsilon_2$
along the path parameterized by
$(R \sin \varphi , R \cos \varphi, 1)$, see~\Fig{fig:scaling_appendix}.
We find that, as $R$ grows, the phase transition occurs
at smaller $\varphi$, which indicates that the
phase boundary is curved instead of straight.

\section{Uniqueness of the ground state of $\hat{H}^{f,X}$} \label{App_Uniqueness}

\begin{figure}[h]
  \centering
  \includegraphics[width=0.4\textwidth]{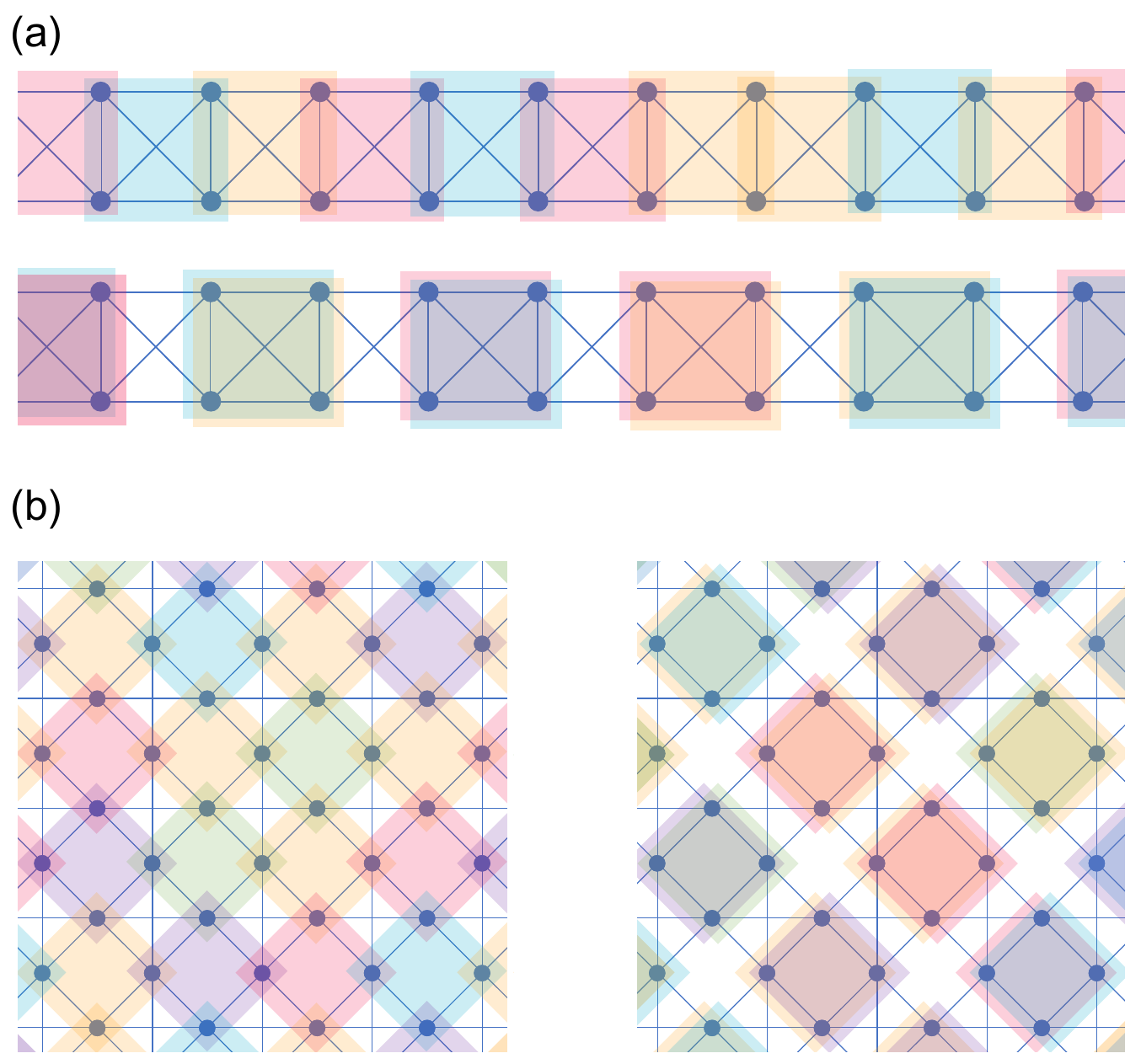}
  \caption{A ``nontrivial" ground state is degenerated with product states, 
  if $\Lambda^{[1]}_{X} = \emptyset$ and the corresponding quantum spin model lives on a bipartite lattice.
  (a) Spin-1 bosons on a Creutz ladder. Each colored square denotes a CLS with spin degree of freedom. 
  Note that the ground stats are superpositions of all allowed spin configurations.
  (b) Spin-2 bosons on a checkerboard lattice. $X'$ in this case is a square lattice.}
  \label{fig:degenerate_GS}
\end{figure}

\begin{figure}[!htbp]
  \centering
  \includegraphics[width=0.45\textwidth]{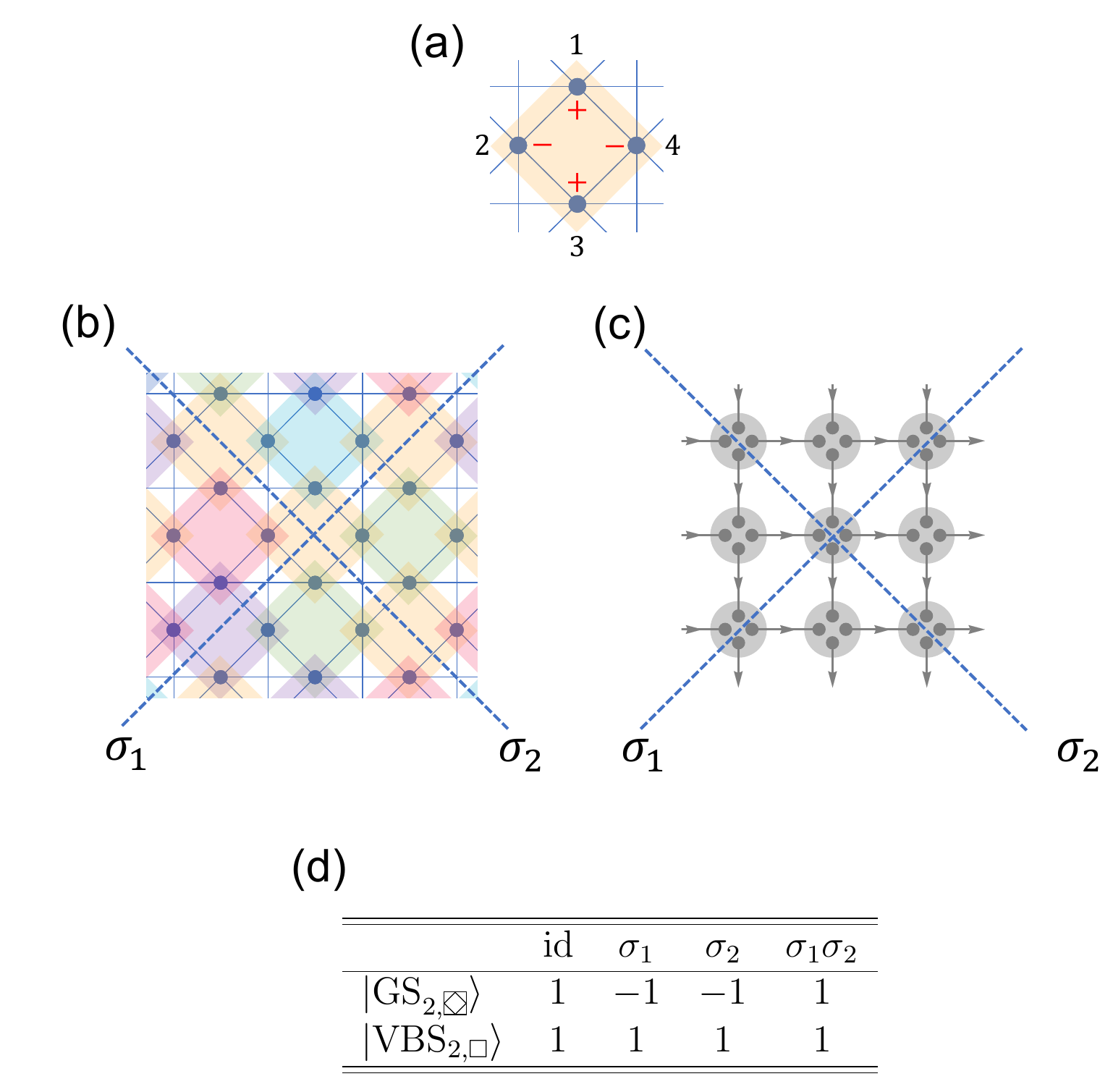}
  \caption{(a) A CLS in the checkerboard lattice. 
  Sign of the amplitude alternates from site 1 to 4.
  (b) An FPS on the checkerboard lattice. 
  ($|\text{GS}_{2,\includegraphics[width=0.011\textwidth]{checkerboard_symbol.pdf}}\rangle$ is $D_2$ symmetric, 
  and it
  is a superposition of FPSs with different spin configurations.)
  We require that the symmetry center of $D_2$ lies
  at the geometric center of a CLS and the two mirror planes
  are placed as in the figure.
  (c) For $|\text{VBS}_{2,\Box}\rangle$,
  the symmetry center
  should lie at a site (spin)
  in order to be compatible with (b).
  (d) 1D representations of $D_2$ 
  associated with
  $|\text{GS}_{2,\includegraphics[width=0.011\textwidth]{checkerboard_symbol.pdf}}\rangle$
  and
  $|\text{VBS}_{2,\Box}\rangle$.
  The former state yields a nontrivial
  representation and is thus in an SPT phase.}
  \label{fig:D2_checkerboard}
\end{figure}

Mathematically, the uniqueness of the ground state of $\hat{H}^{f,X}$ can be proved 
with additional assumptions: $\Lambda^{[1]}_{X} \neq \emptyset$ and $g_{S,\bm{r}}>0$ for $\forall S$ and $\forall \bm{r} \in \Lambda^{[1]}_{X} $.
With the ``completeness relation" $\sum_{S} \hat{P}_{\bm{r}}^{(S)} = \hat{n}_{\bm{r}} ( \hat{n}_{\bm{r}} -1 )/2$ in mind
and following the deduction in \Eq{uniqueness_of_FPS},
one can show that the ground state can only be a linear combination of FPSs.
The uniqueness of the ground state of $\hat{H}^{f,X}$ then follows from the uniqueness of $|\text{VBS}_{f,X'}\rangle$.
The assumption $\Lambda^{[1]}_{X} \neq \emptyset$ is always satisfied in lattices generated by
the cell construction, see, for example, \Fig{fig:1D_FB_lattices}(b) and \Fig{fig:2D_FB_lattices}(a).
However, for the kagome lattice shown in \Fig{fig:2D_FB_lattices}(b), $\Lambda^{[1]}_{X} = \emptyset$.
Nevertheless, we propose the following conjecture:
even in lattice $X$ with $\Lambda^{[1]}_{X} = \emptyset$,
the exact ground state of $\hat{H}_{X}$ is unique when $X'$ is not a bipartite lattice.
For the kagome lattice, $X'$ is a triangular lattice which is not bipartite.
Note that if $X'$ is bipartite and $\Lambda^{[1]}_{X} = \emptyset$,
the ground state of $\hat{H}^{f,X}$ will be degenerate.
For example, for the Creutz ladder in \Fig{fig:degenerate_GS}(a),
let $(\hat{B}_{j,\beta_{j}}^{1, \boxtimes} )^\dagger$ create a CLS of \mbox{spin-1} boson, 
it is easy to see that the following two states both have zero energy:
\begin{subequations}
	\begin{align}
		&\text{Tr} \prod_{j=1}^N \Bigg[\sum_{\beta_j} M^{\beta_j} \left(\hat{B}_{j,\beta_{j}}^{1, \boxtimes} \right)^\dagger  \Bigg] | \text{vac} \rangle,\\
	&\prod_{\ell=1}^{N/2} \Bigg[ \left(\hat{B}_{2\ell,0}^{1, \boxtimes} \right)^\dagger \left(\hat{B}_{2\ell,0}^{1, \boxtimes} \right)^\dagger - 2 \left(\hat{B}_{2\ell,1}^{1, \boxtimes} \right)^\dagger \left(\hat{B}_{2\ell,-1}^{1, \boxtimes} \right)^\dagger \Bigg] | \text{vac} \rangle.
	\end{align}
\end{subequations}
These two states are depicted in \Fig{fig:degenerate_GS}(a).
The first ``nontrivial" state is a linear combination of FPSs, while the second state is a product state.
A similar thing happens in spin-2 bosons loaded on the checkerboard lattice, see \Fig{fig:degenerate_GS}(b).
In such cases, though the Hamiltonians $\hat{H}^{f,X}$ do not exhibit any nontrivial phases due to the degeneracy,
the ``nontrivial" ground state (= the state which is a linear combination of FPSs) can always be regarded as the unique ground states of 
some other (usually more complicated) parent Hamiltonians~\cite{perez2007peps},
and hence the classification of such ``nontrivial" states from the viewpoint of SPT phases still makes sense.
For example,
let $|\text{GS}_{2,\includegraphics[width=0.011\textwidth]{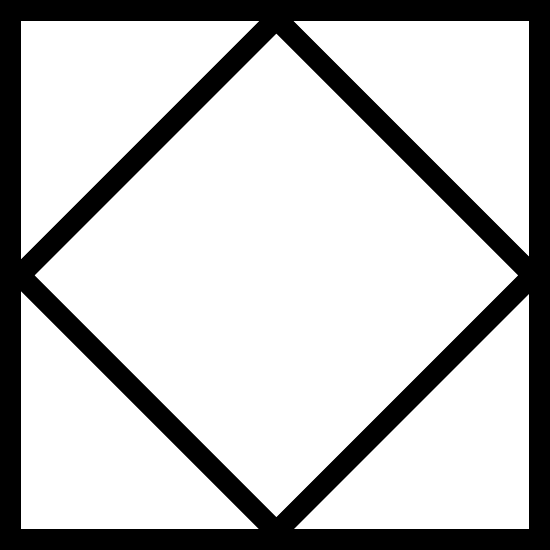}}\rangle$
be the ``nontrivial" ground state of spin-2 bosons
on the checkerboard lattice.
Following
the discussion in Sec.~\ref{Sec_GS_SPT},
by properly chosing the symmetry center and the mirror planes,
one can show that 
$|\text{GS}_{2,\includegraphics[width=0.011\textwidth]{checkerboard_symbol.pdf}}\rangle$
is in an SPT phase protected by $D_2$;
see \Fig{fig:D2_checkerboard}.

\section{Lieb-Schultz-Mattis (LSM) theorems and SPT phases}
\label{App_LSM}

LSM theorems are a class of no-go theorems on
the ``ingappability" of certain quantum systems. 
These theorems
ensure that a spin system with 
a \textit{half-odd-integer} spin
per unit cell can never have 
a unique gapped ground state, 
if certain $\text{on-site}\times\text{spatial}$ symmetry is present.
The original LSM theorem~\cite{LSM1961, LSM1986} is about
1D systems with 
the combination of SO(3)
and translation symmetry
[denote the symmetry group as $\text{SO(3)}\times\text{trn}$].
The theorem is then extended to higher dimensions~\cite{PhysRevLett.84.1535, PhysRevB.69.104431, PhysRevX.10.031008}.

Recently, more symmetries other than $\text{SO(3)}\times\text{trn}$
have been found to render the ingappability~\cite{PhysRevB.83.035107,PhysRevB.93.104425, parameswaran2013topological, watanabe2015filling, ogata2019lieb, ogata2020general, yao2020twisted, PhysRevB.78.054431}.
For example, 
it is now known that the combination of 
time-reversal (TR)
and \textit{site-centered} reflection symmetry
(denote the symmetry group as
$\text{TR}\times D_1$) in 1D systems with 
a half-odd-integer spin per unit cell
also guarantees the ingappability~\cite{PhysRevB.93.104425, ogata2019lieb, ogata2020general}.
Other such symmetries in 1D include
$\mathbb{Z}_2 \times \mathbb{Z}_2 \times \text{trn}$,
$\mathbb{Z}_2 \times \mathbb{Z}_2 \times D_1$,
TR$\times$trn, and so on~\cite{PhysRevB.93.104425, ogata2019lieb, ogata2020general}.
In fact, 
$\mathbb{Z}_2 \times \mathbb{Z}_2 \times\text{trn}$
and $\text{TR} \times \text{trn}$ apply to any $d \geqslant 1$ dimensions~\cite{yao2020twisted}.

\begin{figure}[h]
  \centering
  \includegraphics[width=0.5\textwidth]{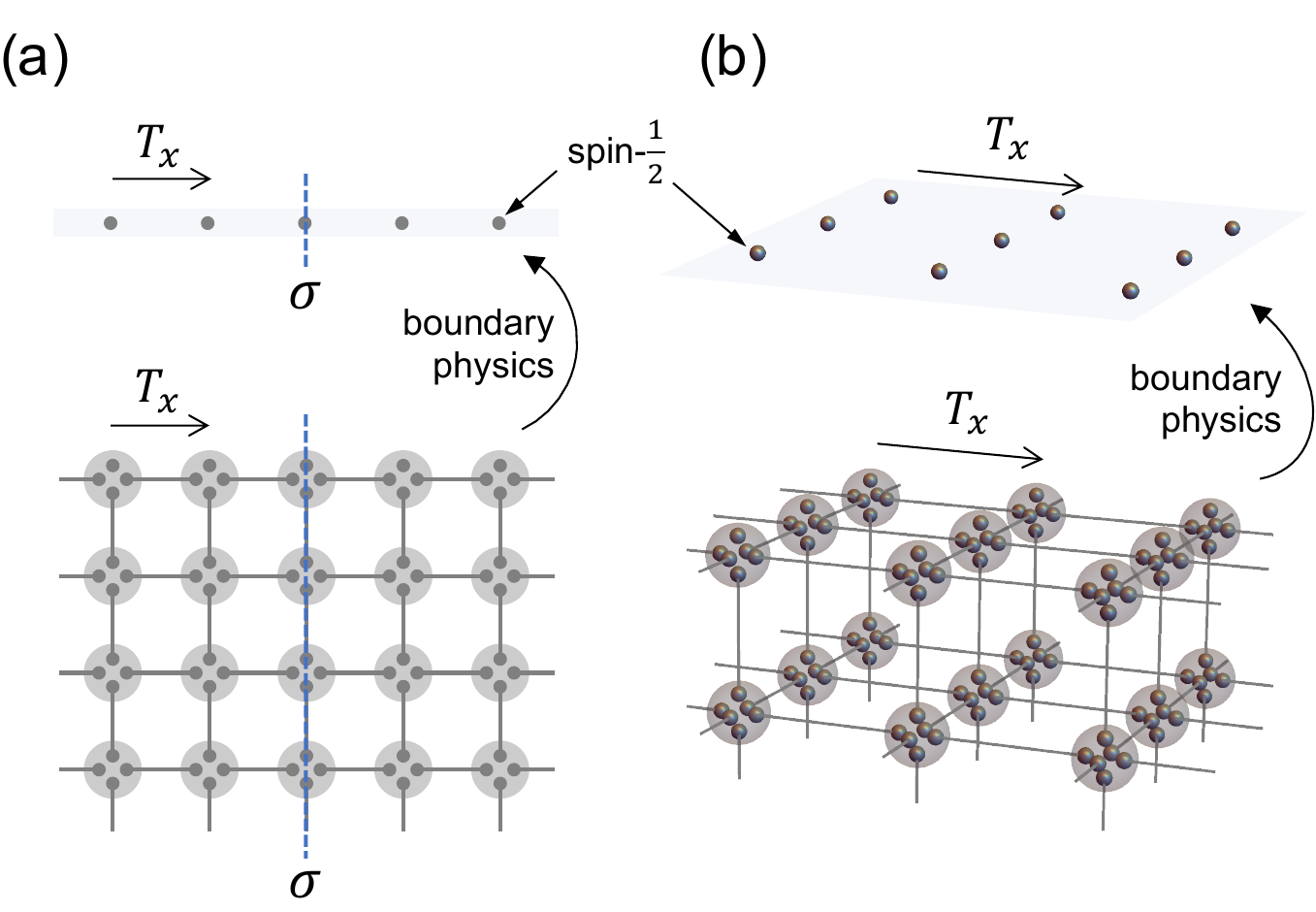}
  \caption{LSM theorems as a special case of 
constraints at the boundaries of SPT phases:
a gapless or degenerate edge state 
ensured by certain symmetry in the boundary
implies an SPT phase protected by the same symmetry in the bulk.
The translation group trn is generated by $T_x$,
while the group $D_1$ is 
generated by the site-centered reflection $\sigma$.
(a) A \mbox{spin-1/2} simple linear chain
can be regarded as the edge of $|\text{VBS}_{2,\Box}\rangle$.
(b) A spin-1/2 system on a square lattice
can be regarded as the surface of 
$|\text{VBS}_{3,\includegraphics[width=0.013\textwidth]{cube_symbol.pdf}}\rangle$.}
  \label{fig:LSM}
\end{figure}

The boundary (edge state) of an SPT phase 
is usually gapless or degenerate~\footnote{As already noted in Appendix~\ref{App_edge},
for SPT phases protected by crystalline symmetry alone
in $d<3$ dimensions, there are no anomalous edge states~\cite{PTBO_2010,PhysRevB.96.205106}.},
coincident with the statements of the LSM theorems.
Indeed, the LSM theorems are, in a precise sense, 
a special case of 
constraints at the boundaries of SPT phases~\cite{PhysRevX.6.041068}.
In other words, a gapless or degenerate edge state 
ensured by certain symmetry in the boundary
implies an SPT phase protected by the same symmetry in the bulk,
which is known as the bulk-boundary correspondence.
For example, a \mbox{spin-1/2} simple linear chain
can be thought of as the edge of $|\text{VBS}_{2,\Box}\rangle$, the spin-2
VBS state on a square lattice.
Similarly, a spin-1/2 system on a square lattice
can be regarded as the surface of 
$|\text{VBS}_{3,\includegraphics[width=0.013\textwidth]{cube_symbol.pdf}}\rangle$,
the spin-3 VBS state on a cubic lattice;
see \Fig{fig:LSM}.
The ingappability of a spin-1/2 
simple linear chain
due to the $\text{TR}\times D_1$ or
$\mathbb{Z}_2 \times \mathbb{Z}_2 \times D_1$
symmetry implies that
$|\text{VBS}_{2,\Box}\rangle$ is in an SPT phase protected
by $\text{TR}\times D_1$ 
or $\mathbb{Z}_2 \times \mathbb{Z}_2 \times D_1$.
Similarly, one can also show that 
$|\text{VBS}_{2,\Box}\rangle$ and
$|\text{VBS}_{3,\includegraphics[width=0.013\textwidth]{cube_symbol.pdf}}\rangle$
are in an SPT phase protected by 
$\mathbb{Z}_2 \times \mathbb{Z}_2 \times \text{trn}$
and 
$\text{TR} \times \text{trn}$,
as summarized in Table~\ref{tab:LSM}.
Note that the edge of $|\text{VBS}_{3,\triangle}\rangle$
is a \mbox{spin-1} chain, so that the LSM theorems 
do not apply.

\FloatBarrier 

\bibliography{Ref_Spin-1}

\end{document}